\documentclass[auth]{aa} 
\usepackage{graphicx}
\usepackage{longtable,rotating}
\usepackage{booktabs}
\usepackage{amsmath}
\usepackage{natbib}
\usepackage{placeins}
\usepackage{color}
\usepackage[normalem]{ulem} 
\usepackage{soul} 

\usepackage{txfonts}
\def\ltsima{$\; \buildrel < \over \sim \;$}
\def\simlt{\lower.5ex\hbox{\ltsima}}

\begin{document}
    \title{Properties of the dense core population in Orion~B as seen by the Herschel\thanks{Herschel is an ESA space observatory
           with science instruments provided by European-led Principal Investigator consortia and with important participation from NASA.} 
           Gould Belt survey
           }

   \subtitle{}

  \author{
           V. K\"onyves\inst{1,2}
          \and
           Ph. Andr\'e\inst{1}
	  \and 
	   D. Arzoumanian\inst{3,4}
	  \and
	   N. Schneider\inst{5}
          \and	  
	   A. Men'shchikov\inst{1}
          \and 
	   S. Bontemps\inst{6}
          \and 
	   B. Ladjelate\inst{7}
          \and 
	   P. Didelon\inst{1}	   
	  \and 
	   S. Pezzuto\inst{8}
	  \and
	   M. Benedettini\inst{8}
          \and 
	   A. Bracco\inst{9}
	  \and
           J. Di Francesco\inst{10,11}
          \and
	   S. Goodwin\inst{12}
	  \and
	   K. L. J. Rygl\inst{13}
	  \and
	   Y. Shimajiri\inst{1}
	  \and 
	   L. Spinoglio\inst{8}
	  \and
	   D. Ward-Thompson\inst{2}
	  \and
	   G. J. White\inst{14,15} 
	  } 

   \institute{Laboratoire d'Astrophysique (AIM), CEA/DRF, CNRS, Universit\'{e} Paris-Saclay, Universit\'{e} Paris Diderot, Sorbonne Paris Cit\'{e}, 91191 Gif-sur-Yvette, France
             \email{vkonyves@uclan.ac.uk, pandre@cea.fr}
         \and
	     Jeremiah Horrocks Institute, University of Central Lancashire, Preston PR1 2HE, UK
         \and
	     Department of Physics, Graduate School of Science, Nagoya University, Nagoya 464-8602, Japan
         \and
	     Instituto de Astrof\'isica e Ci{\^e}ncias do Espa\c{c}o, Universidade do Porto, CAUP, Rua das Estrelas, PT4150-762 Porto, Portugal
	 \and
             I. Physik. Institut, University of Cologne, Cologne, Germany 
	 \and
	     Laboratoire d'Astrophysique de Bordeaux - UMR 5804, CNRS - Universit\'e Bordeaux 1, BP 89, 33270 Floirac, France
	 \and        
             Instituto de Radioastronom\'{i}a Milim\'etrica (IRAM), Av. Divina Pastora 7, N\'ucleo Central, 18012 Granada, Spain	 
	 \and
             Istituto di Astrofisica e Planetologia Spaziali-INAF, Via Fosso del Cavaliere 100, I-00133 Roma, Italy
         \and
	     Laboratoire de Physique de l'\'Ecole Normale Sup\'erieure, ENS, Universit\'e PSL, CNRS, Sorbonne Universit\'e, Universit\'e de Paris, Paris, France
         \and
             Department of Physics and Astronomy, University of Victoria, P.O. Box 355, STN CSC, Victoria, BC, V8W 3P6, Canada
         \and
             National Research Council Canada, 5071 West Saanich Road, Victoria, BC, V9E 2E7, Canada
         \and
	     Department of Physics and Astronomy, University of Sheffield, Hounsfield Road, Sheffield S3 7RH, UK
	 \and
             INAF--Istituto di Radioastronomia, and Italian ALMA Regional Centre, Via Gobetti 101, I-40129 Bologna, Italy
         \and
	     Department of Physics and Astronomy, The Open University, Walton Hall, Milton Keynes, MK7 6AA, UK
	 \and    
	     RAL Space, STFC Rutherford Appleton Laboratory, Chilton, Didcot, Oxfordshire, OX11 0QX, UK
             }

   \date{Received 30 November 2018; Accepted 29 September 2019}

   \abstract{
    We present a detailed study of the Orion~B molecular cloud complex ($d\sim400$\,pc), which was imaged with the PACS and SPIRE photometric cameras 
    at wavelengths from 70~$\mu$m to 500~$\mu$m as part of the {\it Herschel} Gould Belt survey (HGBS). 
    We release new high-resolution maps of column density and dust temperature for the whole complex, 
    derived in the same consistent manner as for other HGBS regions. 
    In the filamentary subregions NGC~2023 and 2024, NGC~2068 and 2071, and L1622, a total of 1768 starless dense cores were identified based on 
    {\it Herschel} data, 490--804 ($\sim28-45$\%) of which are self-gravitating prestellar cores that will likely form stars in the future. 
    A total of 76 protostellar dense cores were also found. 
    The typical lifetime of the prestellar cores was estimated to be $t_{\rm pre}^{\rm OrionB} = 1.7_{-0.6}^{+0.8}\,$Myr. 
    The prestellar core mass function (CMF) derived for the whole sample of prestellar cores peaks at $\sim$0.5\,$M_\odot$ 
    (in d$N$/dlog$M$ format) and is consistent with a power-law with logarithmic slope --1.27 $\pm 0.24$ at the high-mass end, 
    compared to the Salpeter slope of $-1.35$. 
    In the Orion~B region, we confirm the existence of a transition in prestellar core formation efficiency (CFE) around a fiducial value $A_{\rm V}^{\rm bg} \sim 7$\,mag 
    in background visual extinction, which is similar to the trend observed with {\it Herschel} in other regions, such as the Aquila cloud.
    This is not a sharp threshold, however, but a smooth transition between a regime with very low prestellar CFE at $A_{\rm V}^{\rm bg} < 5$ and a regime 
    with higher, roughly constant CFE at $A_{\rm V}^{\rm bg} \gtrsim 10$.
    The total mass in the form of prestellar cores represents only a modest fraction ($\sim 20$\%) of the dense molecular cloud gas above $A_{\rm V}^{\rm bg} \gtrsim 7$\,mag. 
    About 60--80\% of the prestellar cores are closely associated with filaments, and this fraction increases up to $> 90\% $ when a more complete sample of 
    filamentary structures is considered.
    Interestingly, the median separation observed between nearest core neighbors corresponds to the 
    typical inner filament width of $\sim 0.1$~pc, which is commonly observed in nearby molecular clouds, including Orion~B.
    Analysis of the CMF observed as a function of background cloud column density shows that the most massive prestellar cores are spatially segregated 
    in the highest column density areas, and suggests that both higher- and lower-mass prestellar cores may form in denser filaments. 
    }


   \titlerunning{\emph{Herschel} Gould Belt survey results in Orion~B}

   \maketitle
%

\section{Introduction}\label{sec:intro}

Following initial results from the {\it Herschel} Gould Belt survey (HGBS) \citep[see][]{Andre+2010}, the first complete study of dense cores 
based on {\it Herschel} data in the Aquila cloud complex confirmed, with high number statistics, that the prestellar core mass function (CMF) 
is very similar in shape to the initial mass function (IMF) of stellar systems and that there is approximately one-to-one mapping between these 
two mass distributions \citep{Konyves+2015}. 
Similar trends between the prestellar CMF and the stellar IMF can also be seen in the Taurus L1495 \citep{Marsh+2016}, Corona Australis \citep{Bresnahan+2018}, 
Lupus (Benedettini et al. 2018), and Ophiuchus \citep{Ladjelate+2019} clouds. These ``first-generation'' papers from the HGBS include complete 
catalogs of dense cores and also demonstrate that the spatial distribution of dense cores is strongly correlated with the underlying, ubiquitous filamentary 
structure of molecular clouds.
Molecular filaments appear to present a high degree of universality \citep[][]{Arzoumanian+2011, Arzoumanian+2019, KochRosolowsky2015} and play a key role 
in the star formation process. Additional specific properties of filaments based on {\it Herschel} imaging data for closeby ($d < 3\,$kpc) molecular clouds 
are presented in \citet{Hennemann+2012} on the DR21 ridge in Cygnus~X, in \citet{Benedettini+2015} on Lupus, and in \citet{Cox+2016} on the Musca filament.
 
Overall, the combined properties of  filamentary structures and dense cores observed with {\it Herschel} in nearby clouds support a scenario in which the 
formation of solar-type stars occurs in two main steps \citep{Andre+2014}. First, the dissipation of kinetic energy in large-scale magneto-hydrodynamic 
(MHD) compressive flows generates a quasi-universal web of filamentary structures in the cold interstellar medium (ISM); 
second, the densest molecular filaments fragment into prestellar cores by gravitational instability, and then evolve into protostars. 
Furthermore, our {\it Herschel} findings on the prestellar core formation efficiency (CFE) in the Aquila molecular cloud  \citep{Konyves+2015}, 
coupled with the relatively small variations in the star formation efficiency (SFE) per unit mass of dense gas observed between nearby clouds and 
external galaxies \citep[e.g.,][]{Lada+2012, Shimajiri+2017}, suggest that the filamentary picture of star formation proposed for nearby galactic 
clouds may well apply to extragalactic giant molecular clouds (GMCs) as well \citep[][]{Andre+2014, Shimajiri+2017}. 

Here, we present a detailed study of the dense core population in the Orion~B molecular cloud based on HGBS data. 
With little doubt, Orion is the best studied star-forming region in the sky, and also the nearest site with ongoing low- and high-mass 
star formation \citep[for an overview see][]{Bally2008}. The complex is favorably located for us in the outer boundary of the Gould Belt 
where confusion from foreground and background gas is minimized; it lies toward the outer Galaxy and below the galactic plane at an average distance 
of about 400~pc from the Sun \citep[][]{Menten+2007, Lallement+2014, Zucker+2019}, which is the distance adopted in this paper. 

\begin{figure*}[!h]
 \centering
  \begin{minipage}{1.0\linewidth}
  \begin{center} 
   \resizebox{0.85\hsize}{!}{\includegraphics[angle=0]{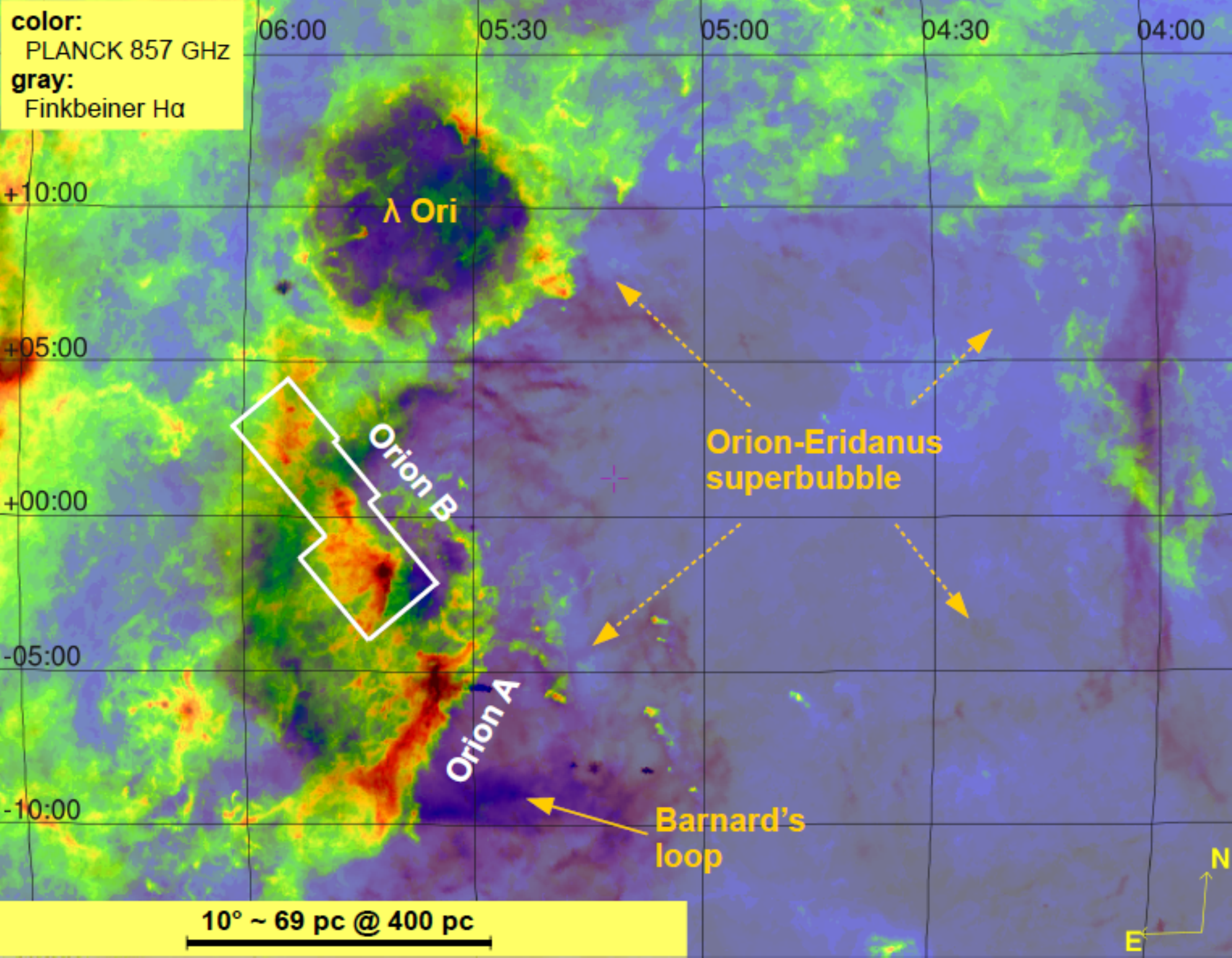}}
  \end{center}
  \end{minipage}
  \caption   
  {Large-scale view of the Orion-Eridanus superbubble (which reaches beyond Barnard's Loop) and star-forming clouds along its wall. 
   The color background shows {\it Planck} 857~GHz (350~$\mu$m) emission (Public Data Release 2, PR2), and the overplotted gray features  
   represent H$\alpha$ emission \citep{Finkbeiner2003}. The region analyzed in this paper is outlined in white. 
   (PR2 is described in the Planck 2015 Release Explanatory Supplement which is available at https://wiki.cosmos.esa.int/planckpla2015/)}
  \label{fig_bigOri}%
\end{figure*}

Most of the famous clouds in the Orion region  (Fig.~\ref{fig_bigOri}) seem to scatter along the wall of the so-called Orion-Eridanus superbubble 
\citep[``Orion's Cloak'', see e.g.,][]{Bally2008, Ochsendorf+2015, Pon+2016}, which was created by some of Orion's most massive members that 
died in supernova explosions. This enormous superbubble spans $\sim$300~pc in diameter \citep{Bally2008} with some nested shells inside, among which, 
for example, the Barnard's Loop bubble and the $\lambda$ Ori bubble \citep[see e.g.,][]{Ochsendorf+2015}.
Both the Orion~A (Polychroni et al., in prep.) and the Orion~B cloud (this paper) have been observed by the HGBS.

\begin{figure*}[!h]
 \centering
  \begin{minipage}{1.0\linewidth}
   \resizebox{0.499\hsize}{!}{\includegraphics[angle=0]{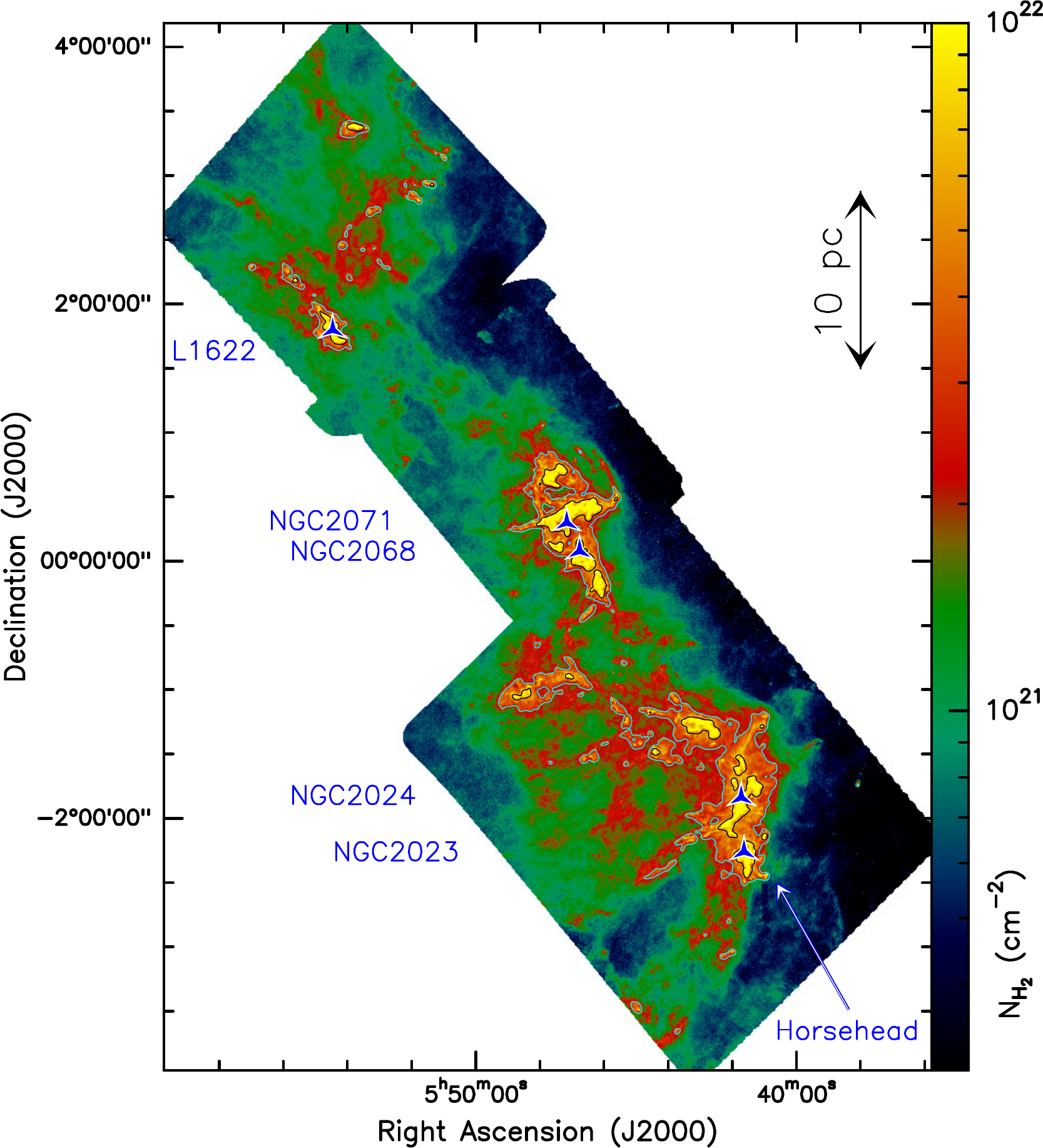}}
   \resizebox{0.491\hsize}{!}{\includegraphics[angle=0]{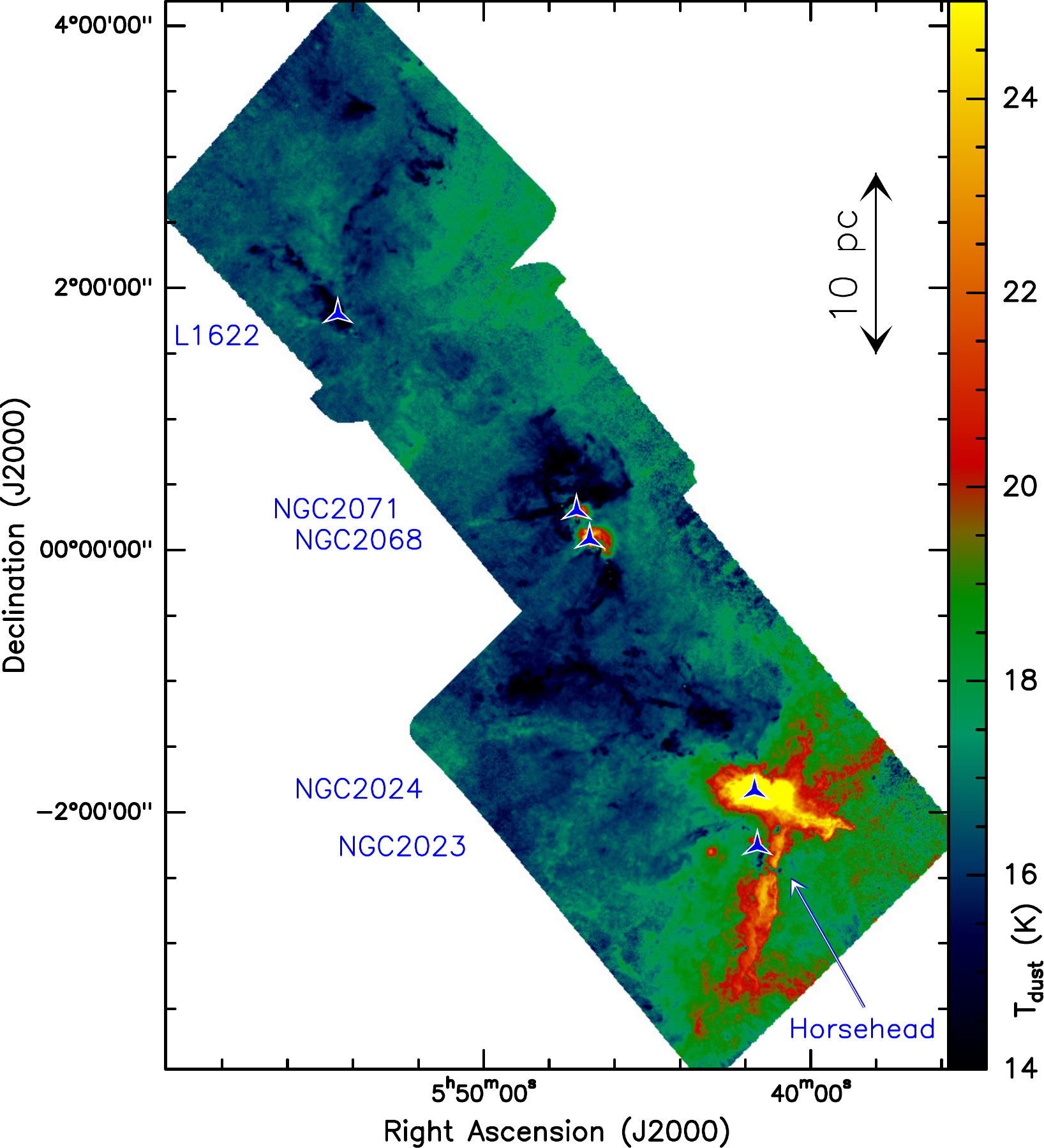}}
  \end{minipage}
   \caption{{\bf Left:} H$_2$ column density map of Orion~B at $18.2\arcsec$ angular resolution, derived from HGBS data.
   The contours correspond to $A_{\rm V} \sim 3$ (cyan) and 7\,mag (black) in a column density map smoothed to a resolution of $2\arcmin$.
   {\bf Right:} Dust temperature map of the Orion~B region at 36.3\arcsec resolution, as derived from the same data set.
   Blue stars mark the NGC~2023 and 2024, NGC~2068 and 2071 HII regions and reflection nebulae, and the L1622 cometary cloud. 
   The position of the Horsehead Nebula is also shown.   
   }
  \label{fig_T_NH2_maps}%
\end{figure*}

Within the Orion~B cloud complex discussed in here, the most prominent features are the clumps of gas and dust associated with the NGC~2024 H~II region, 
the NGC~2023, 2068, and 2071 reflection nebulae, and the Horsehead Nebula, as well as the Lynds~1622 cometary cloud in the north of our observed region 
(see Fig.~\ref{fig_T_NH2_maps}). 
The cometary, trunk-like column density features and their alignments in the northern region of the field strongly suggest that 
L1622 --and the clouds north of it-- interact with the H$\alpha$ shell of Barnard's Loop, thus 
these northern clouds may also be at a closer distance of $d\sim$170--180~pc \citep{Bally2008,Lallement+2014}, in front of the rest of Orion~B. 
However, as the distance to Barnard's Loop is still uncertain \citep[see e.g.,][and references therein]{Ochsendorf+2015}, for simplicity 
and until further studies, we adopt a uniform distance of 400~pc for deriving physical properties in the whole observed field. 
The cores identified in the northern area are nevertheless clearly marked in the catalogs attached to this paper.

From first-look {\it Herschel} GBS results in Orion~B, \citet{Schneider+2013} concluded that the deviation point of the column density probability 
density function (PDF) from a lognormal shape is not universal and may vary from cloud to cloud.
The lognormal part of the PDF can be broadened by external compression, which is consistent with numerical simulations.
The appearance of a power-law tail at the higher density end of the PDFs may be due to the global collapse of filaments and/or to core formation. 

SCUBA-2 observations of Orion~B allowed  \citet{HKirk+2016a} to identify and characterize $\sim$900 dense cores. 
Their clustering and mass segregation properties were studied by \citet{HKirk+2016b} and \citet{Parker2018}.
Using spatially- and spectrally-resolved observations from the IRAM-30m telescope \citep{Pety+2017}, the cloud was segmented in regions of similar 
properties by \citet{Bron+2018} and its turbulence was characterized by \citet{Orkisz+2017}, who compared it with star formation properties derived 
from {\it Spitzer} and {\it Chandra} counts of young stellar objects (YSOs) \citep{Megeath+2012, Megeath+2016}.

The present paper is organized as follows. 
Section~\ref{sec:obs_data} provides details about the {\it Herschel} GBS observations of the Orion~B region and the data reduction.
Section~\ref{sec:res_analys} presents the dust temperature and column density maps derived from {\it Herschel} data, describes the filamentary structure 
in these maps, and explains the extraction and selection of various dense cores and their observed and derived properties. 
In Sect.~\ref{sec:discuss}, we compare the spatial distributions of cores and filaments, discuss the mass segregation of dense cores, 
and interpret the typical core-to-core and core-to-filament separations. 
We also present further observational evidence of a column density transition for prestellar core formation, along with estimates of the typical prestellar 
core lifetime and a comparison of the prestellar CMF in Orion~B with the stellar IMF. Finally, Sect.~\ref{sec:conclusions} concludes the paper. 

\section{Herschel observations and data reduction}\label{sec:obs_data}

The {\it Herschel} Gould Belt survey observations of the Orion~B complex were taken between 25 September 2010 and 13 March 2011 with the {\it Herschel} 
space observatory \citep{Pilbratt2010}. 
The SPIRE and PACS parallel-mode scan maps covered a common $\sim$19~deg$^2$ area with both SPIRE \citep{Griffin2010} and PACS \citep{Poglitsch2010}. 
With one repetition in two orthogonal observing directions, three tiles were imaged (OBSIDs: 1342205074, 1342205075, 1342215982, 1342215983, 1342215984, 
1342215985) with a scanning speed of 60$\arcsec$s$^{-1}$. The total duration of the mapping was 19.5 hours. 
The above strategy is similar for all parallel-mode SPIRE and PACS observations of the HGBS. 

The PACS data reduction was done with the same tools and in the same way as for the Aquila cloud \citep[see Sect.~3 of][]{Konyves+2015}. 
The PACS 70 and 160\,$\mu$m output fits files were produced in Jy/3\arcsec-pixel units. 

The SPIRE 250~$\mu$m, 350~$\mu$m, and 500~$\mu$m data were also reduced as presented in \citet[][]{Konyves+2015} with the exception of an additional 
correction. The 250~$\mu$m data were saturated in the course of the observations at the center of the NGC~2024 region. In order to correct for this 
local saturation, a small patch was re-observed in SPIRE photometer mode (OBSID: 1342239931). 
The re-observed stamp could not be directly mosaicked --or reduced together-- with the original large map, because the edges of the small patch remained 
strongly imprinted in the combined map. 
To get around this problem, we 1) reprojected the re-observed saturation stamp to the same pixel grid as that of the original map, after which we 
2) correlated the original pixel values with those of the re-observed patch in the commonly covered spot (excluding the noisy map edges of the small patch). 
Then, 3) this correlation was fitted with a linear relation and 4) the originally saturated pixels were assigned the linear fit-corrected values. 
No artifact was left in the resulting 250~$\mu$m map. 

The three tiles observed with both PACS and SPIRE were separately reduced, which allowed us to derive more realistic temperature and column density maps 
(see Sect.~\ref{sec:cd_t_maps} below).

\section{Results and analysis}\label{sec:res_analys}

\subsection{Dust temperature and column density maps}\label{sec:cd_t_maps}

The Orion~B region observed with {\it Herschel}'s SPIRE and PACS was covered by three tiles in order to follow the elongated dense structures.
The final maps (see e.g., Fig.~\ref{fig_T_NH2_maps}) span $\sim$8\degr\,in declination. For constructing temperature and column density maps, we decided, 
therefore, to add Planck offsets separately to these three tiles (at 160--500\,$\mu$m), as the offsets may not be the same in the southern and northern 
ends of the maps.   
The derivation of the zero-level offsets for the $Herschel$ maps was made as described in Bracco et al. (in prep.) from comparison with {\it Planck} 
and {\it IRAS} data \citep[see also][]{Bernard+2010}. After adding separate zero-level offsets, the three tiles at a given wavelength were then stitched 
together into one big mosaic. The applied offsets are listed in Table~\ref{tab_planck}.
\begin{table}\small\setlength{\tabcolsep}{4.5pt}
 \caption{Applied zero-level offsets in MJy sr$^{-1}$.}
 \label{tab_planck}      
 {\renewcommand{\arraystretch}{1.2}        
  \begin{tabular}{l|c|c|c|c}  
  \hline
  \hline
		& 160\,$\mu$m	& 250\,$\mu$m	& 350\,$\mu$m	& 500\,$\mu$m  \\
  \hline
  Southern tile	& 67.3		& 60.8		& 32.0		& 13.1         \\
  \hline
  Middle tile	& 57.9		& 48.4		& 25.8		& 10.8         \\
  \hline
  Northern tile	& 53.1		& 51.1		& 27.9		& 11.7         \\
  \hline
  \end{tabular}
 }
\end{table}

We smoothed all $Herschel$ images to the resolution of the 500\,$\mu$m map (36.3$''$) and reprojected them to the same pixel grid. We then 
constructed a dust temperature ($T_{\rm d}$) and a column density ($N_{\rm H{_2}}$) map from pixel-by-pixel spectral energy distribution (SED) 
fitting using a modified blackbody function \citep[cf.][]{Konyves+2015}. As part of a standard procedure, we derived two column density maps: 
a ``low-resolution'' map at the $36.3\arcsec $ resolution of the SPIRE 500~$\mu$m data and a ``high-resolution'' map at the $18.2\arcsec $ resolution 
of the SPIRE 250~$\mu$m data. The detailed procedure used to construct the high-resolution column density map is described in Appendix A of 
\citet{Palmeirim+2013}. 
As in previous HGBS papers, we here adopt the dust opacity law, $\kappa_{\lambda} = 0.1 \times (\lambda/\rm 300\mu m)^{-\beta}$ cm$^2$/g, fixing 
the dust emissivity index $\beta$ to 2 \citep[cf.][]{Hildebrand1983}. 

Figure~\ref{fig_T_NH2_maps} (left and right) show the high-resolution column density map and dust temperature map, respectively, of the Orion~B cloud complex. 
When smoothed to 36.3$\arcsec$ resolution, the high-resolution column density map is consistent with the standard, low-resolution column density map. 
In particular, the median ratio of the two 36.3$\arcsec$ resolution maps is 1 within $\pm 5$\%.
We also checked the reliability of our high-resolution column 
density map against the {\it Planck} optical depth map of the region and near-infrared extinction maps.
The $850\, \mu$m {\it Planck} optical depth map was converted to an H$_2$ column density image using the above dust opacity law.
Then, the {\it Herschel} and {\it Planck} $N_{\rm H{_2}}$ maps were compared, showing good agreement within $\pm 5$\%. 
This confirmed the validity and accuracy of the applied zero-level offsets.  

The high-resolution $N_{\rm H{_2}}$ map was also compared to two near-infrared extinction maps. 
The first extinction map was derived from 2MASS data following \citet[][]{Bontemps+2010} and \citet[][]{Schneider+2011}, with a FWHM spatial resolution 
of $\sim $120$\arcsec$.
The other extinction map was also made from 2MASS point-source data, using the NICEST near-infrared color excess method 
\citep[][and Lombardi et al.~in prep\footnote{http://www.interstellarclouds.org/html}.]{Lombardi2009}. 
For both comparisons, the {\it Herschel} column density map was first smoothed to the resolution of the extinction maps and converted to visual extinction 
units assuming $N_{\rm H_2}\, ({\rm cm}^{-2}) = 0.94 \times 10^{21}\, A_{\rm V} \,({\rm mag }) $ \citep{Bohlin+1978}. 
We then compared the data points of the converted {\it Herschel} maps to those of the $A_{\rm V}$ maps on the same grid.  
The resulting $A_{\rm V, HGBS} / A_{\rm V, Bontemps}$ and 
$A_{\rm V, HGBS} / A_{\rm V, Lombardi}$ ratio maps are shown in Fig.~\ref{fig:ratioHGBSvsSylLom} over the field covered by Fig.~\ref{fig_T_NH2_maps}. 
The median values of both ratios over the whole area are $\sim$0.5. In the extinction range $A_{\rm V} = $\,3--7\,mag, the ratios are also similar, 
$\sim$0.6 and $\sim$0.5, respectively. 
Both extinction maps are in good agreement with each other, while the HGBS column density map indicates lower values by a factor of $< 2$.
This may be the result of a combination of various factors, such as different techniques applied to the {\it Herschel} and 2MASS datasets, 
and less reliable dust opacity assumptions in low extinction regions, 
which are more exposed to the stronger radiation field in Orion~B than in other HGBS regions \citep[cf.][]{Roy+2014}.

\subsection{Filamentary structure of the Orion~B cloud complex}\label{sec:filam}

Following \citet{Andre+2014} and \citet{Konyves+2015}, we define an interstellar filament as any elongated column-density structure in the cold ISM 
which is significantly denser than its surroundings. We adopt a minimum aspect ratio of $\sim 3$ and a minimum column density excess of $\sim 10\% $ 
with respect to the local background, that is, $\Delta N_{\rm H_2}^{\rm fil}/N_{\rm H_2}^{\rm bg} > 0.1 $, when averaged along the length of the structure.

\begin{figure*}[!ht]
 \centering
  \begin{minipage}{1.0\linewidth}
   \resizebox{0.37\hsize}{!}{\includegraphics[angle=0]{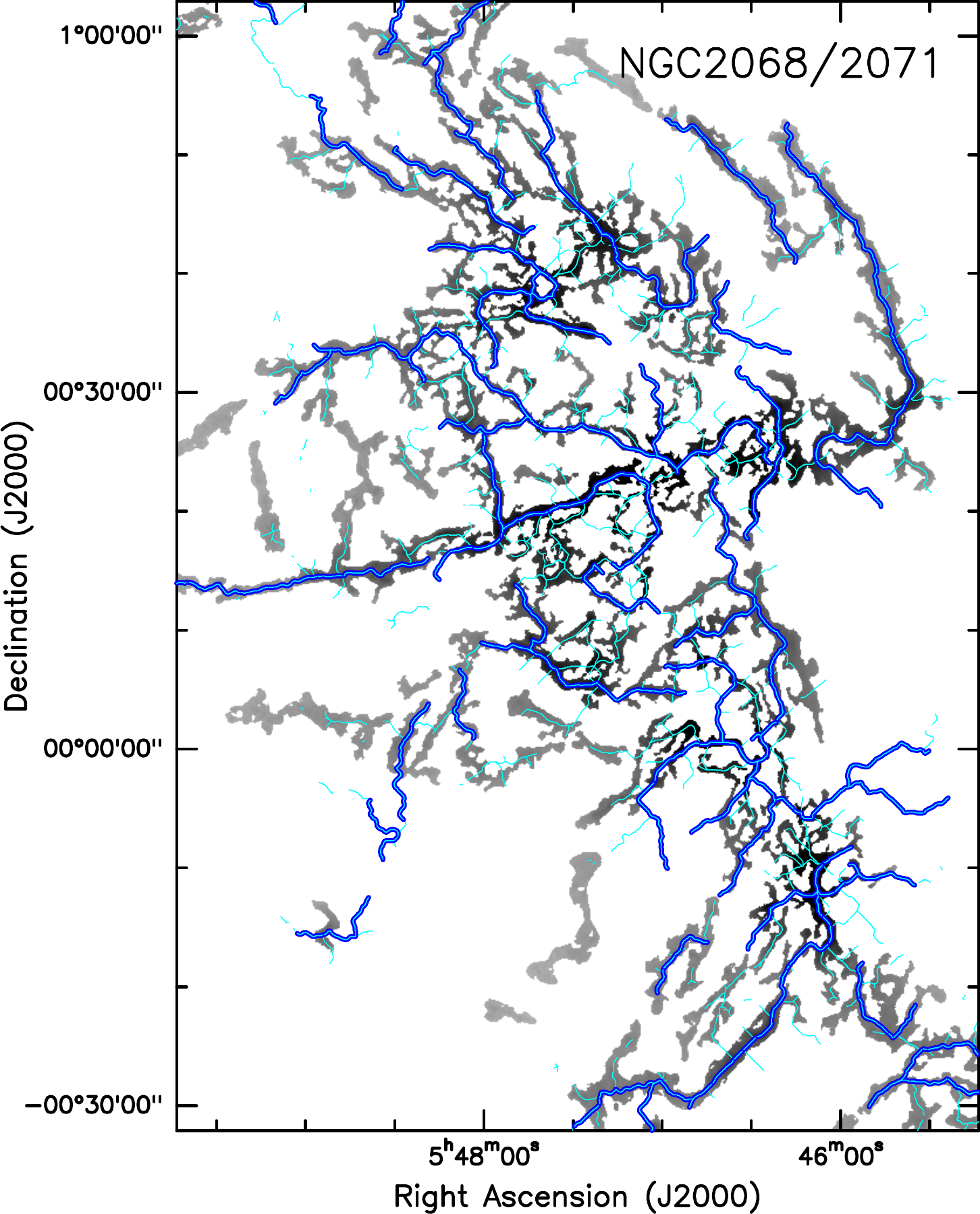}}
   \hspace{2mm}
   \resizebox{0.61\hsize}{!}{\includegraphics[angle=0]{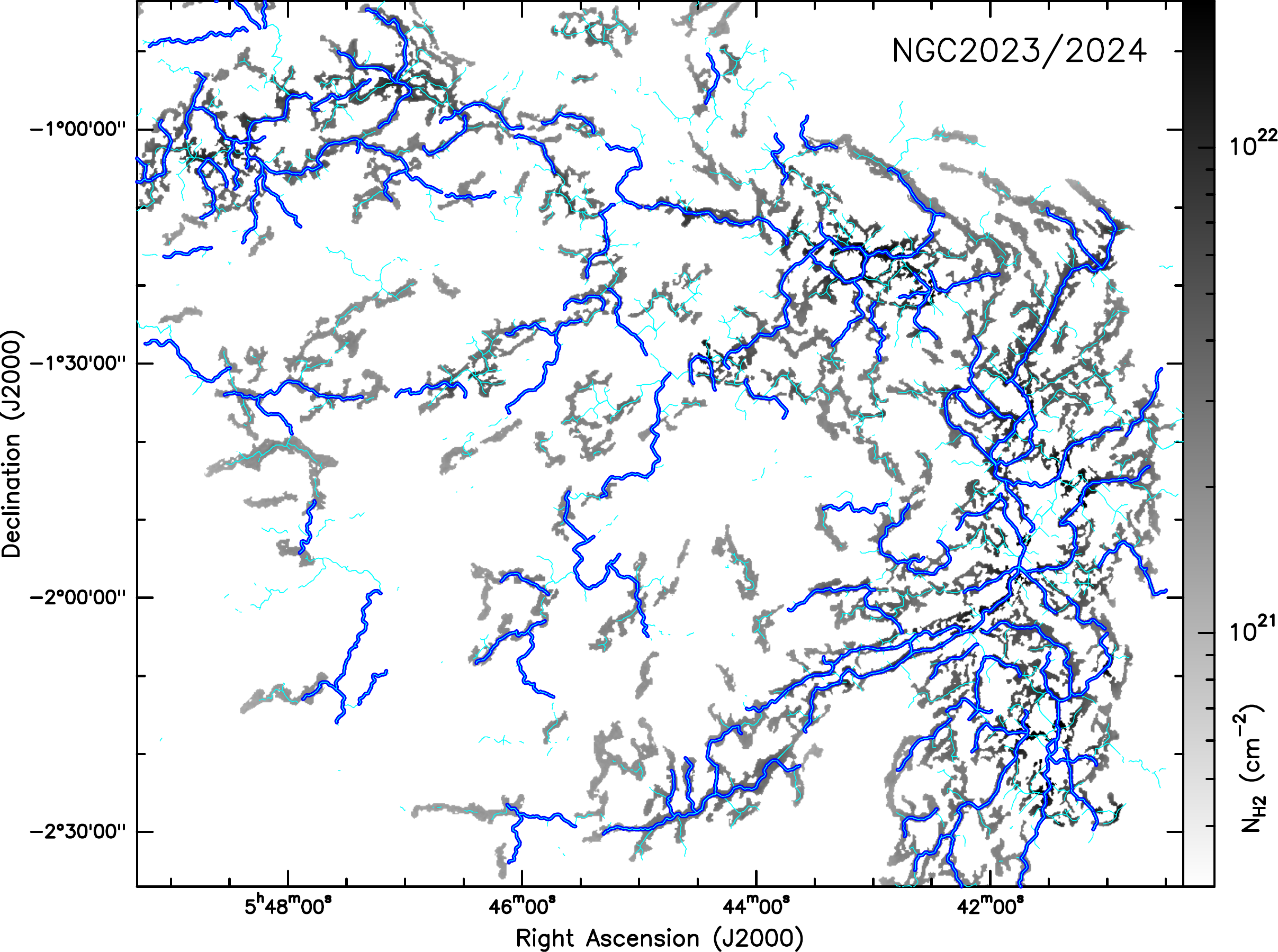}}
  \end{minipage}
   \caption{
   Filament networks identified in the subregions NGC~2068 and 2071 ({\bf left}), and NGC~2023 and 2024 ({\bf right}). 
   The gray background image displays the mask of the filamentary network traced with \textsl{getfilaments} \citep{Menshchikov2013}.
   The overplotted thick blue skeletons mark the robust filament crests identified with the DisPerSE algorithm 
   \citep[][see Sect.~\ref{sec:filam} for details]{Sousbie2011}.  
   Together with the additional thin cyan crests, these filament crests make up the entire raw sample of DisPerSE filaments.
   Within the \textsl{getfilaments} mask, the gray scale is the same in both panels, and reflects column density values in the column density map 
   (Fig.~\ref{fig_T_NH2_maps} left). Transverse angular scales up to $100\arcsec$ are shown, corresponding to $\sim$0.2~pc at $d = 400$~pc. 
   Figure~\ref{fig:fils_coresL1622} (left) shows a similar view for the L1622 cloud.  
   }
  \label{fig:fils}%
\end{figure*}

In order to trace filaments in the {\it Herschel} maps and in the column density map, we used the multi-scale algorithm \textsl{getfilaments} 
\citep{Menshchikov2013}. Briefly, by analyzing highly-filtered spatial decompositions of the images across a wide range of scales, 
\textsl{getfilaments} separates elongated clusters of connected pixels (i.e., filaments) from compact sources and non-filamentary background at 
each spatial scale, and then constructs cumulative mask images of the filamentary structures present in the original data by accumulating 
the filtered signal up to an upper transverse angular scale of interest. 
Figure~\ref{fig:fils} and Fig.~\ref{fig:fils_coresL1622} display grayscale mask images of the filamentary networks identified with 
\textsl{getfilaments} in the high-resolution column density maps of the NGC~2068 and 2071, NGC~2023 and 2024, and L1622.  
Here, we adopted an upper transverse angular scale of 100\arcsec, or $\sim$0.2~pc at $d = 400$~pc, which is consistent with the typical maximum 
FWHM inner width of filaments, considering a factor of $\sim 2$ spread around the nominal filament width of $\sim 0.1$~pc measured by 
\citet[][]{Arzoumanian+2011,Arzoumanian+2019}.    
The filament mask images obtained using an upper angular scale of 50\arcsec or $\sim$0.1~pc are very similar.

As an alternative independent method, we also employed the DisPerSE algorithm \citep{Sousbie2011}, which traces filaments by connecting saddle points 
to local maxima with integral lines following the gradient in a map. 
The filament skeletons overplotted in blue in Fig.~\ref{fig:fils} and Fig.~\ref{fig:fils_coresL1622} (left) were obtained with the DisPerSE method. 
DisPerSE was run on the standard column density map (at 36.3\arcsec resolution) with 3\arcsec/pixel scale, using a ``persistence'' threshold of 
10$^{20}$~cm$^{-2}$, roughly corresponding to the lowest rms level of background fluctuations in the map, and a maximum assembling angle 
of 50$\degr$, as recommended by \citet{Arzoumanian+2019}. 
This led to a raw sample of DisPerSE filaments. Then, to select a robust sample of filaments, raw filamentary structures 
were trimmed below the peak of the column density PDF (see Fig.~\ref{fig:cdPDF} left), and raw features shorter than $10 \times$ the HPBW 
resolution of the map were removed, as suggested by \citet{Arzoumanian+2019}. 
The resulting sample of robust DisPerSE filaments comprises a total of 238 crests shown as thick blue skeletons in Fig.~\ref{fig:fils} and 
Fig.~\ref{fig:fils_coresL1622} (left). The thin cyan crests show the additional, more tentative filaments from the raw sample extracted with DisPerSE.

Although we shall see in the following that physical results derived from either of the two methods are similar, DisPerSE tends to trace longer 
filamentary structures as a result of its assembling step, while \textsl{getfilaments} typically finds shorter structures down to the selected spatial scales. 

\subsection{Distribution of mass in the Orion~B complex}\label{sec:mass_distrib}

\begin{figure*}[]
 \centering
  \begin{minipage}{1.0\linewidth}
    \resizebox{0.50\hsize}{!}{\includegraphics[angle=0]{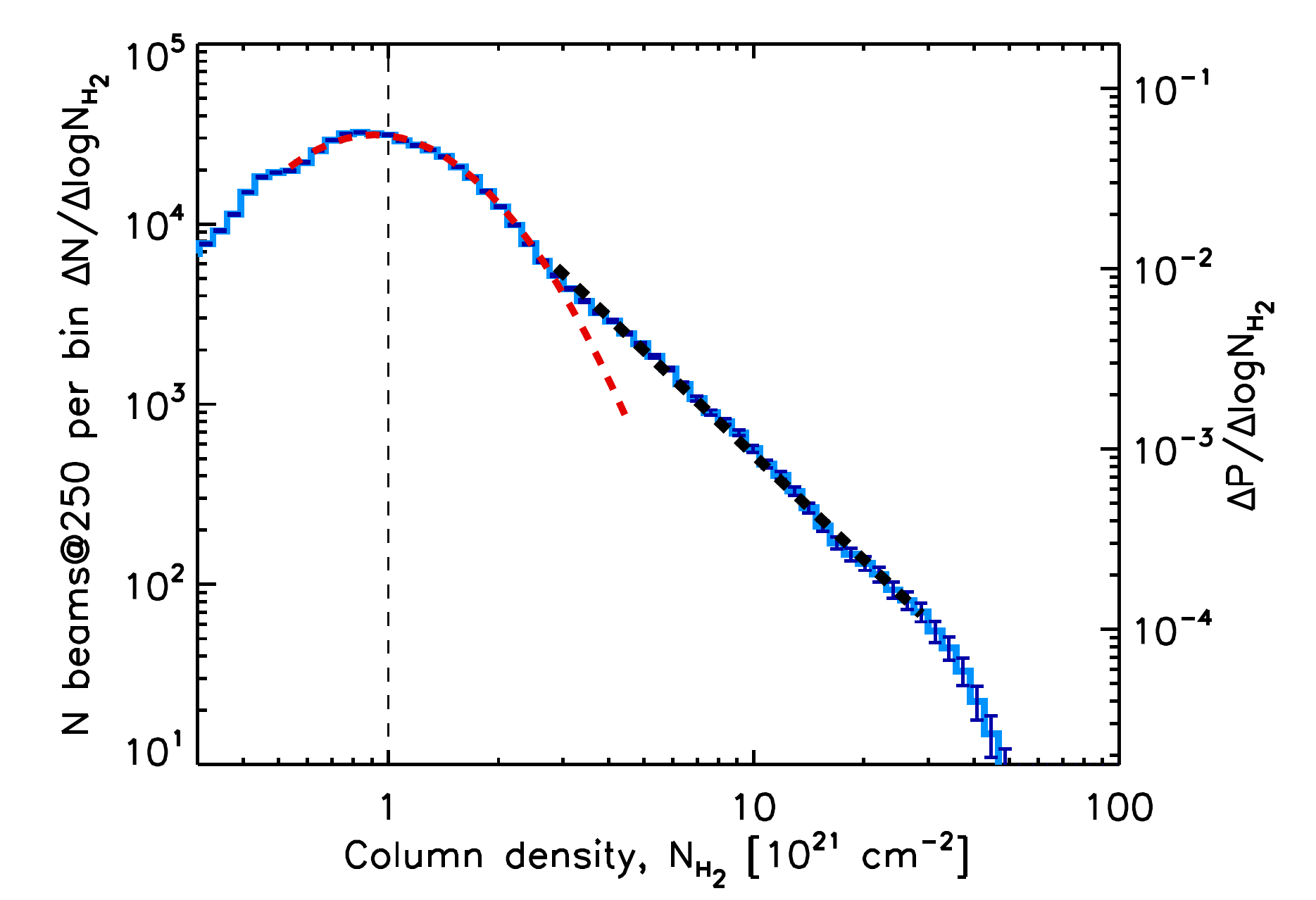}}   
    \hspace{8mm}
    \resizebox{0.50\hsize}{!}{\includegraphics[angle=0]{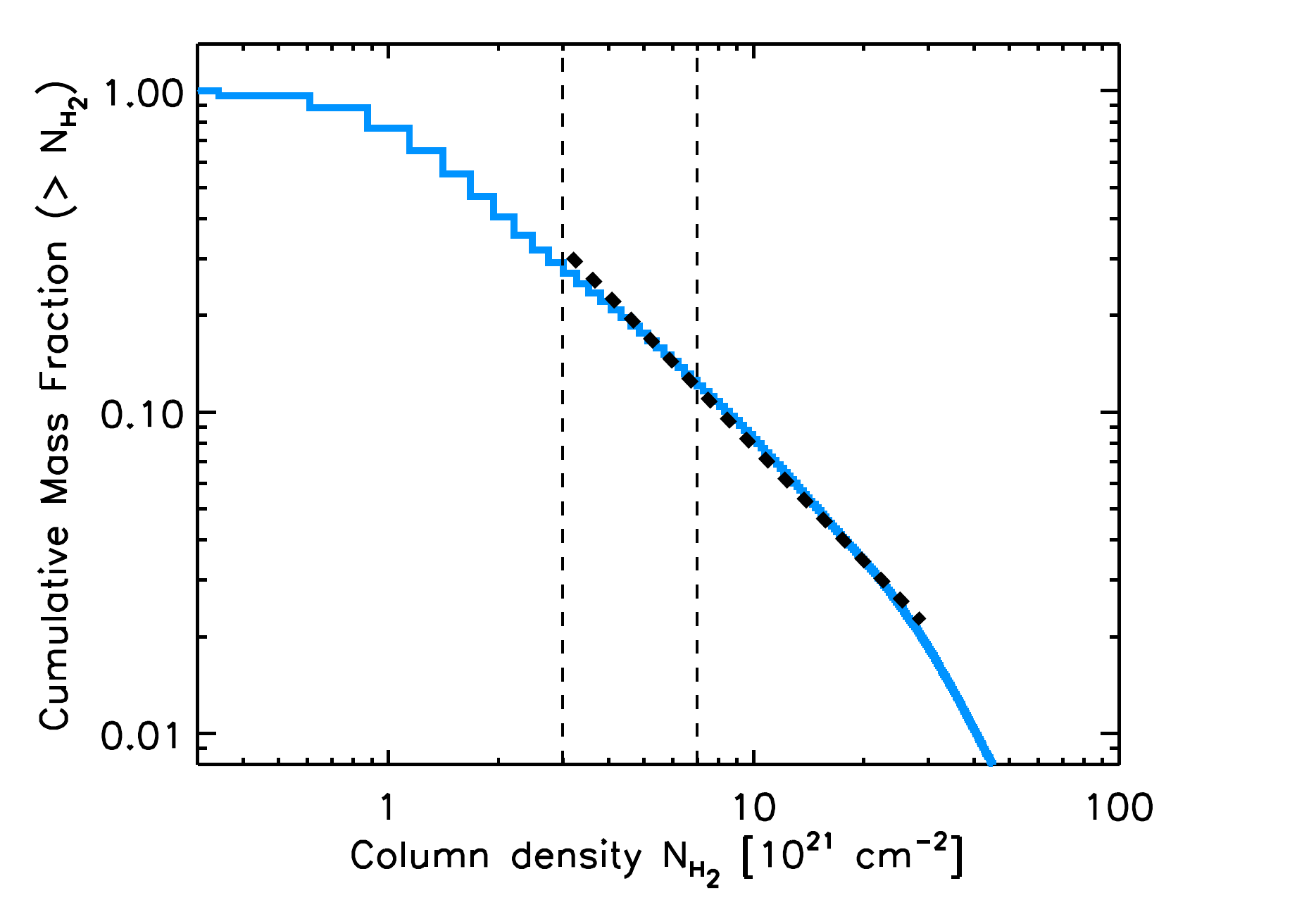}}
  \end{minipage}
  \caption{{\bf Left:} 
   High-resolution $N_{\rm H_2}$ column density PDF of Orion~B, derived from the column density map of Fig.~\ref{fig_T_NH2_maps} left. 
   The lognormal fit (dashed red curve) peaks at $0.92 \times 10^{21}\, {\rm cm}^{-2}$ in $N_{\rm H_2} $, and has a standard deviation of $\sim$0.25 in log$_{10}N_{\rm H_2}$. 
   The power-law fit to the higher column density tail at $A_{\rm V} \gtrsim 3$ (dashed black line) 
   is $\Delta N$/$\Delta$log$N_{\rm H_2} \propto N_{\rm H_2}^{-1.9 \pm 0.2}$. 
   The vertical dashed line at $A_{\rm V} \sim 1$ marks the lowest closed contour in the column density map.
   {\bf Right:} Corresponding (normalized) cumulative mass fraction as a function of column density. 
   The dense material above $A_{\rm V} \sim $~3--7\,mag represents only $\sim$30--13\% of the total cloud mass, respectively, as indicated 
   by the two dashed vertical lines. 
   A power-law fit to the cumulative mass fraction for $A_{\rm V} \gtrsim 3$ gives $M(> N_{\rm H_2}) \propto N_{\rm H_2}^{-1.1 \pm 0.2}$.
   }
  \label{fig:cdPDF}       
\end{figure*}

The probability density function of column density in the Orion~B cloud is shown in Fig.~\ref{fig:cdPDF} (left). 
For a similar $N_{\rm H_2}$-PDF plot from an earlier column density map of Orion~B based on HGBS data, see \citet[][]{Schneider+2013}. 

Throughout the paper we take advantage of the \citet{Bohlin+1978} conversion factor (see Sect.~\ref{sec:cd_t_maps}) 
according to which H$_2$ column density values in units of $10^{21}\,{\rm cm}^{-2}$ approximately correspond to $A_{\rm V}$ in magnitudes.
In our terminology we use the following column density ranges in corresponding visual extinction: ``low'' for $A_{\rm V} < 3$, ``intermediate'' for $A_{\rm V} = 3-7$, 
and ``high'' for $A_{\rm V} > 7$\,mag, which are similar to, but not exactly the same as, those defined by \citet{Pety+2017}.  

The column density PDF in Fig.~\ref{fig:cdPDF} (left) is well fitted by a log-normal distribution at low column densities and by a power-law distribution 
above the corresponding visual extinction of $A_{\rm V} \sim 3$\,mag. 
In the left panel of Fig.~\ref{fig:cdPDF}, the lowest closed contour in the column density map (Fig.~\ref{fig_T_NH2_maps} left) is marked by a 
vertical dashed line at $A_{\rm V} \sim 1$.
According to \citet{Alves+2017}, the lowest closed contour corresponds to the completeness level of a column density PDF, 
down to which the PDF is generally well represented by a pure power-law. 
Here, above the completeness level at $A_{\rm V} \sim 1$, the column density PDF is best fit by a power-law {\it plus} a lognormal. 
Although the break between the two fits at $A_{\rm V} \sim 3$ is not very pronounced, the column density excess in the lognormal part above 
the power-law fit appears to be significant. 

The power-law tail of the PDF is as developed as in the case of the Aquila cloud complex \citep{Konyves+2015}. In Orion~B, however, the best power-law 
fit for $A_{\rm V} > 3$ corresponds to $\Delta N$/$\Delta$log$N_{\rm H_2} \propto N_{\rm H_2}^{-1.9\pm 0.2}$, which is significantly shallower than the logarithmic 
slope of $-2.9\pm 0.1$ found in the Aquila cloud.  

The total mass of the Orion~B clouds was derived from the column density map in Fig.~\ref{fig_T_NH2_maps} (left) as:
$$ M_{\rm cl} = \delta A_{\rm pixel}\, \mu_{\rm H_2}\, m_{\rm H} \sum_{{\rm pixels}} N_{\rm H_2}.$$  
In the above equation, $ \delta A_{\rm pixel}$ is 
the surface area of one pixel at the adopted distance ($d = 400$~pc), 
$\mu_{\rm H_2} = 2.8$ is the mean molecular weight per ${\rm H_2}$ molecule, $m_{\rm H}$ is 
the hydrogen atom mass, and the sum of column density values is over all of the pixels in the map. 
This approach yielded a total cloud mass of $\sim$2.8$\times 10^4$~$M_\odot$. 	

In order to obtain the cumulative mass fraction of gas mass as a function of column density (Fig.~\ref{fig:cdPDF} right), we repeated the above mass calculation 
for the pixels above a given column density. 
The dense molecular gas material represents only $\sim$13\% ($\sim 3.7\times 10^3\, M_\odot$) of the total cloud mass above a visual extinction 
$A_{\rm V} \sim 7\,$mag. 
As in other clouds, the low and intermediate column density regions at $A_{\rm V} < 3$--7~mag in the map of Fig.~\ref{fig_T_NH2_maps} (left) occupy most ($> 95\% $) of the 
surface area imaged with {\it Herschel} and account for most ($\sim 74-88\% $) of the cloud mass. 

In agreement with the typical ratios $A_{\rm V, HGBS} / A_{\rm V, ext} \ga 0.5$ found in Sect.~\ref{sec:cd_t_maps} from comparing 
the HGBS column density map with extinction maps, we note that 
the total mass reported here for the Orion~B clouds is a factor of $\sim$\,1.8 lower than the total mass 
indicated by the extinction maps ($\sim$$5 \times 10^4\,$$M_\odot$). 
\subsection{Multiwavelength core extraction with \textsl{getsources}}\label{sec:getsources}
Compact sources were extracted simultaneously from the PACS and SPIRE images with the  \textsl{getsources} 
algorithm \citep{Menshchikov2012}\footnote{The HGBS first-generation catalog of cores presented in this paper (see Appendix~\ref{sec:appendix_catalog}) 
was produced with the ``November 2013'' major release of \textsl{getsources} (v1.140127), which is publicly available from 
http://gouldbelt-herschel.cea.fr/getsources.}.
A  complete summary of the source extraction method can be found in Sect.~4.4 of \citet{Konyves+2015}, and full technical details are provided
in \citet{Menshchikov2012}. 
Briefly, the extraction process consists of a detection and a measurement stage.
At the detection stage, \textsl{getsources} analyzes fine spatial decompositions of all the observed wavebands over a wide range of scales. 
This decomposition filters out irrelevant spatial scales and improves source detectability, especially in crowded regions and for extended sources. 
The multi-wavelength design combines data over all wavebands and produces a wavelength-independent detection catalog.
At the measurement stage, properties of detected sources are measured in the original observed images at each wavelength. 

For the production of HGBS first-generation catalogs of starless and protostellar cores, two sets of dedicated \textsl{getsources} extractions 
are performed, optimized for the detection of dense cores and YSOs/protostars, respectively.  
In the first, ``core'' set of extractions, all of the {\it Herschel} data tracing column density are combined at the detection stage, to improve the 
detectability of dense cores.
The detection image is thus combined from the 160\,$\mu$m, 250\,$\mu$m, 350\,$\mu$m, and 500\,$\mu$m maps, together with the high-resolution column 
density image (see Sect.~\ref{sec:cd_t_maps}) used as an additional ``wavelength''. The 160~$\mu$m component of the detection image is 
``temperature-corrected'' to reduce the effects of strong, anisotropic temperature gradients present in parts of the observed fields 
\citep[see][for details]{Konyves+2015}.

A second, ``protostellar'' set of \textsl{getsources} extractions is performed to trace the presence of self-luminous YSOs/protostars and discriminate 
between protostellar and starless cores. In this second set of extractions, the only {\it Herschel} data used at the detection stage are the PACS 
70\,$\mu$m data. This allows us to exploit the result that the level of point-like 70~$\mu$m emission traces the internal luminosity (and therefore the presence) 
of a protostar very well \citep[e.g.,][]{Dunham+2008}.

At the measurement stage of both sets of extractions, source properties are measured at the detected positions of either dense cores or YSOs/protostars,  
using the observed, background-subtracted, and deblended images at all five {\it Herschel} wavelengths, plus the high-resolution column density map. 
The only difference from the procedure described in  \citet[][]{Konyves+2015} is that, following improvements in the preparation of maps for source extraction, 
a new method of background flattening, called \textsl{getimages}  \citep{Menshchikov2017}, was used before performing the protostellar set of \textsl{getsources} extractions.
Tests have shown that this additional step drastically reduces the large number of spurious 70~$\mu$m sources in areas of bright, structured PDR 
(photodissociation region) emission around HII regions. Before extracting protostellar sources, \textsl{getimages} subtracts the large-scale background 
emission and equalizes the levels of small-scale background fluctuations in the images using a median-filtering approach.

\begin{figure*}[]   
 \centering
  \includegraphics[scale=0.9, angle=0]{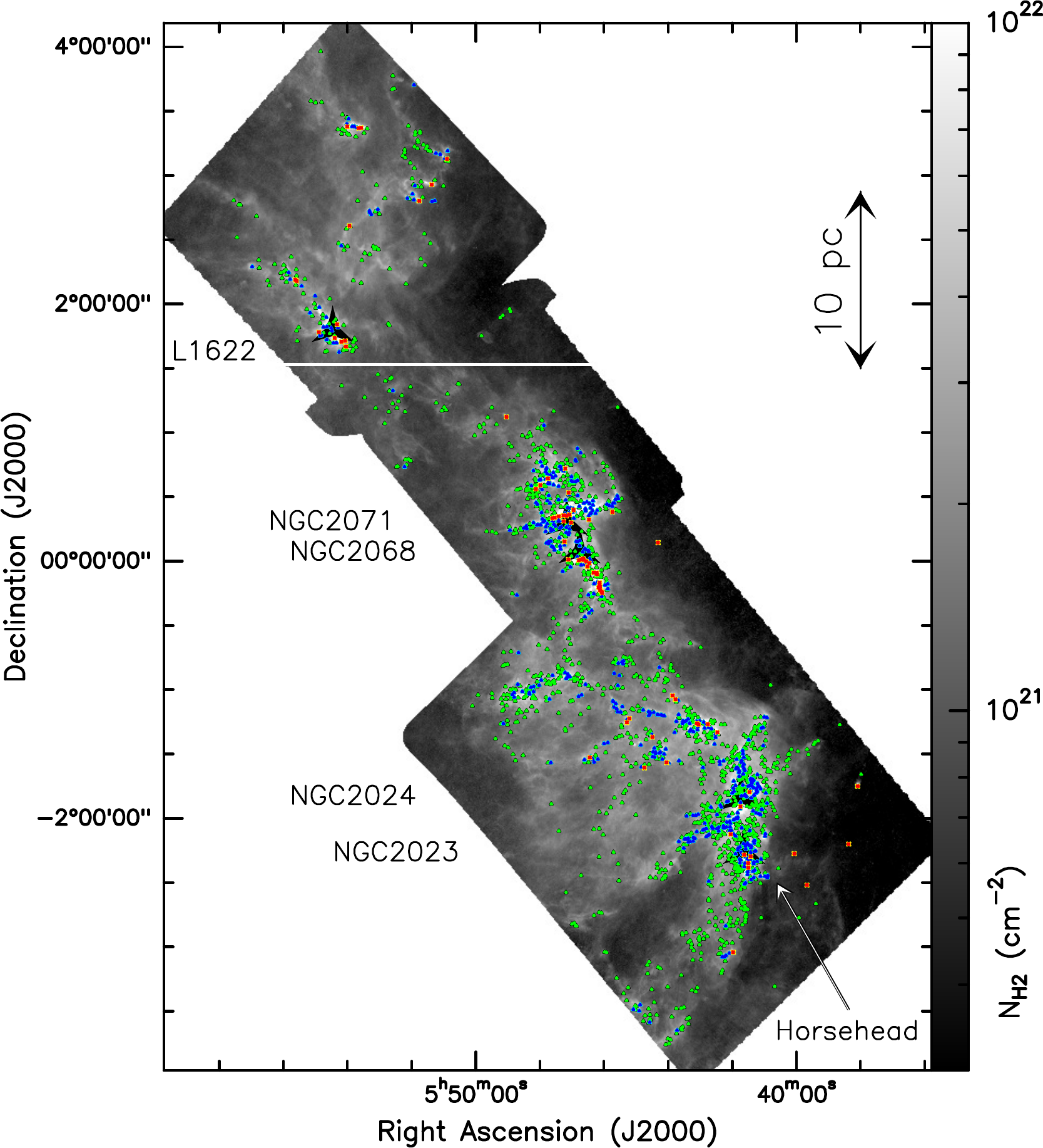}
  \caption{
   High-resolution ($\sim$18$''$) column density map of Orion~B, derived from SPIRE and PACS observations. The positions of the 1768 identified starless 
   cores (green triangles), 490 {\it robust} prestellar cores (blue triangles), and 76 embedded protostellar cores (red symbols) are overplotted. 
   The vast majority of the cores lie in high column density areas. 
   The white declination line at Dec$_{\rm 2000}=1$\degr 31$'$55$''$ marks the lower boundary of the 
   northern region where cores may be associated with the Barnard's Loop and at a more uncertain distance ($d\sim \,$170--400~pc). 
   See Sects.~\ref{sec:intro} and ~\ref{sec:post_selection} for details.
   }  
  \label{fig:map_cores}       
\end{figure*}

\subsection{Selection and classification of reliable core detections}\label{sec:core_selection}

\subsubsection{Initial core selection}\label{sec:init_selection}

The criteria used to select reliable dense cores and YSOs from the raw source catalogs delivered by \textsl{getsources} are described in detail in 
Sect.~4.5 of \citet{Konyves+2015}. 

First we select candidate dense cores (starless and protostellar) from the ``core'' extractions using criteria based on several output 
parameters provided by \textsl{getsources}: detection significance, signal-to-noise ratio (S/N), and ``goodness'' parameter. 
Then, we select candidate YSOs from the second set of extractions based on their corresponding detection significance, S/N, 
flux density, goodness, and source size and elongation parameters.    

Afterward, we cross-match the selected dense cores with the candidate YSOs/protostars and distinguish between starless cores and 
young (proto)stellar objects based on the absence or presence of compact 70\,$\mu$m emission. YSOs must be detected as compact emission sources 
above the 5$\sigma$ level at 70\,$\mu$m, while starless cores must remain undetected in emission (or detected in absorption) at 
70\,$\mu$m \citep{Konyves+2010}. 

In addition to Aquila \citep{Konyves+2015}, this whole method has been so far applied to studies of various regions of the HGBS: the Taurus L1495 
cloud \citep{Marsh+2016}, Corona Australis \citep{Bresnahan+2018}, the Lupus complex \citep{Benedettini+2018}, Perseus \citep{Pezzuto+2019}, 
Ophiuchus \citep{Ladjelate+2019}, the Cepheus Flare (Di Francesco et al., in prep.), further parts of Taurus (Kirk et al., in prep.), and 
the Orion~A clouds (Polychroni et al., in prep.).

\subsubsection{Post-selection checks}\label{sec:post_selection}

All of the candidate cores initially selected using the criteria defined in detail in Sect.~4.5 of \citet{Konyves+2015} were subsequently run through an 
automated post-selection step.
Given the large number ($\sim 2000$) of compact sources detected with {\it Herschel} in the Orion~B region, we tried to define automatic post-selection 
criteria based on source morphology in the SPIRE and PACS and column density images (see blow-up maps in Figs.~\ref{fig_zooms1} \& \ref{fig_zooms2}), 
in order to bypass the need for detailed visual inspection. 
A final visual check of all candidate cores in the present catalog (online Table~\ref{tab_obs_cat}) was nevertheless performed.
With our automated post-selection script described in detail in Appendix~\ref{sec:appendix_postSel}, we revisited the location of each formerly selected core 
in all of the {\it Herschel} maps (including the column density plane) and performed a number of pixel value checks within and around the core's FWHM contour 
to make sure that it corresponds to a true intensity peak or enhancement in the analyzed maps.   
This post-selection procedure allowed us to remove dubious sources from the final catalog of cores in online Table~\ref{tab_obs_cat}.

To eliminate extragalactic contaminants from our scientific discussion, we cross-matched all selected sources with the NASA Extragalactic 
Database\footnote{https://ned.ipac.caltech.edu/forms/nearposn.html} (NED). We found matches within 6\arcsec \, for four classified external galaxies, 
which were removed from our scientific analysis. 
As Orion~B lies toward the outer Galaxy and below the Galactic plane, we can expect several extragalactic contaminants in the low column density 
portions of the field, beyond the four sources listed in the NED database. Using {\it Herschel}'s SPIRE data from the HerMES extragalactic survey 
\citep{Oliver+2012}, \citet{Oliver+2010} determined that the surface number density of extragalactic sources at the level $S_{\rm 250\mu m} = 100$\,mJy 
is about 12.8$\pm$3.5~deg$^{-2}$. 
In the $\sim$19~deg$^2$ area observed with {\it Herschel} in Orion~B, our catalog includes three candidate cores with 250~$\mu$m integrated flux 
densities below $100$\,mJy and 26 candidate cores with 250~$\mu$m integrated flux densities below $200$\,mJy, mostly in low background emission spots.
Rougly 1/3 of these candidate cores with low flux densities at 250~$\mu$m have Unclassified Extragalactic Candidate matches within 6\arcsec. 
These 26 tentative cores are listed in our observed core catalog (online Table~\ref{tab_obs_cat}), but excluded from the scientific analysis. 
%

\begin{figure*}[!ht]
 \begin{center}
  \begin{minipage}{0.47\linewidth}
   \resizebox{1.0\hsize}{!}{\includegraphics[angle=0]{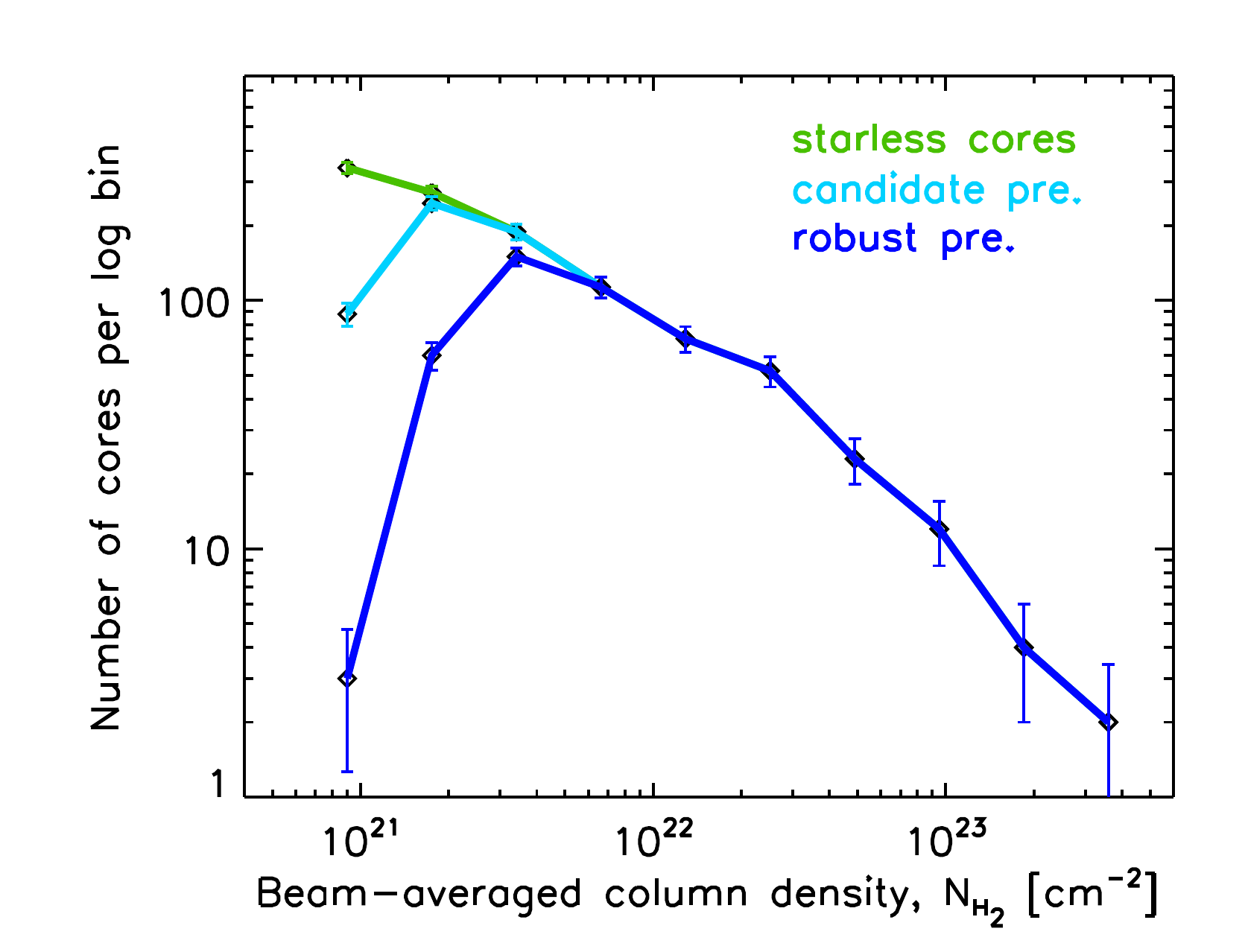}}
  \end{minipage}
  \hspace{0.4cm}
  \begin{minipage}{0.47\linewidth}
   \resizebox{1.0\hsize}{!}{\includegraphics[angle=0]{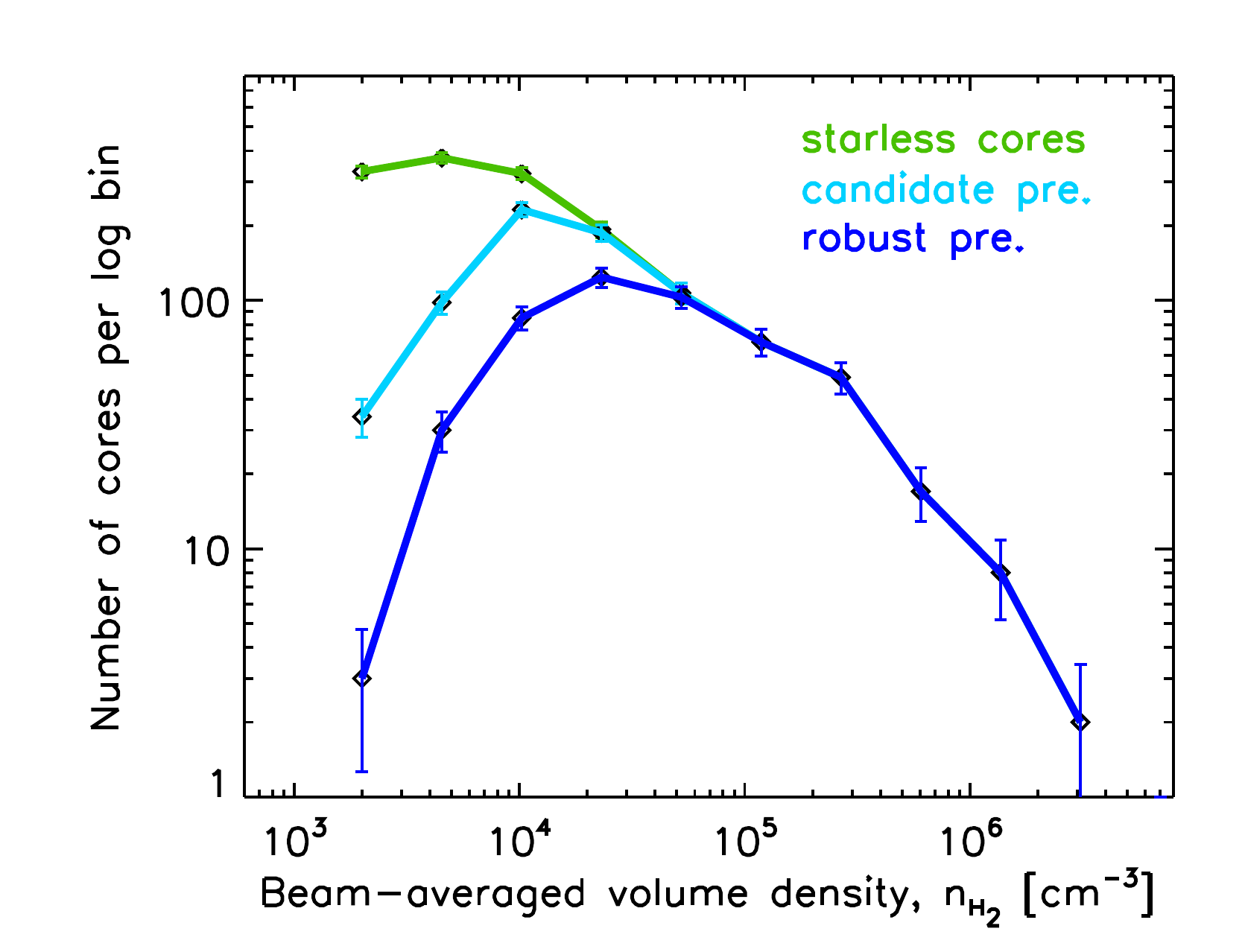}}
  \end{minipage} 
 \end{center}
 \caption{Distributions of beam-averaged column densities {\bf (left)} and beam-averaged volume densities 
  {\bf (right)} at resolution of SPIRE 500~$\mu$m observations for 1768 starless cores (green), 804 {\it candidate} prestellar 
  cores (light blue), and 490 {\it robust} prestellar cores (dark blue curve) in Orion~B.}
 \label{fig:dens}%
\end{figure*}
Likewise, we checked likely associations between the selected {\it Herschel} cores and objects in the SIMBAD database or the {\it Spitzer} database 
of Orion~by \citet{Megeath+2012}. Any match is reported in the online catalog of Table~\ref{tab_obs_cat}.
In particular, 30 associations with a {\it Spitzer} source were found using a 6$\arcsec$ matching radius, where we used the same selection of YSOs 
from \citet{Megeath+2012} as \citet{Shimajiri+2017}.   

Overall, our \textsl{getsources} selection and classification procedure resulted in a final sample of 1844 dense cores in Orion~B 
(not counting $\sim 26$ possible extragalactic contaminants), comprising 1768 starless cores and 76 protostellar cores (see Fig.~\ref{fig:map_cores}). 
The observed properties of all these post-selected {\it Herschel} cores are given in the accompanying online catalog with a few example 
lines in Table~\ref{tab_obs_cat}. The corresponding derived properties of the 1844 dense cores are 
provided in online Table~\ref{tab_der_cat}.

\subsection{Derived core properties}\label{sec:deriv_core_prop}

For deriving core physical properties we used a SED fitting procedure similar to that described in Sect.~\ref{sec:cd_t_maps} for the production of 
the column density map. 
The SED data points for the properties of each core were given by their integrated flux densities measured by \textsl{getsources}, and the data points 
were weighted by 1/$\sigma_{\rm err}^2$, where $\sigma_{\rm err}$ is the flux measurement error estimated by \textsl{getsources} for each data point. 
The modified blackbody fits of the observed SEDs were obtained in IDL with the MPCURVEFIT routine \citep{Markwardt2009}. 
These fits provided direct mass and line-of-sight-averaged dust temperature estimates for most of the selected cores. 
We derived core masses with the same dust opacity assumptions as in Sect.~\ref{sec:cd_t_maps} and a distance $d = 400$~pc for Orion~B. 
The angular FWHM size estimates, measured by \textsl{getsources} at $18.2\arcsec $ resolution in the high-resolution column density map, were converted 
to a physical core radius. In our catalog (see online Table~\ref{tab_der_cat}), we provide estimates for both the deconvolved and the observed radius 
of each core (estimated as the geometrical average between the major and minor FWHM sizes).  

For each core, the peak (or central beam) column density, average column density, central-beam volume density, and the average volume density were then 
derived based on their mass and radius estimates \citep[see Sect. 4.6. in][]{Konyves+2015}.
The column- and volume density distributions for the different population of cores are shown in Fig.~\ref{fig:dens}.
In online Table~\ref{tab_der_cat}, all the above derived properties are provided for the whole sample of selected {\it Herschel} cores.

In order to assess the robustness of our SED fits for each source, we compared two successive runs of the fitting procedure with somewhat different 
weighting schemes. 
In the first run the \textsl{getsources} detection errors were used to weigh the SED points and the 70~$\mu$m data were included in the fit, while the 
more conservative measurement errors were used without fitting the 70~$\mu$m data points in the second run.
The following criteria were applied for accepting the SED fits for a given source: 1) significant flux measurements exist in at least three {\it Herschel} 
bands, 2) the source has larger integrated flux density at 350~$\mu$m than at 500~$\mu$m, and 3) less than a factor of 2 difference is found between 
the core mass estimates from the two SED fits. 

The starless core masses with rejected SED fit results from the above procedure were estimated directly from the measured integrated flux density
at the longest significant wavelength. Optically thin dust emission was assumed at the median dust temperature that we found for 
starless cores with reliable SED fits. Such cores have more uncertain properties for which we assigned a ``no SED fit'' comment in online Table~\ref{tab_der_cat}.

The accuracy of our core mass estimates is affected by two systematic effects:
1) uncertainties in the dust opacity law induce mass uncertainties of up to a factor $\sim \,$1.5--2 \citep[cf.][]{Roy+2014},
2) the SED fits only provide line-of-sight averaged dust temperatures.
In reality, however, starless cores are externally heated objects, and they can show a significant drop in dust temperature toward their centers
\citep[e.g.,][]{Nielbock+2012, Roy+2014}.  
The mass of a starless core may be underestimated in such a situation, as the average SED temperature may overestimate the mass-averaged dust 
temperature.
For low column density cores, such as B68 \citep{Roy+2014}, this effect is less than $\sim 20$\%, but for high-density cores, with 
average column densities above $\sim 10^{23}$ cm$^{\rm -2}$, it increases to up to a factor of $\sim 2$. 
Here, we did not attempt to use special techniques \citep[cf.][]{Marsh+2014, Roy+2014} to retrieve the intrinsic temperature structure and 
derive more accurate mass estimates. 

The Monte-Carlo simulations that we used to estimate the completeness of the survey (see Sect.~\ref{sec:completeness} below), however show that
the SED masses of starless cores (in online Table~\ref{tab_der_cat}) are likely underestimated by only $\sim 25$\% on average compared to the intrinsic 
core masses. This is mainly because the SED dust temperatures slightly overestimate the intrinsic mass-averaged temperatures within 
starless cores. 
We have not corrected the column- and volume densities in Table~\ref{tab_der_cat} and in Fig.~\ref{fig:dens} for this small effect.
 
We may also compare the results of the present core census with the SCUBA-2  findings of \citet{HKirk+2016a}, used in both \citet{HKirk+2016b} and \citet{Parker2018}. 
Adopting a matching separation of $< 6$\arcsec, we found 192 matches between HGBS starless cores and SCUBA-2 cores
and 47 matches between HGBS protostellar cores and SCUBA-2 objects.
The relatively low number of matches is the result of the significantly smaller area coverage of the SCUBA-2 data, the higher sensitivity 
and broader wavelength coverage of the {\it Herschel} data, and very different source-extraction techniques.   
As the assumptions about dust emissivity properties differ in the HGBS and SCUBA-2 studies, we re-derived two sets of mass estimates for the SCUBA-2 cores from the 
published 850~$\mu$m total fluxes: 
1) the first set of mass estimates ($M_{\rm SCUBA-2, Td}$) was derived using the standard HGBS dust opacity law (see Sect.~\ref{sec:cd_t_maps}) and the dust 
temperatures estimated from {\it Herschel} SED fitting ($T_{\rm d}^{\rm SED}$); 
2) the second set ($M_{\rm SCUBA-2, 20K}$) was derived using the same HGBS dust opacity law and a uniform dust temperature of 20~K for all cores 
(as assumed by \citealp{HKirk+2016a}).
The resulting comparison of core mass estimates is summarized in Fig.~\ref{fig:massComp}, which plots the distributions of the ratios 
$M_{\rm SCUBA-2, Td}/M_{\rm HGBS}$ and $M_{\rm SCUBA-2, 20K}/M_{\rm HGBS}$ for all SCUBA-2--HGBS core matches, separately for starless and protostellar cores. 
In each panel of Fig.~\ref{fig:massComp}, the corresponding median ratio is indicated by a red dashed line.   
Overall, a good correlation exists between the re-derived SCUBA-2 masses and the HGBS masses. 
The $M_{\rm SCUBA-2, Td}$ mass estimates for the matched starless cores are typically only $\sim$25\% higher, and the $M_{\rm SCUBA-2, 20K}$ estimates $\sim$60\% lower, 
than the $M_{\rm HGBS}$ masses.
The most likely reason for the relatively low $M_{\rm SCUBA-2, 20K}/M_{\rm HGBS}$ ratios is that the typical dust temperature of starless cores is significantly lower 
than 20~K, implying that $M_{\rm SCUBA-2, 20K}$ typically underestimates the true mass of a starless core.    
For the 47 matched protostellar cores, the $M_{\rm SCUBA-2, 20K}/M_{\rm HGBS}$ ratio is close to 1 on average, indicating a very good agreement 
between the $M_{\rm SCUBA-2, 20K}$ and  $M_{\rm HGBS}$ mass estimates, while the $M_{\rm SCUBA-2, Td}$ masses are typically a factor of $\sim$2 higher than the 
$M_{\rm HGBS}$ masses. 
The latter trend most likely arises from the fact that the dust temperature $T_{\rm d}^{\rm SED}$, derived from our standard SED fitting between 160 and 500\,$\mu$m, 
typically overestimates the ``mass-averaged'' temperature of a protostellar core. 
We stress, however, that $T_{\rm d}^{\rm SED}$ is more reliable for starless cores and that protostellar cores are not the main focus of the present paper.

\subsection{Selecting self-gravitating prestellar cores}\label{sec:bound_selection}

The subset of starless cores which are self-gravitating are likely to evolve to protostars in the future and can be classified as {\it prestellar} cores 
\citep[cf.][]{Ward+2007}. 
In the absence of spectroscopic observations for most of the $Herschel$ cores, the self-gravitating status of a core can be assessed using the 
Bonnor-Ebert (BE) mass ratio $\alpha_{\rm BE} = M_{\rm BE,crit} / M_{\rm obs} $, where $M_{\rm obs} $ is the core mass as derived from {\it Herschel} 
data and $ M_{\rm BE,crit} $ is the thermal value of the critical BE mass. 
The first criterion adopted here to select self-gravitating cores is $\alpha_{\rm BE} \leq 2$.
This simplified approach, using the thermal BE mass and thus neglecting the nonthermal component of the velocity dispersion, 
is justified by observations of nearby cores in dense gas tracers \citep[e.g.,][]{Myers1983,Tafalla+2002,Andre+2007}. 
These studies show that the thermal component dominates at least in the case of low-mass cores. 
The thermal BE mass ($M_{\rm BE}$) was estimated for each object from the deconvolved core radius ($R_{\rm deconv}$) that we measured in the high-resolution 
column density map (see Sect.~\ref{sec:deriv_core_prop}) assuming 10~K as a typical internal core 
temperature\footnote{We adopt the same typical internal core temperature of 10~K in Orion~B as in the Aquila cloud \citep[cf. Sect.~4.7 of][]{Konyves+2015} 
since the dust temperature maps derived from {\it Herschel} data suggest that the average radiation field is similar in both cloud complexes, with the possible 
exception of the vicinity of the bright HII region NGC~2024 \citep[see Table 6 of][]{Shimajiri+2017}. 
We also stress that radiative transfer models suggest that the internal temperature of starless dense cores depends only weakly on the ambient radiation 
field, in contrast to the temperature of the core outer layers.
}.
\begin{figure}[!h]
 \begin{center}
  \resizebox{1.0\hsize}{!}{\includegraphics[angle=0]{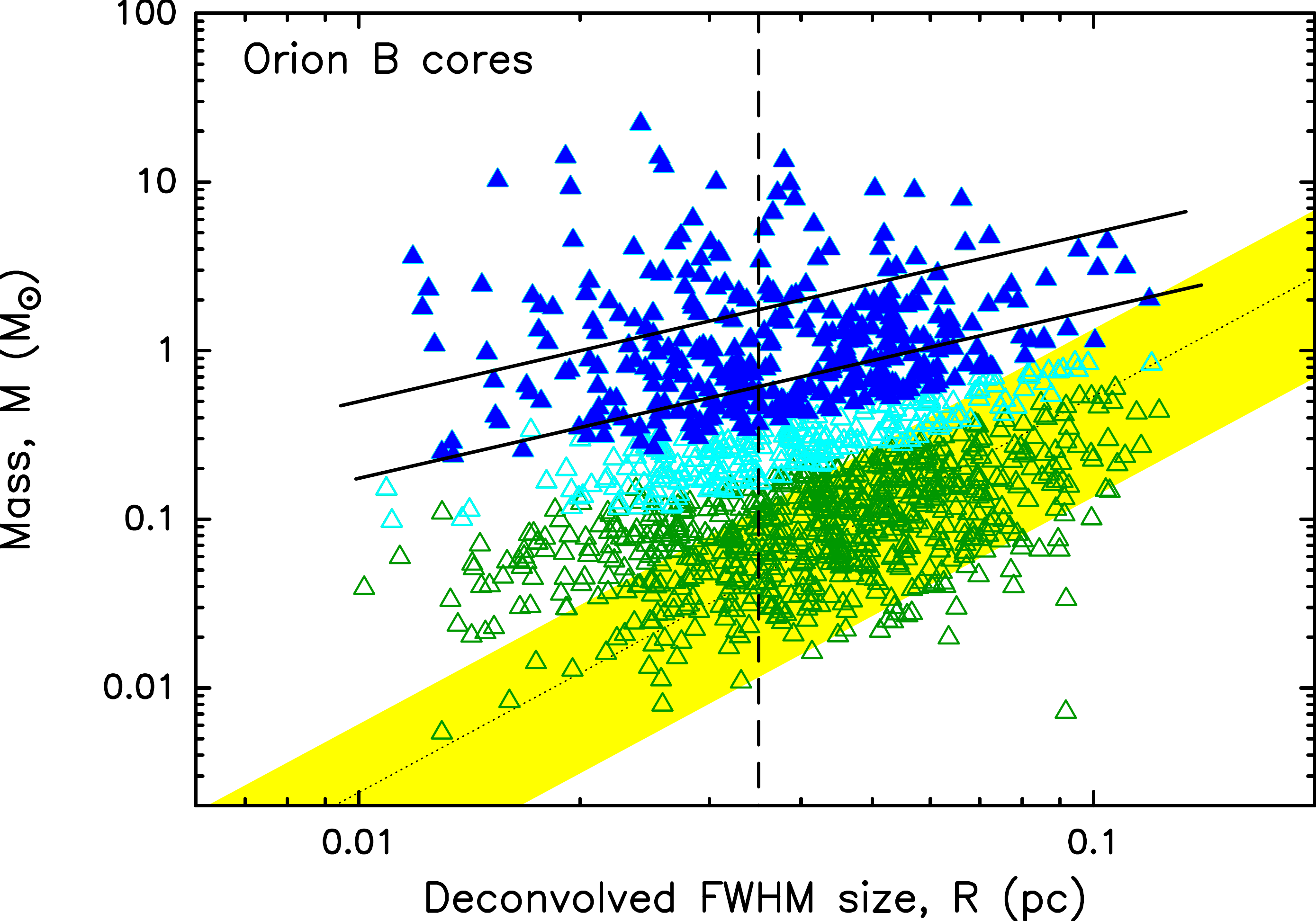}}
 \end{center}
  \caption{
   Mass versus size diagram for all 1768 starless cores (triangles) identified with $Herschel$ in Orion~B, 
   including the 804 {\it candidate} prestellar cores (light or dark blue triangles) and 490 {\it robust} prestellar cores (filled triangles in dark blue).
   The core FWHM sizes were measured with \textsl{getsources} in the high-resolution column density map (Fig.~\ref{fig_T_NH2_maps} left)
   and deconvolved from an $18.2\arcsec$ (HPBW) Gaussian beam. The corresponding physical $\overline{HPBW}$ resolution at $d = 400$~pc is 
   marked by the vertical dashed line. 
   Models of critical isothermal Bonnor-Ebert spheres at $T = 7$~K and $T = 20$~K are plotted as black solid lines for comparison. 
   The diagonal dotted line and yellow band display the observed mass--size correlation of diffuse CO clumps \citep[]{ElmegreenFalgarone1996}. 
   }
  \label{fig:massSize}%
\end{figure}
This selection of self-gravitating cores is illustrated in the mass versus size diagram of Fig.~\ref{fig:massSize}, where source positions can be 
used to distinguish between {\it robust} prestellar cores (filled blue triangles) and unbound starless cores (open green 
triangles). 

In addition, a second, less restrictive selection of {\it candidate} prestellar cores (open light blue triangles in Fig.~\ref{fig:massSize}) was obtained by 
deriving an empirical lower envelope for self-gravitating objects in the mass versus size diagram, based on the Monte-Carlo simulations used in 
Sect.~\ref{sec:completeness} to derive the completeness of the survey.
The advantage of this less restrictive selection criterion, also employed by \citet{Konyves+2015} in the Aquila cloud, is that it provides a larger, 
more complete sample of {\it candidate} prestellar cores (at the cost of including a larger fraction of misclassified unbound cores). 

Here, the resulting final number of {\it robust} prestellar cores based on the first criterion ($\alpha_{\rm BE} \leq 2$) was 490, and the number of
{\it candidate} prestellar cores based on the empirical criterion was 804, out of a grand total of 1768 starless cores in the entire Orion~B field. 
The spatial distribution of all starless cores and {\it robust} prestellar cores is plotted in Figure~\ref{fig:map_cores}.
We note that all of the {\it robust} prestellar cores belong to the wider sample of {\it candidate} prestellar cores.

These two samples of prestellar cores reflect the uncertainties in the classification of gravitationally bound/unbound objects, which are fairly large for 
unresolved or marginally-resolved cores, as the intrinsic core radius (and therefore the intrinsic BE mass) is more uncertain for such cores.
In the following discussion (Sect.~\ref{sec:discuss}) below, we consider both samples of prestellar cores. 
The 314 {\it candidate} prestellar cores which are not {\it robust} are clearly marked with a ``tentative bound'' comment in online Table~\ref{tab_der_cat}. 

In the northern region around L1622, which may be associated with Barnard's Loop, 
there are 179 starless cores ($\sim$10\% of the whole sample of starless cores), including 
67 {\it candidate} prestellar cores (or $\sim$8\% of all {\it candidate} prestellar cores), 41 {\it robust} prestellar cores 
($\sim$8\%), and 15 protostellar cores ($\sim$20\% of all protostellar cores). 
As mentioned in Sect.~\ref{sec:intro}, these cores are at a more uncertain distance.
By default and for simplicity, these northern objects are assumed to be at the same distance of 400~pc as the rest of the cores \citep[cf.][]{Ochsendorf+2015}.
The cores lying north of the declination line Dec$_{\rm 2000}=1$\degr 31$'$55$''$ in Fig.~\ref{fig:map_cores} are nevertheless marked with an ``N region'' comment 
in the online catalogs (Tables~\ref{tab_obs_cat} and ~\ref{tab_der_cat}). 

In the mass versus size diagram of  Fig.~\ref{fig:massSize} for the entire population of starless cores, there is a spread of deconvolved 
sizes ($\overline{FWHM}$) between $\sim 0.01$~pc and $\sim 0.1$~pc and a range of core masses 
between $\sim 0.02\, M_\odot$ and $\sim 20\, M_\odot$. 
The fraction of self-gravitating starless objects is $\sim 28\%$ or $\sim 45\%$ considering the samples of {\it robust} or {\it candidate} prestellar 
cores, respectively. 
Most of the unbound starless cores identified with {\it Herschel} in Orion~B are located in the same area of the mass vs. size diagram as typical CO clumps 
(marked by the yellow band in Fig.~\ref{fig:massSize}), which are known to be mostly unbound structures \citep[e.g.,][]{ElmegreenFalgarone1996,Kramer+1998}. 
It can also be seen in Fig.~\ref{fig:massSize} that self-gravitating prestellar cores are typically more than an order of magnitude denser than unbound CO clumps. 

Finally, we note that the distributions of beam-averaged column densities and volume densities for starless cores, {\it candidate} prestellar cores, 
and {\it robust} prestellar cores all merge at high (column) densities in Fig.~\ref{fig:dens}, suggesting that all {\it Herschel} starless cores 
with $N^{\rm peak}_{\rm H_2} \ga 7\times 10^{21}\, {\rm cm}^{-2}$ or $n^{\rm peak}_{\rm H_2} \ga 4\times 10^{4}\, {\rm cm}^{-3}$ 
are dense enough and sufficiently strongly centrally concentrated to be self-gravitating and prestellar in nature.

\begin{figure}[!h]
 \begin{center}
  \begin{minipage}{1.0\linewidth}
   \resizebox{1.0\hsize}{!}{\includegraphics[angle=0]{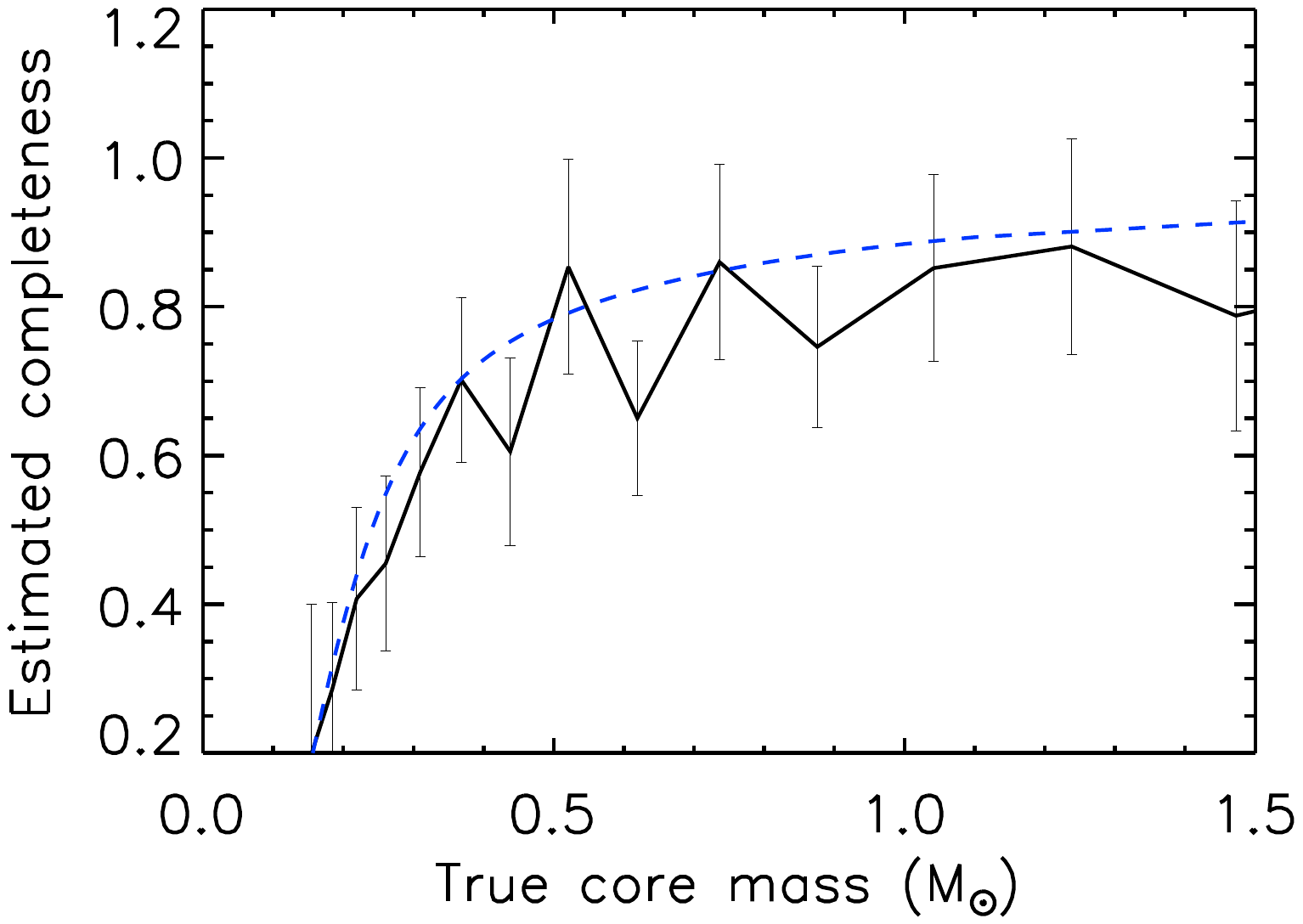}}
  \end{minipage}
 \end{center}
  \caption{
   Completeness curve of {\it candidate} prestellar cores as function of true core mass (solid line), which was 
   estimated from Monte-Carlo simulations (see Sect.~\ref{sec:completeness}).
   For comparison, the global completeness curve of the model introduced at the end of Appendix~\ref{sec:appendix_sim} is shown by the dashed blue line.
   }
  \label{fig:complete}%
\end{figure}

\subsection{Completeness of the prestellar core survey}\label{sec:completeness}

Following the Monte-Carlo procedure used by \citet[][]{Konyves+2015} to estimate the completeness limit of the HGBS prestellar core census in the Aquila cloud,
we performed similar simulations in Orion~B. 
A population of 711 model Bonnor-Ebert-like cores were inserted in clean-background images at the {\it Herschel} wavelengths, from which 
we also derived a synthetic column density plane as described in Sect.~\ref{sec:cd_t_maps}.
Then, core extractions were performed on the resulting synthetic images with \textsl{getsources} in the same way as for the observed data 
(see Sect.~\ref{sec:getsources}).  
The results of these simulations suggest that our sample of {\it candidate} prestellar cores  in Orion~B is $> 80\% $ complete above a {\it true} core mass of 
$\sim$0.5~$M_\odot$ (cf. Fig.~\ref{fig:complete}), which translates to $\sim$0.4~$M_\odot$ in {\it observed} core mass. Indeed,  the observed masses 
of starless cores tend to be underestimated by $\sim \, $25\% due to their internal temperature gradient (see Fig.~\ref{fig:completeMT}--left and
Appendix~\ref{sec:appendix_sim} for further details).

In Appendix~B of \citet{Konyves+2015} it was shown that the completeness level of a census of prestellar cores 
such as the present survey is background dependent, namely that the completeness level decreases 
as background cloud column density and cirrus noise increase.
A similar model adapted to the case of Orion~B  quantifies this trend (see Fig.~\ref{fig_completeness} in Appendix~\ref{sec:appendix_sim}) 
and shows that the global completeness of the present census of prestellar cores  
is consistent with that inferred from the Monte-Carlo simulations (see Fig.~\ref{fig:complete} and Appendix~\ref{sec:appendix_sim}).

\section{Discussion}\label{sec:discuss}

\subsection{Spatial distribution of dense cores}\label{sec:spatial_distrib}

\subsubsection{Connection between dense cores and filaments}\label{sec:cores_filams}

As already pointed out in earlier HGBS papers \citep[e.g.,][]{Andre+2010, Konyves+2015, Marsh+2016, Bresnahan+2018}, 
there is a very close correspondence between the spatial distribution of compact dense cores and 
the networks of filaments identified in the {\it Herschel} column density maps of the Gould Belt regions. 
Furthermore, prestellar cores and embedded protostars are preferentially found within the densest filaments with 
transcritical or supercritical masses per unit length 
(i.e., $M_{\rm line} \ga M_{\rm line, crit} \equiv  2\, c_s^2/G$) \citep[e.g.,][]{Andre+2010,Andre+2014}. 

\begin{figure*}[!ht]
 \centering
  \begin{minipage}{1.0\linewidth}
   \resizebox{0.37\hsize}{!}{\includegraphics[angle=0]{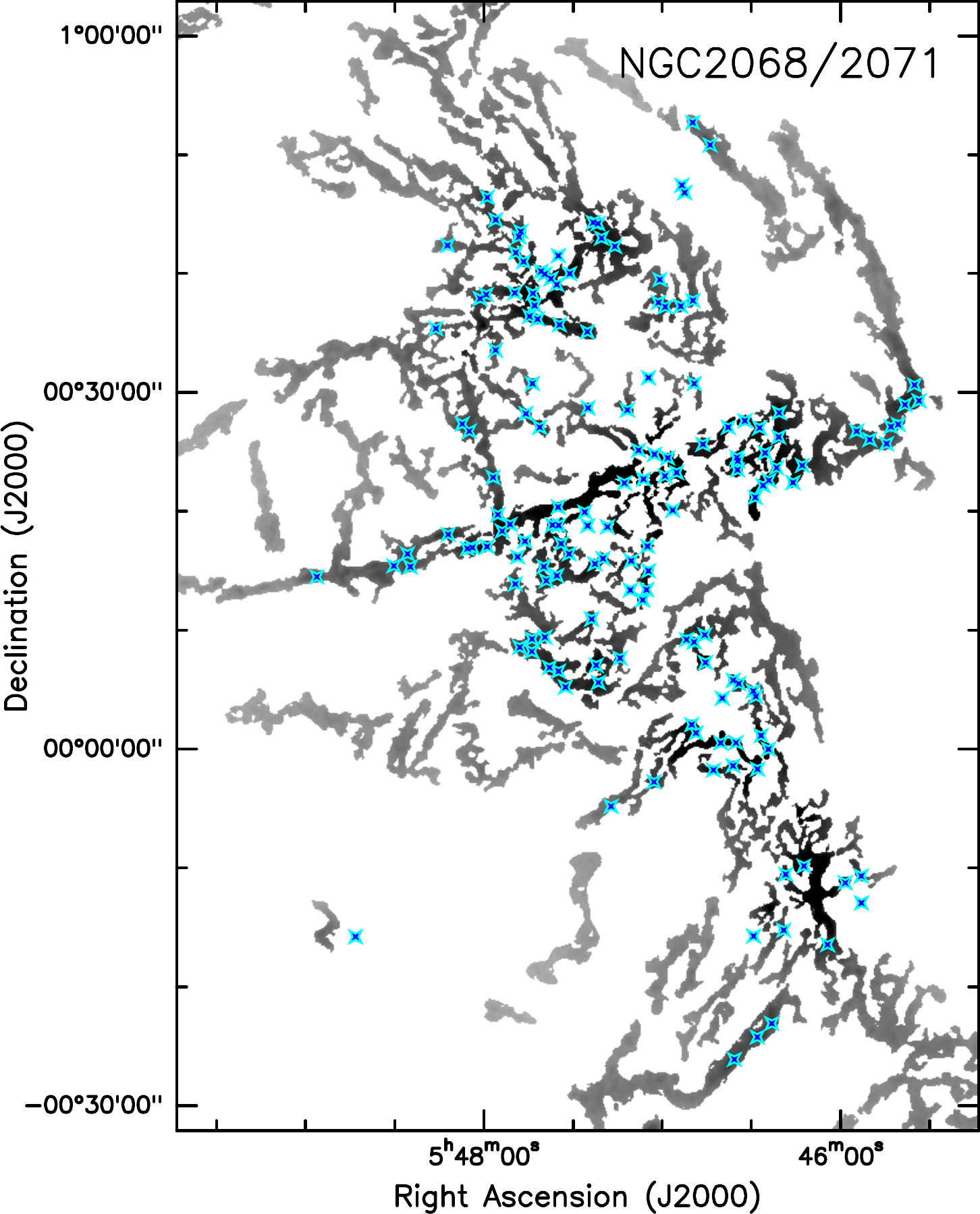}}
   \hspace{2mm}
   \resizebox{0.61\hsize}{!}{\includegraphics[angle=0]{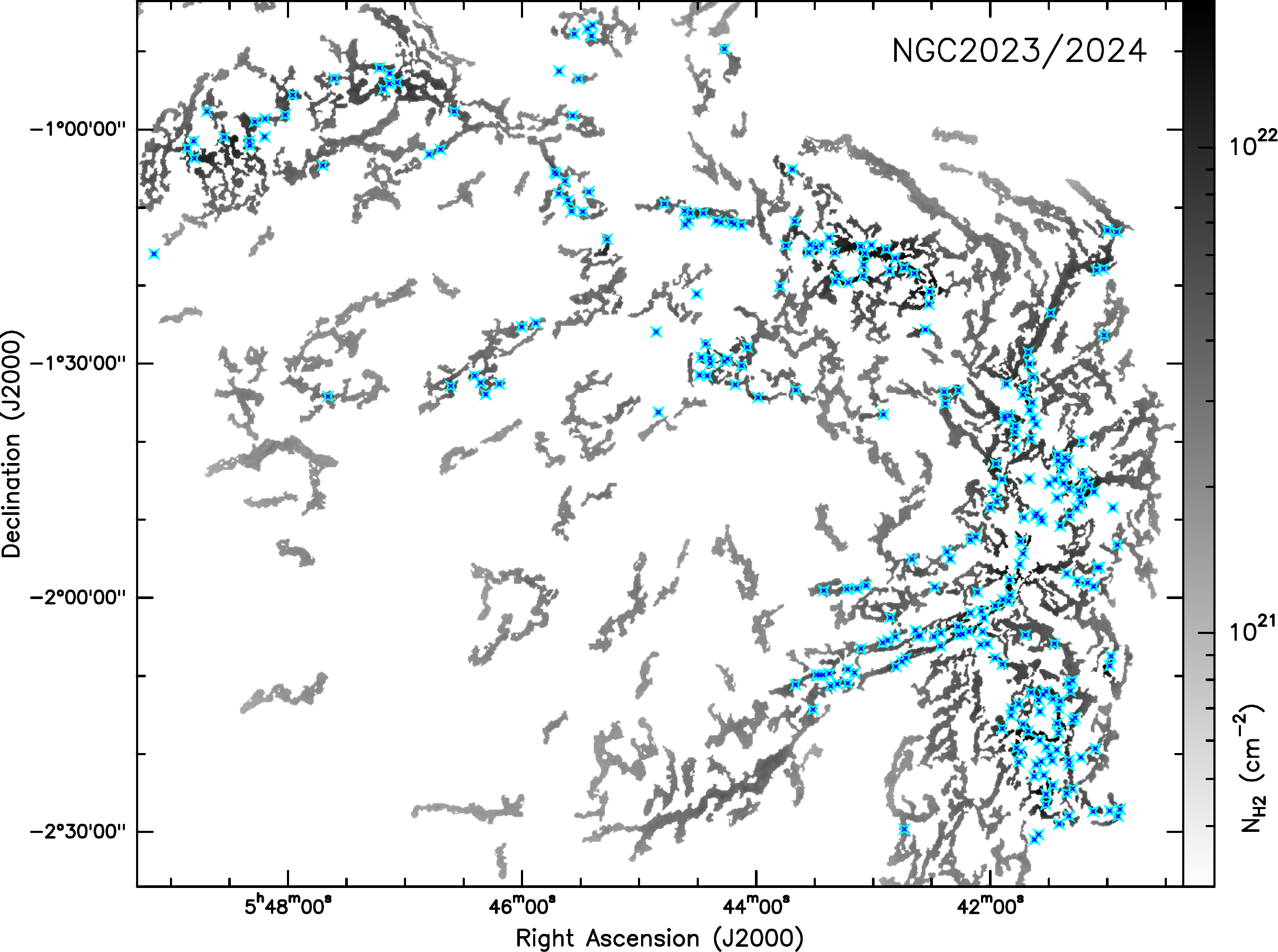}}    
  \end{minipage}
   \caption{Spatial distributions of {\it robust} prestellar cores (blue symbols) overplotted on the  filament networks (gray background images)
   traced with \textsl{getfilaments} in NGC~2068 and 2071 ({\bf left}), and NGC~2023 and 2024 ({\bf right}).  
   Transverse angular scales up to $100\arcsec$ across the filament footprints are shown here, corresponding to $\sim$0.2~pc at $d = 400$~pc. 
   In both panels, the color scale displayed within the filament mask 
   corresponds to column density values in the original column density map (Fig.~\ref{fig_T_NH2_maps} left). 
   See Fig.~\ref{fig:fils_coresL1622} (right) for a similar view in the case of L1622. 
   }
  \label{fig:fils_cores}%
\end{figure*}

The two panels of Fig.~\ref{fig:fils_cores} provide close-up views of two subfields of Orion~B, where the projected locations of {\it robust} 
prestellar cores are overlaid on the filament masks (shown up to $100\arcsec$ transverse angular scales) extracted with \textsl{getfilaments} 
(see Sect.~\ref{sec:filam}). More than 80\% of the {\it robust} prestellar cores are found inside this filament sample in the whole Orion~B region.
The connection between cores and filaments can be quantified in more detail based on the census of cores presented in Sects.~\ref{sec:getsources}, 
\ref{sec:core_selection}, \& \ref{sec:bound_selection} and the census of filaments described in Sect.~\ref{sec:filam}. 
The locations of extracted cores were correlated with various filament samples to illustrate the robustness of this connection. 
We consider three of these filament samples, also displayed in Fig.~\ref{fig:fils}:
\begin{itemize}
 \item \textsl{getfilaments} mask images including transverse angular scales up to $50\arcsec$ or $100\arcsec$, which correspond to transverse physical scales 
       up to $\sim$0.1~pc or $\sim$0.2~pc at $d = 400$~pc. Such $100\arcsec$ mask images are shown in grayscale in, for example, Fig.~\ref{fig:fils}.
 \item Maps of robust filament crests extracted with the DisPerSE algorithm and cleaned with the criteria described in Sect.~\ref{sec:filam}
       (thick blue skeletons overplotted in Fig.~\ref{fig:fils}). 
       Corresponding mask images were prepared from the 1-pixel-wide crests using Gaussian smoothing kernels with three different FWHM widths: 0.05~pc, 0.1~pc, and 0.2~pc.
 \item Maps of raw filament crests extracted with the DisPerSE algorithm, prior to cleaning and post-selection (see Sect.~\ref{sec:filam}). 
       In Fig.~\ref{fig:fils}, the combination of blue and cyan filament crests together make up the DisPerSE raw  sample. 
       As for DisPerSE robust crests, mask images were prepared using three Gaussian smoothing kernels with FWHMs of 0.05~pc, 0.1~pc, and 0.2~pc.
\end{itemize}
\begin{table}\small\setlength{\tabcolsep}{4.5pt}
 \caption{Fractions of cores associated with filaments in Orion~B.}
 \label{tab_ONfil1}      
  {\renewcommand{\arraystretch}{1.2}        
   \begin{tabular}{l|c|c|c}  
   \hline
    \hline
 							& {\it robust}	& {\it candidate}	&		\\
 							& prestellar	& prestellar	& starless	\\
    \hline
    \textsl{getfilaments} scales up to $\sim$0.1~pc	& 87\%		& 83\%		& 72 \%         \\
    \hline
    \textsl{getfilaments} scales up to $\sim$0.2~pc	& 90\%		& 87\%		& 76\%          \\
    \hline\hline
    robust DisPerSE filaments, 0.05~pc			& 60\%		& 55\%		& 43\%          \\
    \hline
    robust DisPerSE filaments, 0.1~pc			& 65\%		& 60\%		& 47\%          \\
    \hline
    robust DisPerSE filaments, 0.2~pc			& 71\%		& 66\%		& 53\%          \\
    \hline
    raw DisPerSE filaments, 0.05~pc			& 93\%		& 89\%		& 79\%          \\
    \hline
    raw DisPerSE filaments, 0.1~pc			& 97\%		& 94\%		& 85\%          \\
    \hline
    raw DisPerSE filaments, 0.2~pc			& 99\%		& 97\%		& 90\%          \\
    \hline
   \end{tabular}
  }
\tablefoot{This table gives fractions of prestellar and starless cores found within various \textsl{getfilaments} and DisPerSE 
mask images. For the \textsl{getfilaments} masks, upper transverse scales of either $\sim$0.1~pc or $\sim$0.2~pc 
(i.e., $\sim$\,$50\arcsec$ or $\sim$\,$100\arcsec$, respectively), were considered.
The DisPerSE filament footprints were assumed to be 0.05, 0.1, or 0.2~pc-wide around the nominal filament crests, and both the robust and the raw 
sample of DisPerSE filaments (see Sect.~\ref{sec:filam} for details) were considered. 
} 
\end{table}

Table~\ref{tab_ONfil1} reports the fractions of cores lying within the footprints of each of the above-mentioned filament samples, 
assuming that the intrinsic inner width of each filament is within a factor of 2 of the typical 0.1~pc value found by \citet[][]{Arzoumanian+2011,Arzoumanian+2019} 
in nearby clouds, including Orion~B.

\begin{figure}[!h]
 \centering
  \begin{minipage}{1.0\linewidth}
   \resizebox{1.0\hsize}{!}{\includegraphics[angle=0]{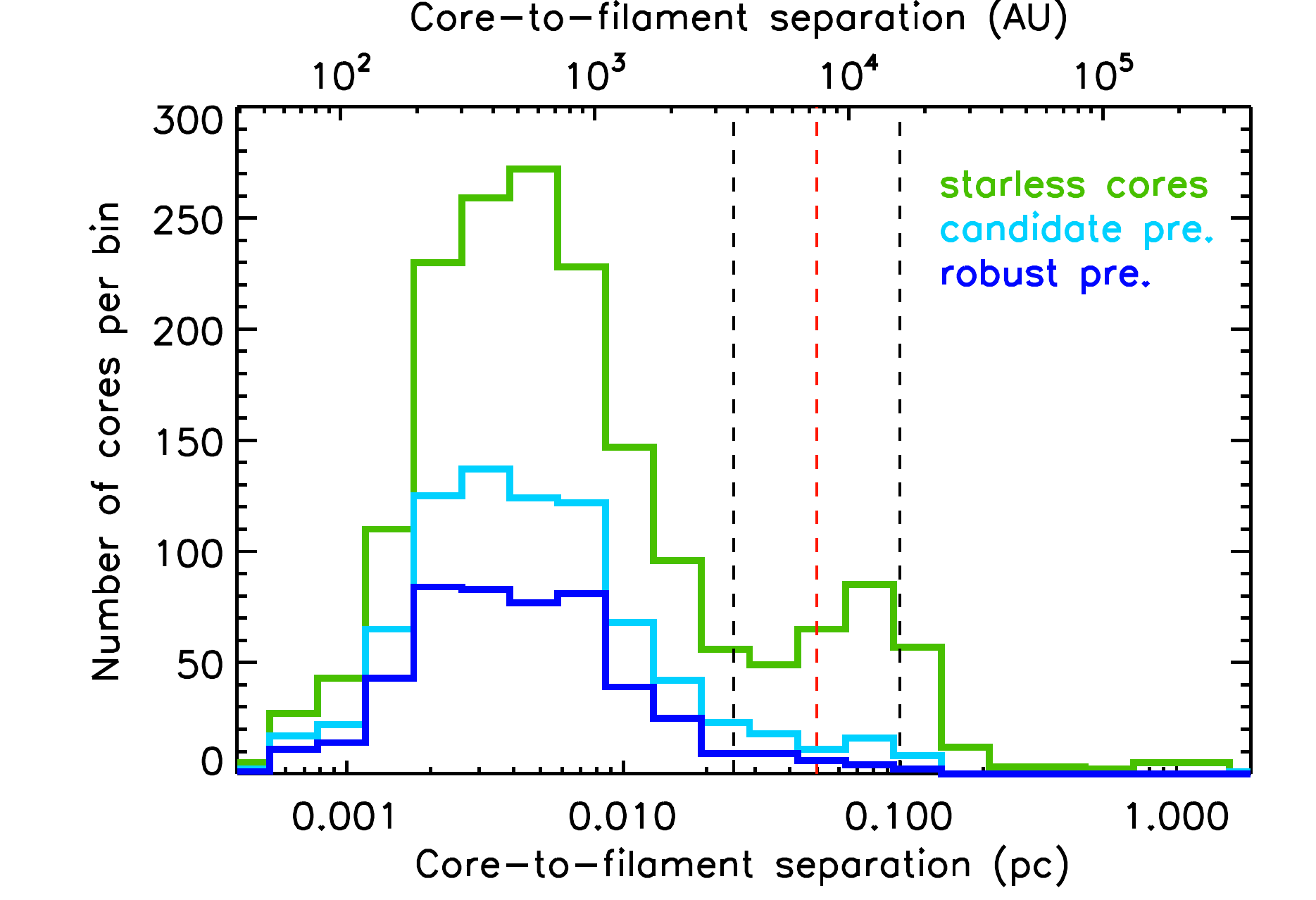}}
  \end{minipage}
   \caption{ 
    Distribution of separations between dense cores and the nearest filament crests in the DisPerSE raw sample of filaments. 
    The green histogram is for all 1768 starless cores, the light blue histogram for the 804 {\it candidate} prestellar cores,
    and the dark blue histogram for the 490 {\it robust} prestellar cores. 
    The vertical lines mark the half widths (i.e., on either side of the filament crests) of a 0.1\,pc-wide (red), 0.5$\times$0.1\,pc-wide (inner black), 
    and a 2$\times$0.1\,pc-wide (outer black) filament.         
    }
   \label{fig:dist2fil}%
\end{figure}

We also derived the distribution of core separations to the nearest filament crests using various core samples (starless, {\it candidate} and {\it robust} prestellar) 
and the crests of both the robust and the raw sample of DisPerSE filaments.  
The results shown in Fig.~\ref{fig:dist2fil} were obtained with the raw (uncleaned) sample of DisPerSE filaments. The median separation to the nearest 
filament is 0.006~pc for the {\it robust} (dark blue histogram) and {\it candidate} (light blue) prestellar core samples, and 0.007~pc for the starless core sample (green).

The fraction of cores within the nominal filament width (0.1~pc) of the raw DisPerSE sample, marked by the red vertical line in Fig.~\ref{fig:dist2fil}, 
is also presented in Table~\ref{tab_ONfil1}. 
Based on the numbers given in Table~\ref{tab_ONfil1}, we confirm that a high fraction of prestellar cores are closely associated with filaments in Orion~B, 
as already observed in Aquila \citep{Konyves+2015}, Taurus L1495 \citep{Marsh+2016}, and Corona Australis \citep{Bresnahan+2018}, for example. 

The tight connection between cores and filaments is found independently of the detailed algorithm employed to trace filamentary structures 
(see Table~\ref{tab_ONfil1}). In particular, while one may think that the association between cores and filaments is partly an artificial consequence 
of the way DisPerSE constructs filament crests (by joining saddle points to local maxima), we stress that the same connection is seen when using \textsl{getfilaments} 
to determine filament footprints completely independently of column density maxima (see Sect.~\ref{sec:filam}). 

When the DisPerSE raw samples of filaments are considered (which are the most complete samples of elongated structures used in Table~\ref{tab_ONfil1}), 
the distributions show that most ($\ga 90\%$) starless or prestellar dense cores lie very close to identified filamentary structures. 
This suggests that the earliest phases of star formation are necessarily embedded in elongated structures (of any density), which may provide an optimal 
basis for cores to gain mass from their immediate environment. 
Such an increase in the number of cores residing in filaments when considering a more complete filament sample is also found in the Aquila cloud. 
  
\begin{figure}[!!!h]
 \centering
  \resizebox{1.0\hsize}{!}{\includegraphics[angle=0]{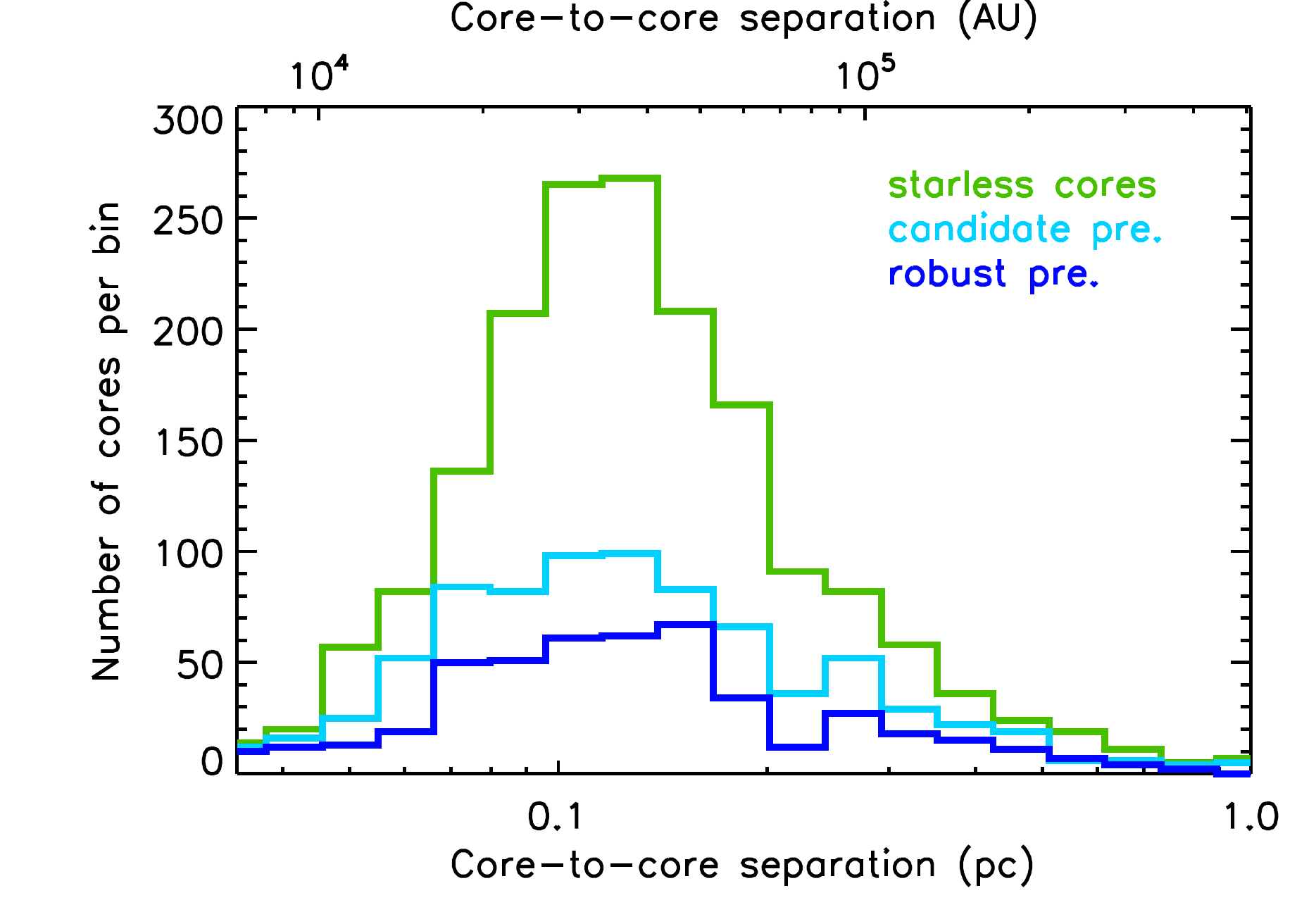}}
  \caption{ 
   Distribution of nearest-neighbor core separations in Orion~B. 
   The green, light-, and dark blue histograms show the distributions of nearest-neighbor separations 
   for starless cores (either bound or unbound), {\it candidate} prestellar cores, and {\it robust} prestellar cores, respectively. 
   In each case, only the separations between cores of the same category were considered. 
   The median core separation is $\sim 0.14\, $pc for all three core samples.
   }
  \label{fig:coreSeparOriB}%
\end{figure} 

\subsubsection{Nearest neighbor core separation}\label{sec:coreSepar}

The distributions of nearest-neighbor core separations are plotted in Fig.~\ref{fig:coreSeparOriB} for all starless cores (green histogram) 
and both {\it candidate} and {\it robust} prestellar cores (light and dark blue histograms, respectively). 
When deriving these distributions, each set of dense cores was analyzed separately, and only separations between cores of the same 
category were considered.
The median nearest-neighbor core separation is similar, $\sim$0.14 pc for all three samples. Interestingly, the histograms show steep rises 
(at $\sim$0.07\,pc) and falls (at $\sim$0.2\,pc), which bracket the characteristic $\sim$0.1\,pc inner width of nearby filaments \citep{Arzoumanian+2019}.

Similar distributions of nearest-neighbor core separations are found for the samples of Aquila cores from \citet{Konyves+2015}, for which the median 
nearest-neighbor core separation is 0.09 pc, and the most significantly rising and falling histogram bins are at separations of $\sim$0.06\,pc and 
$\sim$0.1\,pc, respectively.

\subsection{Mass segregation of prestellar cores}\label{sec:massSegreg}

Investigating how dense cores of different masses are spatially distributed (and possibly segregated) is of vital interest for understanding 
the formation and early dynamical evolution of embedded protostellar clusters.

\begin{figure}[!h]
 \centering
  \begin{minipage}{1.0\linewidth}
   \centering
    \resizebox{0.9\hsize}{!}{\includegraphics[angle=0]{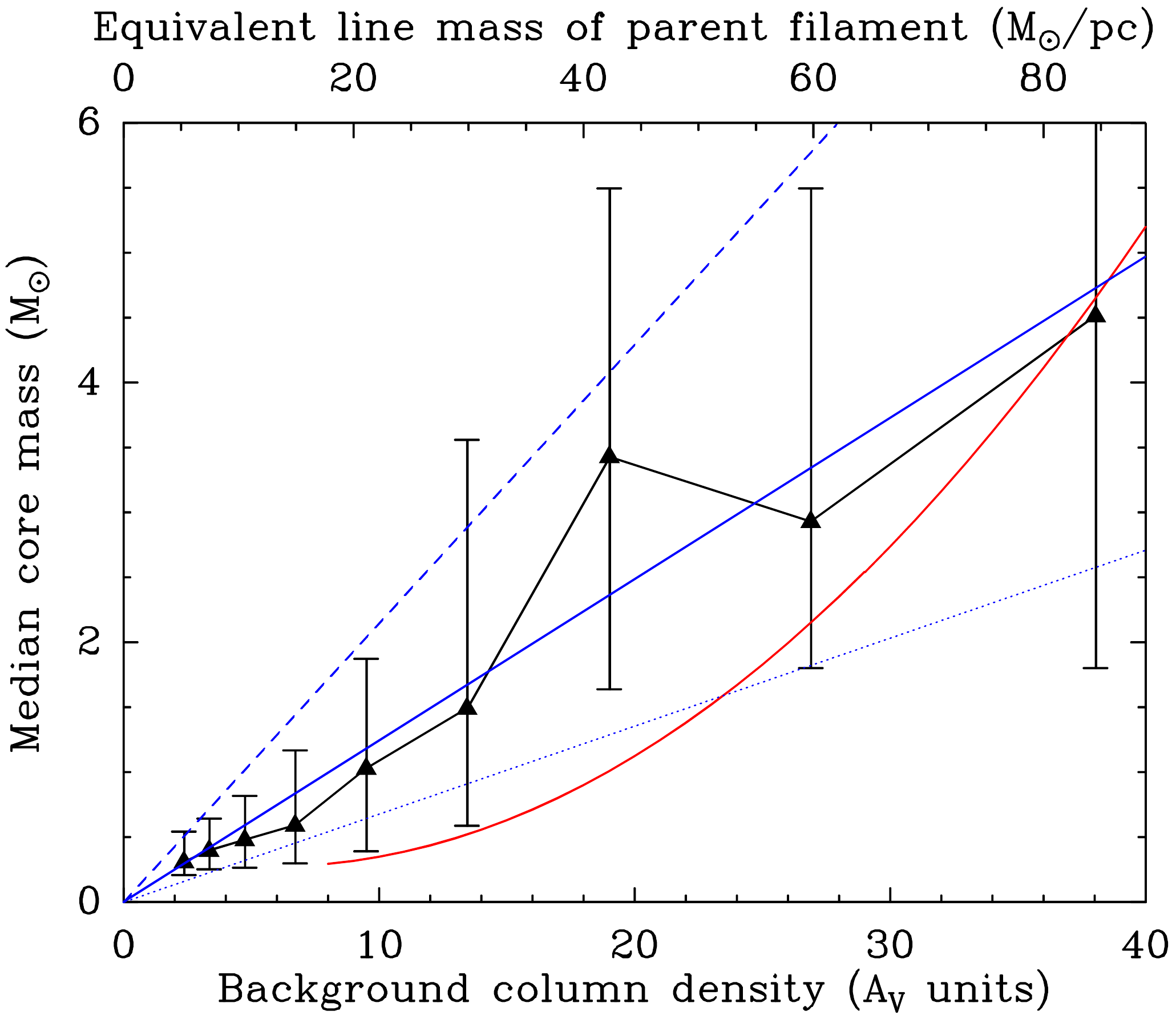}}
    \resizebox{0.9\hsize}{!}{\includegraphics[angle=0]{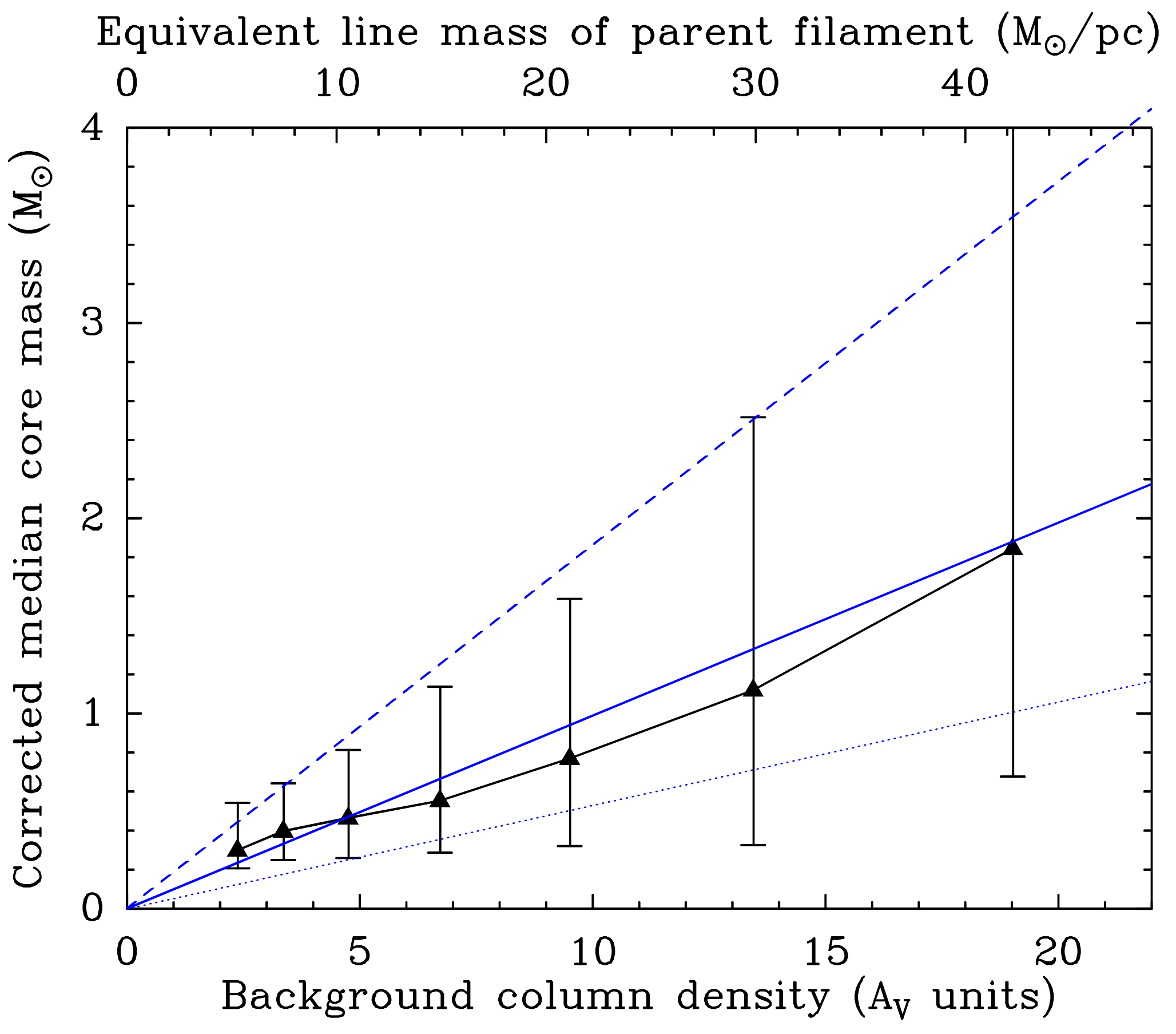}}
   \end{minipage}
   \caption{ 
   {\bf Top:} Median prestellar core mass versus background column density (black triangles). The error bars correspond to the inter-quartile
   range in observed masses for each bin of background column density. The solid, dashed and dotted blue lines are linear fits to the observed
   median, upper-quartile, and lower-quartile masses as function of background column density, respectively. The red curve shows how the mass
   at the 50\% completeness level varies with background column density according to the model completeness curves of  
   Appendix~\ref{sec:appendix_sim} (see Fig.~\ref{fig_completeness}). 
   The upper x-axis provides an approximate mass-per-unit-length scale, assuming the local background around each core 
   is dominated by the contribution of a parent filament with a characteristic width $W_{\rm fil} = 0.1\,$pc (see end of Sect.~\ref{sec:threshold}).
   {\bf Bottom:} Same as upper panel after correcting the distribution of observed core masses 
   for incompleteness effects before estimating the median core mass and inter-quartile range in each column density bin (see Sect.~\ref{sec:massSegreg}). 
   Both the median prestellar core mass and the dispersion in core masses increase roughly linearly with background column density.    
   }
  \label{fig:medMcore_bgCD}%
\end{figure} 

Figure~\ref{fig:medMcore_bgCD} shows the median mass of {\it candidate} prestellar cores as a function of ambient column density in the parent cloud. 
The upper panel displays the median core mass as observed in each column density bin before any correction for incompleteness effects, while the lower 
panel displays median core masses corrected for incompleteness. The incompleteness correction was achieved as follows, using the model completeness curves 
presented in Appendix~\ref{sec:appendix_sim} as a function of background column density (see Fig.~\ref{fig_completeness}). 
In each bin of background column densities as estimated by \textsl{getsources}, each observed core mass was weighted by the inverse of the model completeness 
fraction before computing the weighted median mass and interquartile range for that bin. It can be seen in Fig.~\ref{fig:medMcore_bgCD} that both the median 
prestellar core mass and the dispersion in core masses increase with background column density.
Moreover, this increase in core mass and mass spread is robust and not dominated by incompleteness effects which make low-mass cores somewhat more difficult 
to detect at higher background column densities (see lower panel of Fig.~\ref{fig:medMcore_bgCD}).

Taking differential incompleteness effects into account (cf. Fig.~\ref{fig_completeness}), this suggests that both higher- and lower-mass prestellar cores 
may form in higher density clumps and filaments, a trend also seen in the Aquila cloud \citep[cf.][]{Andre+2017}. 
A simple interpretation of this trend is that the effective Jeans mass increases roughly linearly with background column density, because the velocity 
dispersion is observed to scale roughly as the square root of the column density \citep[see, e.g.,][]{Heyer+2009, Arzoumanian+2013}, while the thermal 
Jeans mass decreases roughly linearly with background column density \citep[see Eq.~1 of][]{Andre+2019}.

\subsection{Prestellar core mass function}\label{sec:CMF}

\begin{figure}[!h]
 \centering
  \resizebox{1.0\hsize}{!}{\includegraphics[angle=0]{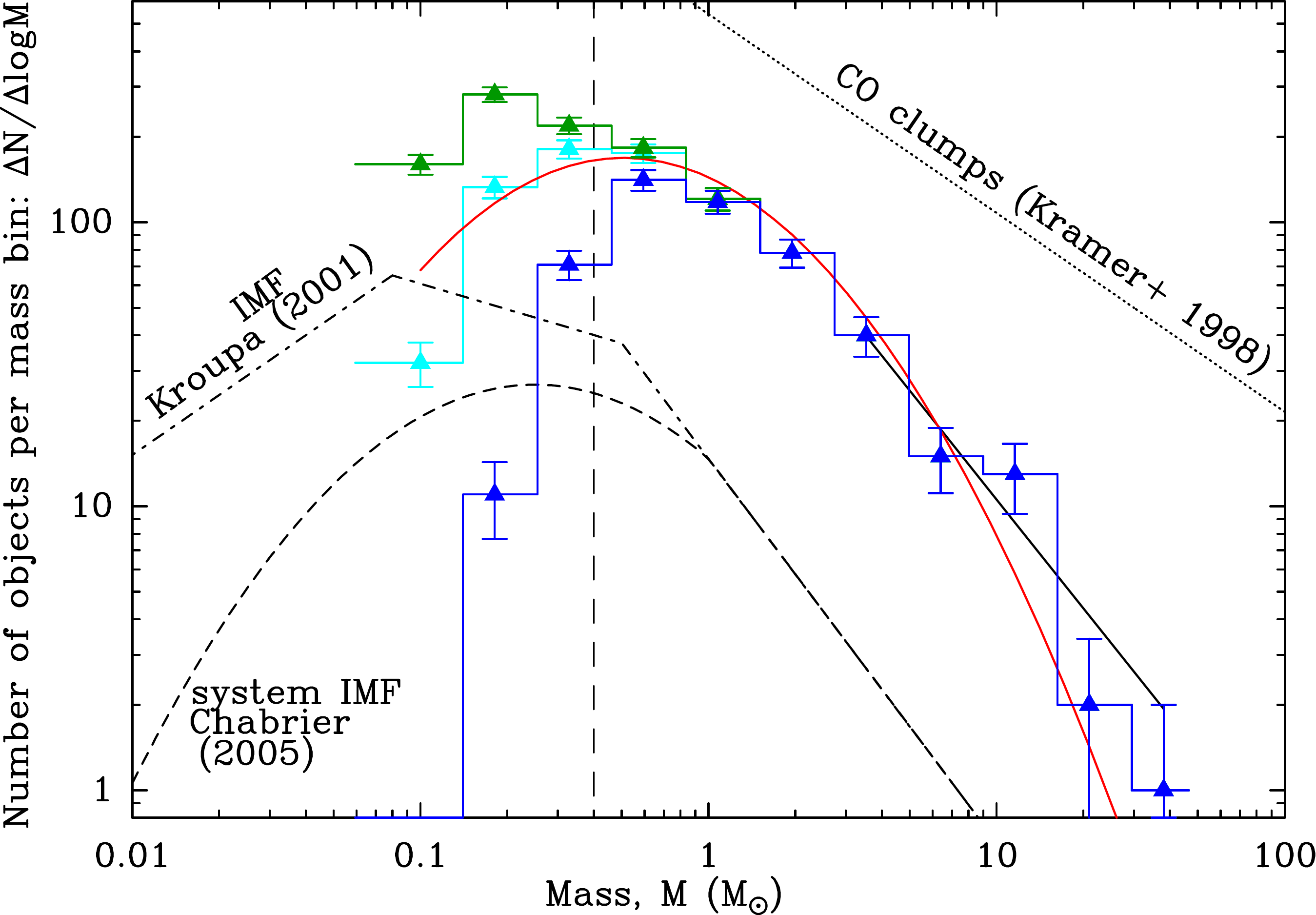}}
  \caption{ 
   Differential core mass function ($\Delta N$/$\Delta$log$M$) of the 1768 starless cores (green histogram), 
   804 {\it candidate} prestellar cores (light blue histogram), and 490 {\it robust} prestellar cores (dark blue histogram) 
   identified with {\it Herschel} in the whole Orion~B field. 
   The error bars correspond to $\sqrt{N}$ statistical uncertainties, and the vertical dashed line shows the completeness limit of
   the prestellar sample at $\sim$\,0.4\,$M_\odot$. 
   A lognormal fit to the CMF of {\it candidate} prestellar cores (solid red curve), as well as a power-law fit to the high-mass end of the CMF 
   (black solid line) are superimposed. 
   The lognormal fit peaks at 0.5\,$M_\odot $, and has a standard deviation of $\sim$0.52 in log$_{10}M$. 
   The power-law fit has a slope of --1.27 $\pm 0.24$ (compared to a Salpeter slope of $-1.35$ in this format).
   The IMF of single stars \citep[corrected for binaries -- e.g.,][]{Kroupa2001},  the IMF of multiple systems \citep[e.g.,][]{Chabrier2005}, 
   and the typical mass distribution of CO clumps \citep[e.g.,][]{Kramer+1998} are also shown for comparison.
   }
  \label{fig:CMF}%
\end{figure}

The core mass functions derived for the samples of 1768 starless cores, 804 {\it candidate} prestellar cores, and 490 {\it robust} prestellar cores identified 
in the whole Orion~B cloud are shown as differential mass distributions in Fig.~\ref{fig:CMF} (green, light blue, and dark blue histograms, respectively). 
We note that all three distributions merge at high masses, which simply reflects the fact that all of the starless cores identified above $\sim 1\, M_\odot$ 
are dense and massive enough to be self-gravitating, hence prestellar in nature.

The high-mass end of the prestellar CMF above $2\, M_\odot$ is well fit by a power-law mass function, 
$\Delta N$/$\Delta$ log$M$ $\propto$ $M^{-1.27 \pm 0.24}$ (see Fig.~\ref{fig:CMF}), which is consistent with the Salpeter power-law IMF 
($\Delta N$/$\Delta$ log$M_\star$ $\propto$ $M_\star^{-1.35}$ in the same format -- Salpeter 1955). 
The black solid line in Fig.~\ref{fig:CMF} is a least-squares fit. 
A non-parametric Kolmogorov-Smirnov (K-S) test \citep[see, e.g.][]{Press+1992} confirms that the CMF above $2\, M_\odot $ is consistent with 
a power-law mass function, $\Delta N$/$\Delta $log$M$ $\propto$ $M^{-1.33 \pm 0.13}$, at a significance level of 95\%. 

The observed peak of the prestellar CMF at $\sim 0.4$--$0.5\,M_\odot$ is similar to that seen in the Aquila region (cf. Fig.~16 of \citealp{Konyves+2015})
but is here extremely close to the estimated $80\% $ completeness limit of 0.4\,$M_\odot$ in observed core mass (cf. Sect.~\ref{sec:completeness} and 
Appendix~\ref{sec:appendix_sim}). It may therefore not be significant.

\begin{figure}[!h]
 \centering
  \resizebox{1.0\hsize}{!}{\includegraphics[angle=0]{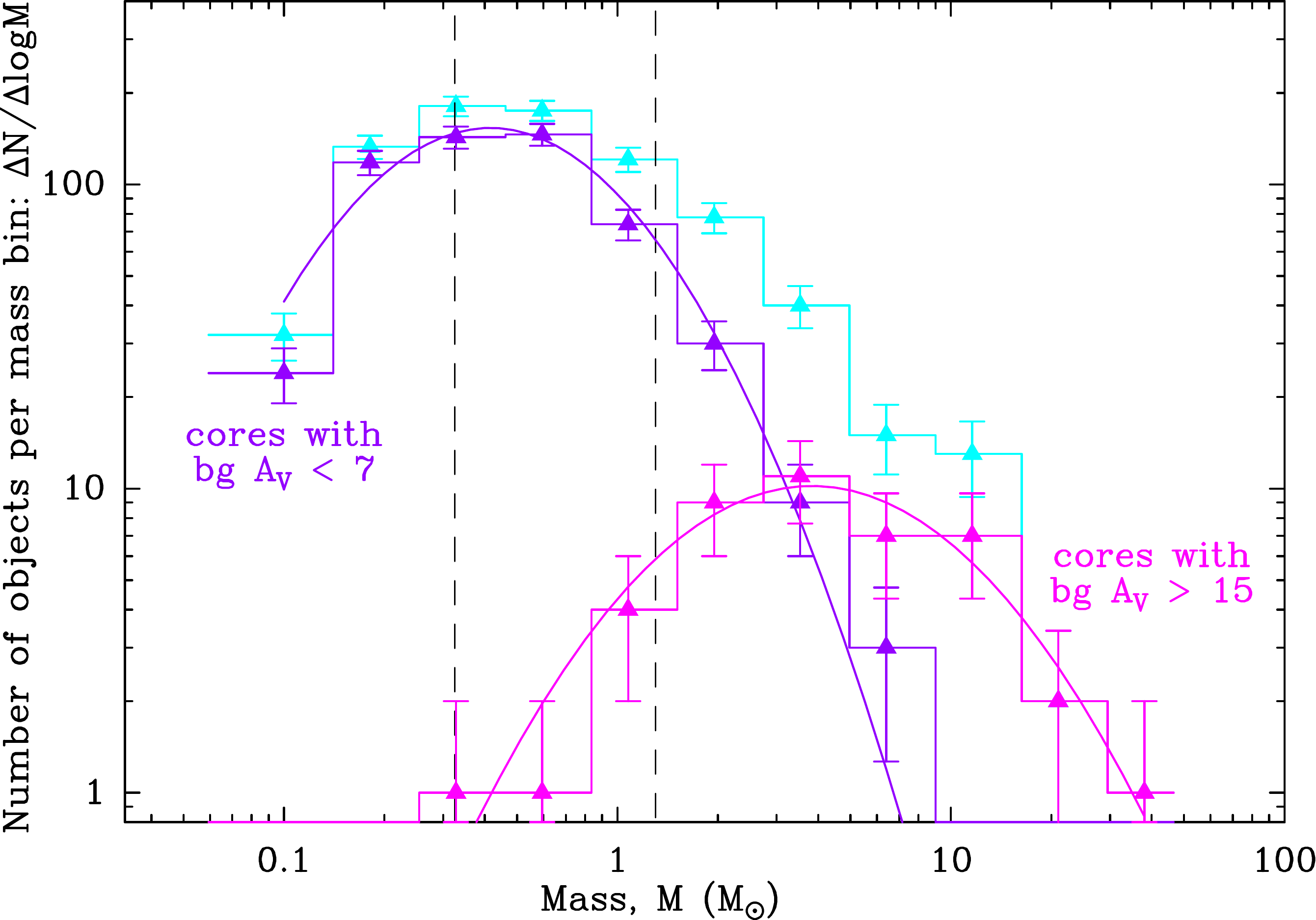}}
  \caption{Comparison of the differential CMF ($\Delta N$/$\Delta $log$M$) derived for the 558 {\it candidate} prestellar cores lying in areas of 
  low background column density ($A_{\rm V}^{\rm bg} < 7$, purple histogram) with the CMF of the 43 {\it candidate} prestellar cores lying in areas 
  of high column density ($A_{\rm V}^{\rm bg} \geq 15$, magenta histogram). 
  Lognormal fits are shown in both cases, and have peaks at 0.4\,$M_\odot$ and 3.8\,$M_\odot$, respectively. 
  Vertical dashed lines show the 80\% completeness limits of the low and high column density samples at 0.33\,$M_\odot$ and 1.3\,$M_\odot$, respectively. 
  The overplotted histogram in light blue is the CMF of all 804 {\it candidate} prestellar cores (same as in Fig.~\ref{fig:CMF}).  
   }
  \label{fig:CMF_splitAv}%
\end{figure}

In order to further test the mass segregation result discussed in Sect.~\ref{sec:massSegreg} above, Figure~\ref{fig:CMF_splitAv} compares the CMF of 
{\it candidate} prestellar cores observed at low and intermediate background column densities ($A_{\rm V}^{\rm bg} < 7$, purple histogram) with that observed 
at high background column densities ($A_{\rm V}^{\rm bg} \geq 15$, magenta histogram).
It can be seen that the prestellar CMF observed at high background $A_{\rm V}$ values peaks at an order of magnitude higher mass than the prestellar CMF observed 
at low and intermediate background $A_{\rm V}$ values, that is, $\sim 3.8\, M_\odot $ versus $\sim 0.4\, M_\odot $, respectively.
Indeed, the majority of prestellar cores found at low and intermediate $A_{\rm V}^{\rm bg}$ have significantly lower masses than the majority of prestellar cores 
found at high $A_{\rm V}^{\rm bg}$, which corresponds to the findings of Sect.~\ref{sec:massSegreg} (see Fig.~\ref{fig:medMcore_bgCD}). 
The split in background visual extinction at $A_{\rm V}^{\rm bg} < 7$ and $A_{\rm V}^{\rm bg} \geq 15$ was somewhat arbitrary, but similar --albeit less significant-- 
results are obtained with other {\it low} and {\it high} background extinction limits. 

Interestingly, a similar difference between the prestellar CMFs observed at high and low background $A_{\rm V}$ values is also seen in the Aquila cloud 
\citep[][]{Andre+2017}, and a related result was also found in Orion~A, where \citet{Polychroni+2013} reported two distinct mass distributions for cores 
``on'' and ``off'' prominent filaments, that is, at high and low background $A_{\rm V}$, respectively.  
We also note that, based on recent ALMA observations, \citet{Shimajiri+2019} found a CMF peaking at $\sim 10\, M_\odot $ in 
the very dense, highly supercritical filament of NGC~6334 (with $M_{\rm line} \approx 500$--$1000\, M_\odot /{\rm pc} \gg M_{\rm line, crit} $, corresponding to 
$A_{\rm V}^{\rm bg} \geq 100$), 
which is again consistent with the same trend.

\subsection{Mass budget in the cloud}\label{sec:budget}

Using our {\it Herschel} census of prestellar cores and filaments, we can derive a detailed mass budget for the Orion~B cloud. 
In the intermediate column density regions ($A_{\rm V}^{\rm bg} \sim 3-7$), a large fraction, $\sim$60\%, of the gas mass is in the form of 
filaments but only $\sim 3-5\% $ of the cloud mass is in the form of prestellar cores.
In higher column density regions ($A_{\rm V}^{\rm bg} \gtrsim 7$), a similarly high fraction of the cloud mass is in the form of filaments 
and a fraction $f_{\rm pre} \sim 22\% $ of the mass is in the form of prestellar cores. 

\begin{figure}[!h]
 \begin{center}
  \resizebox{1.0\hsize}{!}{\includegraphics[angle=0]{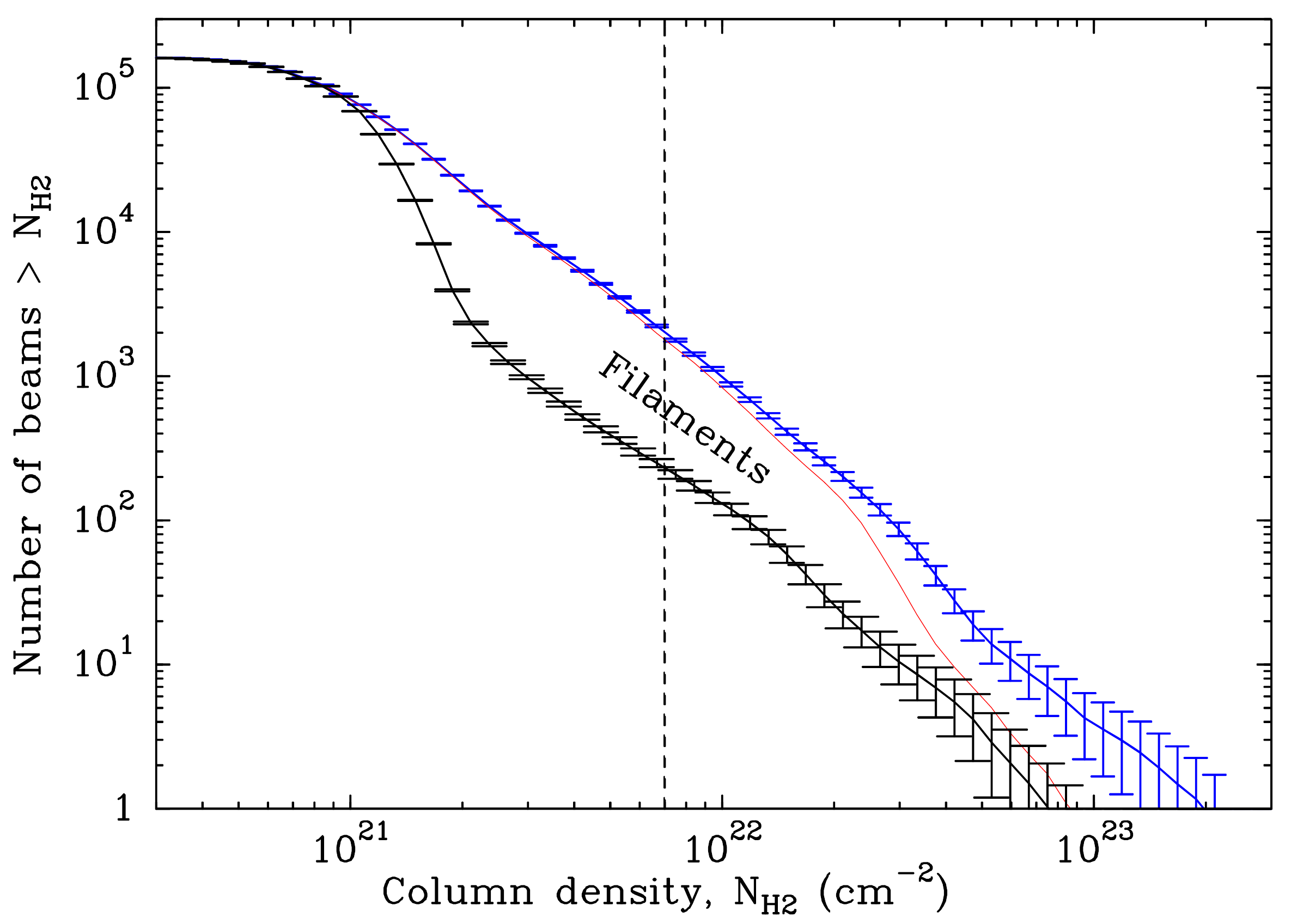}}
 \end{center}
  \caption{Comparison of the global column density PDF in Orion~B (blue histogram) to the column density PDF measured after subtraction of dense cores 
   (red solid line) and the PDF measured after subtraction of both dense cores and filaments (black solid line with statistical error bars).
   The blue histogram is identical to that shown in Fig.~\ref{fig:cdPDF} (left). 
   The vertical dashed line marks the approximate separation between low-to-intermediate and high column density values at $A_{\rm V}^{\rm bg} \sim$ 7. 
   }
  \label{fig:budget_pdf}%
\end{figure}

In an attempt to quantify further the relative contributions of cores and filaments to the cloud material as a function of column density, we compare in 
Fig.~\ref{fig:budget_pdf} the column density PDFs observed for the cloud before any component subtraction (blue histogram, identical to the PDF shown in 
Fig.~\ref{fig:cdPDF} left), after subtraction of dense cores (red solid line), and after subtraction of both dense cores and filaments (black solid line). 
To generate this plot, we used \textsl{getsources} to create a column density map of the cloud after subtracting the contribution of all compact cores, 
and \textsl{getfilaments} to construct another column density map after also subtracting the contribution of filaments. Although there are admittedly 
rather large uncertainties involved in this two-step subtraction process, the result clearly suggests that filaments dominate the mass budget in Orion~B 
throughout the whole column density range. 
In the Lupus region, \citet{Benedettini+2015} also found that the power-law tail of their PDF is dominated by filaments. 

\begin{figure}[]
 \centering
  \resizebox{1.0\hsize}{!}{\includegraphics[angle=0]{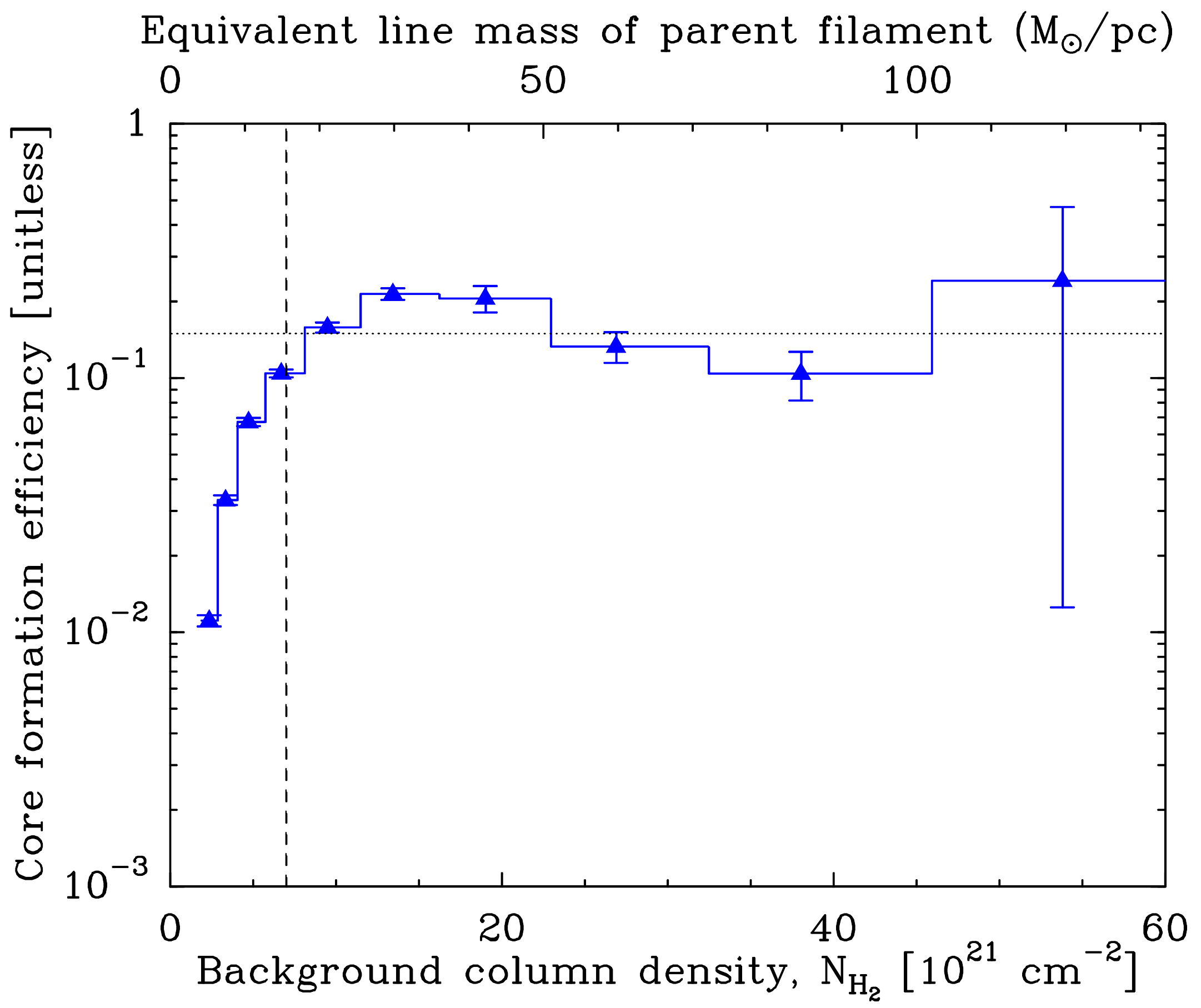}}
  \caption{
   Differential core formation efficiency (CFE) as function of ambient cloud column density expressed in $A_{\rm V}$ units (blue histogram with error bars), 
   obtained by dividing the mass in the form of {\it candidate} prestellar cores in a given column density bin 
   by the cloud mass observed in the same column density bin.
   The upper x-axis provides an approximate mass-per-unit-length scale, assuming the local background around each core is dominated by the contribution 
   of a parent filament with a characteristic width $W_{\rm fil} = 0.1\,$pc (see end of Sect.~\ref{sec:threshold}).
   The vertical dashed line marks the fiducial threshold at $A_{\rm V}^{\rm bg} \sim$ 7. The horizontal dotted line marks the rough asymptotic value 
   of $\sim \,$15\% for the CFE at $A_{\rm V} > 15$. 
   }
  \label{fig:CFE}       
\end{figure}

\subsection{Column density transition toward prestellar core formation}\label{sec:threshold}

The existence of thresholds for star formation have been suspected for a long time, based on the results of ground-based millimeter and submillimeter surveys 
for cores in, for instance, the Taurus, Ophiuchus, and Perseus clouds \citep[e.g.,][]{Onishi+1998,Johnstone+2004,Hatchell+2005}. Following these early claims, the 
HGBS results in Aquila and Taurus provided a much stronger case for a column density threshold for prestellar core formation \citep{Andre+2010, Konyves+2015, Marsh+2016} 
thanks to the excellent column density sensitivity and high spatial dynamic range of the {\it Herschel} imaging data, making it possible to probe prestellar 
cores and the parent background cloud simultaneously. 
The {\it Herschel} results on the characteristic inner width of filaments \citep{Arzoumanian+2011,Arzoumanian+2019} and the tight connection observed between 
cores and filaments  (cf. Sect.~\ref{sec:cores_filams}) also led to a simple interpretation of this threshold in terms of the critical mass per unit length 
of nearly isothermal cylinder-like filaments \citep{Andre+2014}.

\begin{figure}[]
 \centering 
  \begin{minipage}{1.0\linewidth}
   \centering 
    \resizebox{0.98\hsize}{!}{\includegraphics[angle=0]{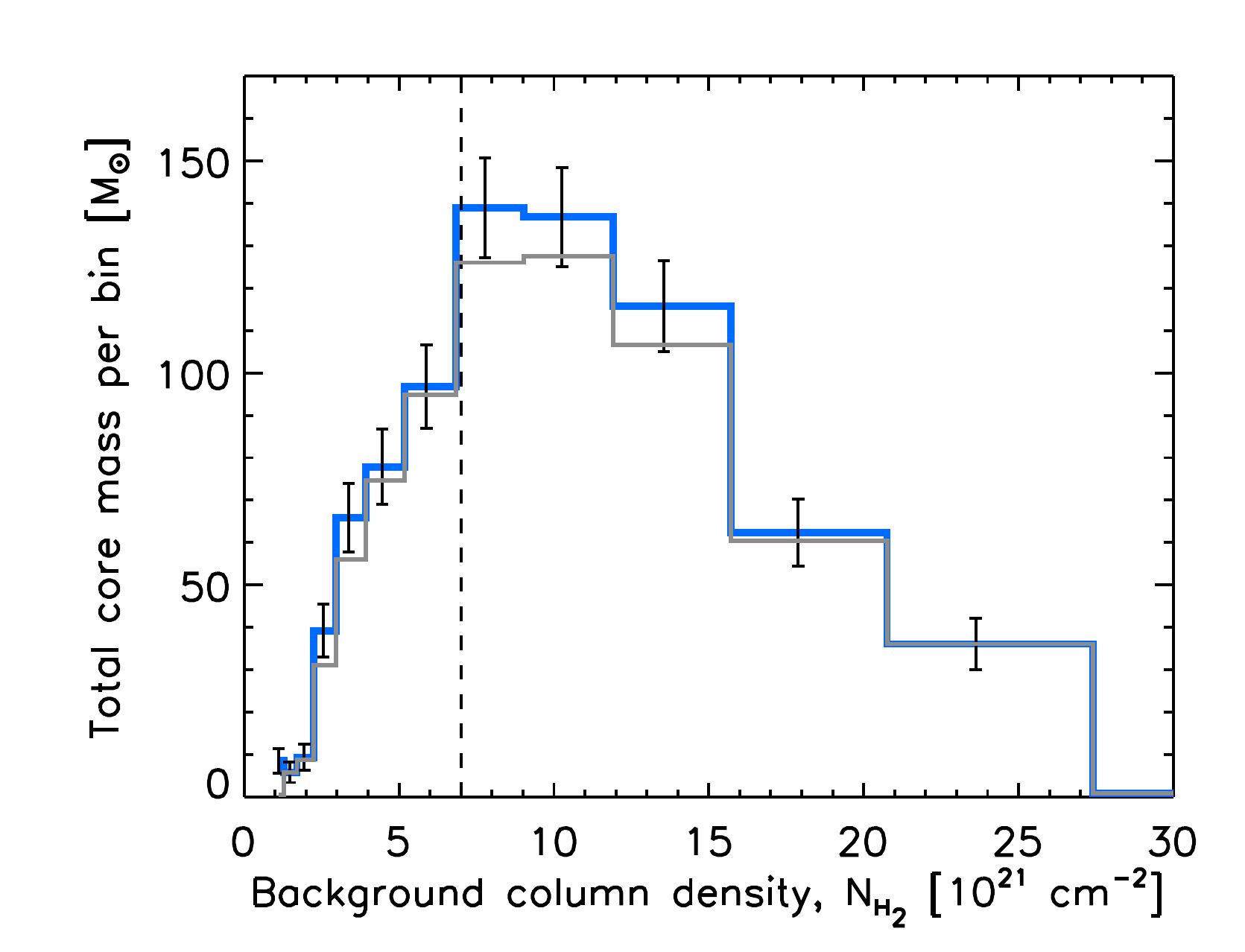}}
    \resizebox{0.98\hsize}{!}{\includegraphics[angle=0]{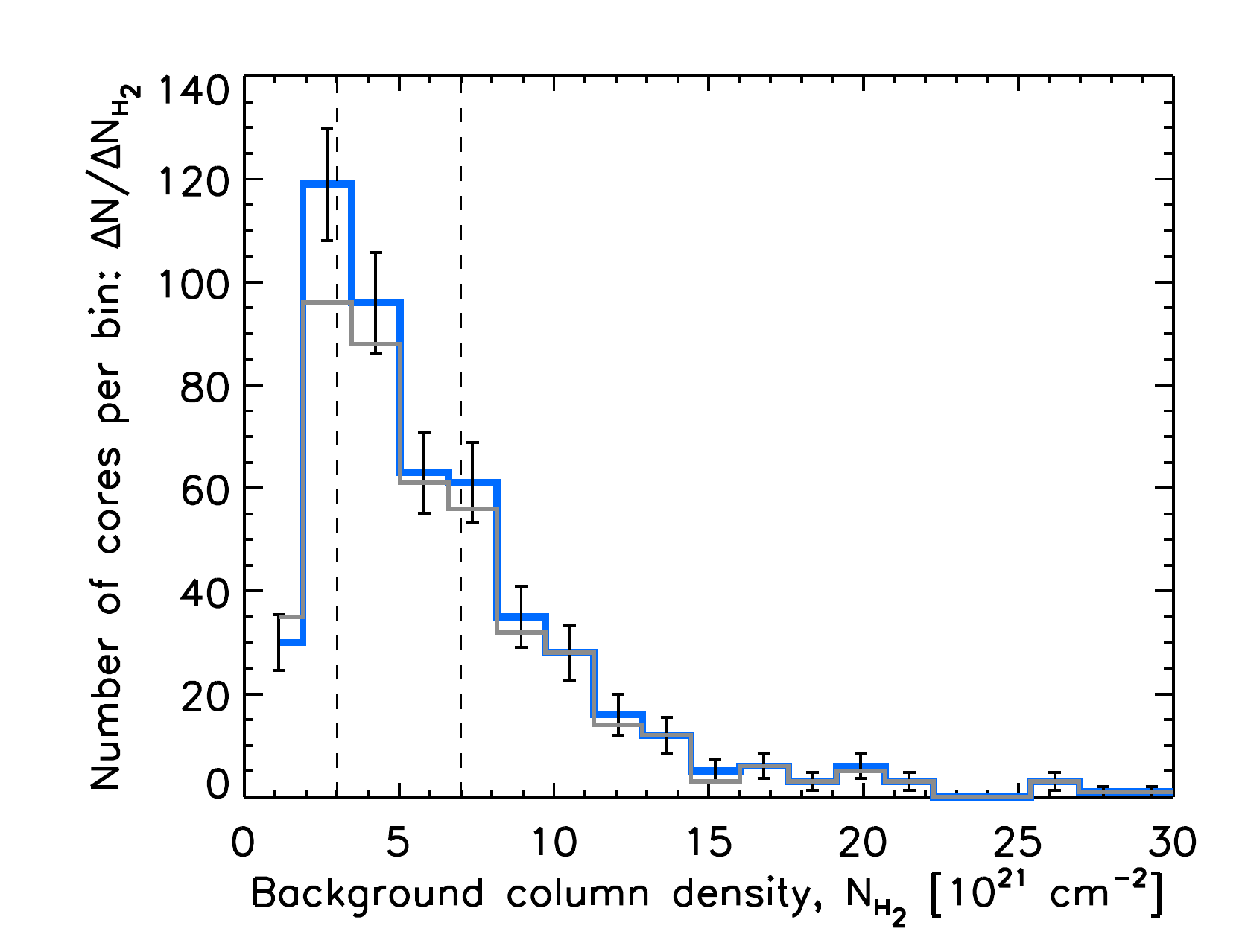}}
  \end{minipage}
  \caption{{\bf Top:} Total mass in form of prestellar cores as function of ambient cloud column density in Orion~B (blue histogram). 
   {\bf Bottom:} Distribution of background column densities for the 490 {\it robust} prestellar cores (blue histogram). The vertical dashed lines mark 
   $A_{\rm V}^{\rm bg} \sim$ 3 and 7\,mag.
   The overlaid gray histograms show the same distributions excluding the cores lying in the northern region L1622, 
   which are at a more uncertain distance (see Sect.~\ref{sec:intro} and Sect.~\ref{sec:post_selection} for details).
   }
  \label{fig:Mcore_bgCD}       
\end{figure}

To investigate whether a column density threshold for core formation is present in Orion~B, and following the study of \citet{Konyves+2015} in Aquila, 
we plot in Fig.~\ref{fig:CFE} the differential prestellar core formation efficiency 
$ {\rm CFE_{obs}}(A_{\rm V}) = \Delta M_{\rm cores}(A_{\rm V})/\Delta M_{\rm cloud}(A_{\rm V}) $ 
as a function of the background cloud column density in the Orion~B complex. 
This was obtained by dividing the mass $\Delta M_{\rm cores}(A_{\rm V})$ of the {\it candidate} prestellar cores identified with {\it Herschel} in a given bin of 
background column densities (expressed in $A_{\rm V}$ units) by the total cloud mass $\Delta M_{\rm cloud}(A_{\rm V}) $ in the same bin.
It can be seen in Fig.~\ref{fig:CFE}  that $ {\rm CFE_{obs}}(A_{\rm V})$ exhibits a steep rise in the vicinity of $A_{\rm V}^{\rm bg} \sim 7$ and resembles a 
smooth step function, fairly similar to that observed in the Aquila cloud \citep[see Fig.~12 of][]{Konyves+2015}. 
The {\it Herschel} observations indicate a sharp transition between a regime of very low $ {\rm CFE_{obs}}$ at 
$A_{\rm V}^{\rm bg} < 5$ and a regime where $ {\rm CFE_{obs}}(A_{\rm V})$ is an order of magnitude higher and fluctuates around a roughly constant value of 
$\sim 15\%$--$20\%$. 
The combination of this sharp transition with the column density PDF of Fig.~\ref{fig:cdPDF} implies that, at least in terms of core mass, most prestellar 
core formation occurs just above $A_{\rm V}^{\rm bg} \sim 7$ (see top panel of Fig.~\ref{fig:Mcore_bgCD}). 
Based on Fig.~\ref{fig:CFE} and Fig.~\ref{fig:Mcore_bgCD} (top panel), we argue for the presence of a true physical ``threshold'' for the formation of prestellar 
cores in Orion~B around a fiducial value $A_{\rm V}^{\rm bg} \sim 7$, which we interpret here as well as resulting from the mass-per-unit-length threshold 
above which molecular filaments become gravitationally unstable. 
It should be stressed, however, $A_{\rm V}^{\rm bg} \sim 7$ is only a {\it fiducial} value for this threshold, which for various reasons should rather be 
viewed as a smooth transition \citep[see discussion in][]{Andre+2014}. 
We also note some environmental variations in the details of this transition between Orion~B and Aquila. In Orion~B, the raw number of prestellar cores per 
column density bin peaks at $A_{\rm V}^{\rm bg} \sim 3$ (see bottom panel of Fig.~\ref{fig:Mcore_bgCD}) or at most at $A_{\rm V}^{\rm bg} \sim 5$--6 
if we take into account a correction factor $A_{\rm V, HGBS} / A_{\rm V, ext} \sim 0.5$--0.6 between the $A_{\rm V}^{\rm bg}$ values derived from the 
HGBS column density map and those derived from near-infrared extinction maps in the range $A_{\rm V} = 3-7$\,mag (see Sect.~\ref{sec:cd_t_maps}). 
In Aquila, the raw number of prestellar cores peaks at a significantly higher background column density $A_{\rm V}^{\rm bg} \sim 8$ 
(see Fig.~11a of \citealp{Konyves+2015}). 
Despite this difference, the shapes of the prestellar core formation efficiency as a function of background $A_{\rm V}$, ${\rm CFE_{obs}}(A_{\rm V})$, are 
remarkably similar in the two clouds (compare Fig.~\ref{fig:CFE} here with Fig.~12 of \citealp{Konyves+2015}). 

Core formation efficiency is discussed as a function of background column density or $A_{\rm V}^{\rm bg}$ here, as column density is the most 
directly observable quantity, but we can also provide a rough conversion in terms of background volume density and mass per unit length. In the 0.1-pc 
filament paradigm of core formation \citep{Andre+2014}, the local background is dominated by the parent filament of each core and there is a direct 
correspondence between the average values of the surface density $\Sigma_{\rm fil} \equiv \mu_{\rm H_2}\, m_{\rm H}\, N^{\rm fil}_{\rm H_2} $, volume density 
$\rho_{\rm fil} \equiv \mu_{\rm H_2}\, m_{\rm H}\, n^{\rm fil}_{\rm H_2} $, and line mass $M_{\rm line}$ of the parent filament. 
For a cylindrical parent filament of width $W_{\rm fil} $, one has: 
$< \Sigma_{\rm fil}>  = M_{\rm line} / W_{\rm fil}$ and $< \rho_{\rm fil}>  = 4\, M_{\rm line} / (\pi \, W_{\rm fil}^2)  = 4\, < \Sigma_{\rm fil}>$$/(\pi \, W_{\rm fil}) $. 
In particular, adopting a uniform filament width $W_{\rm fil} = 0.1\,$pc, the fiducial column density threshold at $A_{\rm V}^{\rm bg} \sim 7$ 
corresponds to a threshold line mass $M_{\rm line, th} \sim 15\, M_\odot/$pc, very close to the thermal value of the critical mass per unit length 
$M_{\rm line, crit} \sim 16\, M_\odot/$pc for $T \sim 10\,$K gas filaments \citep{Ostriker1964,InutsukaMiyama1997}, and to a threshold volume density 
$n^{\rm th}_{\rm H_2}  \sim 3 \times 10^4\, {\rm cm}^{-3}$.  
Alternatively, under the extreme assumption that the background cloud structure is completely isotropic, one can derive another conversion between column 
density and typical volume density \citep[cf. Fig.~A.1d of][]{Bron+2018}, which leads to a significantly lower value of only $\sim 10^3\, {\rm cm}^{-3}$ for 
the threshold volume density. 
Obviously, only an order-of-magnitude estimate of the threshold volume density can be derived. 
Because the {\it Herschel} data indicate that the local background around each core is highly anisotropic (e.g., filamentary), we nevertheless tend to favor 
our higher estimate of the threshold volume density (i.e., $ \sim 3 \times 10^4\, {\rm cm}^{-3}$). 

\subsection{Lifetime of prestellar cores in Orion~B}\label{sec:lifetime}

As our {\it Herschel} survey provides an essentially complete census of prestellar cores in the Orion~B clouds, the core statistics can be used to 
set constraints on the typical lifetime of prestellar cores and the timescale of the core formation process. We may derive a rough lifetime 
by comparing the number of prestellar cores found here with {\it Herschel} to the number of Class~II YSOs detected in the mid-IR  by {\it Spitzer} 
in the same region. The same approach was first used by  \citet{Beichman+1986} based on {\it IRAS} data.
Our findings will hold, assuming that 1) all prestellar cores will evolve into YSOs in the future, and 2) star formation proceeds at a roughly constant rate, 
at least when averaged over the entire Orion~B cloud complex.

The three HGBS  tiles imaged with {\it Herschel} around NGC~2023 and 2024, NGC~2068 and 2071, and L1622 (see, e.g., Fig.~\ref{fig_T_NH2_maps}) were also 
observed by IRAC as part of the {\it Spitzer} Orion survey \citep{Megeath+2012} and thus provide a common field where {\it Herschel} and {\it Spitzer} 
source counts can be directly compared.
The various numbers of dense cores and YSOs identified by the two surveys in this common field are as follows:
\begin{itemize}
  \item {\it Herschel} starless cores: 1032; above the mass completeness limit of 0.4~$M_\odot$: 338
  \item {\it Herschel} {\it candidate} prestellar cores: 538; above 0.4~$M_\odot$: 333
  \item {\it Herschel} {\it robust} prestellar cores: 361; above 0.4~$M_\odot$: 312
  \item {\it Spitzer} Class II YSOs: 312; corrected for incompleteness: 427
  \item {\it Spitzer} entire sample of Class~0/I--flat spectrum--Class~II YSOs: 428; corrected for incompleteness: 586
\end{itemize}

For our lifetime estimates, we used the numbers of  {\it Spitzer} sources corrected for incompleteness with the help of {\it Chandra} data and 
other techniques as described in detail in \citet{Megeath+2016}. In practice, the corrected numbers of {\it Spitzer} sources were derived by multiplying 
the uncorrected number counts by a factor of 1.37, based on Table 1 of \citet[][]{Megeath+2016}. 

If we adopt a reference lifetime of 2~Myr for Class II YSOs \citep{Evans+2009}, the comparison between {\it Herschel} and {\it Spitzer} source 
counts leads to lifetimes ranging from $2 \times 312/586 \approx 1.1\,$Myr to  $2 \times 538/427 \approx 2.5\,$Myr for the prestellar core phase, 
depending on whether we consider only {\it robust} prestellar cores above our mass completeness limit of 0.4~$M_\odot$ or all {\it candidate} prestellar cores, 
and all {\it Spitzer} YSOs up to Class~II or only YSOs nominally classified as Class~IIs. 
Our best estimate of the prestellar core lifetime in Orion~B is thus $t_{\rm pre}^{\rm OrionB} = 1.7_{-0.6}^{+0.8}\,$Myr, derived from the ratio 361/427 
of the total number of {\it Herschel} prestellar cores to the number of {\it Spitzer} Class II YSOs corrected for incompleteness. 
This is slightly longer than, but consistent within errors with, the prestellar lifetime estimate derived by \citet{Konyves+2015} in the Aquila cloud, 
$t_{\rm pre}^{\rm Aquila} = 1.2 \pm 0.3 \,$Myr.

As a result of non-uniform incompleteness effects between the {\it Spitzer} and the {\it Herschel} sample, the uncertainties in these lifetimes are admittedly 
rather large (at least a factor of $\sim 2$).
Additional sources of errors possibly affecting the above lifetime estimate are:
1) \citet{Megeath+2012} used infrared spectral index values uncorrected for extinction (in Aquila, the compared {\it Spitzer} sources had been dereddened), 
2) the classification of cores and YSOs is somewhat uncertain, and 3) the two initial assumptions mentioned above (i.e., prestellar nature of the cores 
and steady state) may not entirely hold.

\section{Summary and conclusions}\label{sec:conclusions}

Using SPIRE and PACS parallel-mode data from the {\it Herschel} Gould Belt survey, 
we obtained an extensive census of dense cores in the Orion~B star-forming region. 
We analyzed both the physical properties of the cores and the connection with interstellar 
filaments. Our main results and conclusions can be summarized as follows:
         
\begin{enumerate}
  \item As in other HGBS clouds, the high-resolution ($\sim 18\arcsec $ or $\sim \,$0.03~pc) {\it Herschel} column density map of Orion~B 
  shows a highly filamentary distribution of matter  and the column density PDF features a prominent power-law tail above $A_{\rm V} \sim 3$~mag.
  \item Based on multi-scale, multi-wavelength core extraction with the \textsl{getsources} algorithm we identified 1768 starless dense cores, 
  804 {\it candidate} prestellar cores, 490 {\it robust} prestellar cores, and 76 protostellar cores in the $\sim$19~deg$^2$ field that was imaged 
  with both SPIRE and PACS at five wavelengths between 70~$\mu$m and 500~$\mu$m.
  The $\sim 80\% $ mass completeness level 
  for {\it candidate} prestellar cores was estimated to be at an observed core mass of $\sim 0.4\, M_\odot $. 
  The {\it candidate} prestellar cores have an estimated median mass $\sim 0.5\, M_\odot $, a median deconvolved FWHM diameter $\sim 0.03$~pc, a median average 
  column density $\sim 4\times 10^{21}\, {\rm cm}^{-2}$,  
  and a median average volume density $\sim 3\times 10^4\, {\rm cm}^{-3}$.
  \item The identified dense cores are closely associated with filamentary structures. A very high fraction of prestellar cores (60--90\%) were found within 
  the 0.1\,pc inner portion of the filaments extracted with both DisPerSE and \textsl{getfilaments}. Considering the deep raw sample of filaments extracted 
  with DisPerSE, we find that almost all dense cores (> 90\%) lie within --stronger or fainter-- elongated filamentary structures. 
  Filaments may provide a more optimal environment for the growth of prestellar cores than spheroidal clumps.
  \item The most massive prestellar cores in our sample (with masses between $M \sim 2\, M_\odot$ and $\sim 10\, M_\odot$)
  tend to be spatially segregated in the highest column density portions of the Orion~B cloud,  at equivalent visual extinctions $A_{\rm V}^{\rm bg} \ga 15$--30. 
  In particular, the distribution of median prestellar core mass as a function of background column density (Fig.~\ref{fig:medMcore_bgCD}) shows that both the 
  median core mass and the dispersion in core masses increase roughly linearly with background column density. 
  This suggests that the prestellar core mass function (CMF) may not be strictly universal but may depend on the local  column density in the ambient cloud. 
  This also suggests that intermediate-mass and high-mass protostars may form only (or at least preferentially) in the dense inner portions of molecular clouds,
  supporting the notion of primordial mass segregation in stellar protoclusters.
  \item The global differential CMF derived for the whole sample of 804 {\it candidate} prestellar cores (or 490 {\it robust} prestellar cores) peaks at $\sim$0.5\,$M_\odot$ 
  (in $\Delta N$/$\Delta$log$M$ format), which is just above the estimated $80\% $ completeness limit of $\sim$0.4\,$M_\odot$ in Orion~B.
  The best power-law fit to the high-mass end of the prestellar CMF has a slope of --1.27 $\pm 0.24$, compared to the Salpeter slope of $-1.35$ (Fig.~\ref{fig:CMF}). 
  \item Splitting the global sample of Orion~B prestellar cores in a subsample of cores observed at high background column densities ($A_{\rm V}^{\rm bg} \geq 15$) 
  and a subsample of cores observed at low and intermediate background column densities ($A_{\rm V}^{\rm bg} < 7$), we found that the CMF of the high-$A_{\rm V}$ subsample 
  peaks at an order of magnitude higher mass ($\sim 3.8\,M_\odot$) than the lower $A_{\rm V}$ subsample ($\sim 0.4\, M_\odot$) (Fig.~\ref{fig:CMF_splitAv}). This result, which cannot be explained by 
  differential incompleteness effects, further illustrates the presence of significant mass segregation in the prestellar core population of the cloud. 
  \item The {\it Herschel} data in Orion~B confirm the presence of a column density transition around $A_{\rm V}^{\rm bg} \sim 5-10$, for the formation of prestellar 
  cores (Fig.~\ref{fig:CFE}). 
  Coupled to the close connection observed between cores and filaments (point 3), this is consistent with the view that dense filamentary structures with line masses 
  near or above the thermal value of the critical mass per unit length of isothermal cylinders provide the most favorable local environment for prestellar cores to grow 
  within molecular clouds. 
  \item The typical lifetime of {\it Herschel} prestellar cores in Orion~B was estimated to be 
 $t_{\rm pre}^{\rm OrionB} = 1.7_{-0.6}^{+0.8}\,$Myr, which is  consistent within error bars
 with the prestellar core lifetime $t_{\rm pre}^{\rm Aquila} = 1.2 \pm 0.3 \,$Myr derived by \citet{Konyves+2015} in the Aquila cloud. 
\end{enumerate}

\begin{acknowledgements}
SPIRE has been developed by a consortium of institutes led by Cardiff Univ. (UK) 
and including: Univ. Lethbridge (Canada); NAOC (China); CEA, LAM (France); 
IFSI, Univ. Padua (Italy); IAC (Spain); Stockholm Observatory (Sweden); 
Imperial College London, RAL, UCL-MSSL, UKATC, Univ. Sussex (UK); 
and Caltech, JPL, NHSC, Univ. Colorado (USA). This development has been 
supported by national funding agencies: CSA (Canada); NAOC (China); 
CEA, CNES, CNRS (France); ASI (Italy); MCINN (Spain); SNSB (Sweden); 
STFC, UKSA (UK); and NASA (USA). 
PACS has been developed by a consortium of institutes led by MPE
(Germany) and including UVIE (Austria); KUL, CSL, IMEC (Belgium); CEA,
OAMP (France); MPIA (Germany); IFSI, OAP/AOT, OAA/CAISMI, LENS, SISSA
(Italy); IAC (Spain). This development has been supported by the funding
agencies BMVIT (Austria), ESA-PRODEX (Belgium), CEA/CNES (France),
DLR (Germany), ASI (Italy), and CICT/MCT (Spain). 
V. K. would like to thank Richard Allison, and Richard Parker for their 
help and discussions on mass segregation with the MST method. 
This work has received support from the European Research Council 
under the European Union's Seventh Framework Programme 
(ERC Advanced Grant Agreements no. 291294 -- ORISTARS -- and no. 267934 -- MISTIC) 
and from the French National Research Agency (Grant no. ANR--11--BS56--0010 -- STARFICH). 
D. A. acknowledges support by FCT/MCTES through national funds (PIDDAC) by the grant UID/FIS/04434/2019.
N. S. and S. B. acknowledge support by the French ANR and the German DFG through 
the project GENESIS (ANR--16--CE92--0035--01/DFG1591/2--1).
G. J. W. gratefully acknowledges support from The Leverhulme Trust. We also acknowledge support 
from the French national programs of CNRS/INSU on stellar and ISM physics (PNPS and PCMI). 
This research has made use of the SIMBAD database, operated at CDS, Strasbourg (France), 
and of the NASA/IPAC Extragalactic Database (NED), operated by the Jet Propulsion Laboratory, 
California Institute of Technology, under contract with the National Aeronautics and Space Administration.
\end{acknowledgements}

\bibliographystyle{aa}
\bibliography{konyves_HGBS_oriB_1stgen_proofCorr}

\newpage

\begin{appendix} 

\section{Catalog of dense cores identified with {\it Herschel} in the Orion~B cloud complex}\label{sec:appendix_catalog}

With our {\it Herschel} SPIRE and PACS parallel-mode imaging survey of the Orion~B cloud complex, we identified a total of 1844 dense cores, including 
1768 starless cores and 76 protostellar cores. 
The master catalog listing the observed properties of all of these {\it Herschel} cores is available in online Table~\ref{tab_obs_cat}. A template of 
this online catalog is provided below to illustrate its form and content.

The derived properties (physical radius, mass, SED dust temperature, peak column density at the resolution of the 500~$\mu$m data, average column 
density, peak volume density, and average density) are given in online Table~\ref{tab_der_cat} for each core. 
A portion of this online table is also provided below.

The end of the catalog of observed core properties also includes 26 additional tentative cores 
which are possible extragalactic contaminants (see Section~\ref{sec:post_selection}). 
The derived properties of these objects are not provided in online Table~\ref{tab_der_cat}.

\clearpage

\begin{sidewaystable*}[htb]\tiny\setlength{\tabcolsep}{2.5pt}
\caption{Catalog of dense cores identified in HGBS maps of Orion~B (template, full catalog only provided online).} 
\label{tab_obs_cat}
{\renewcommand{\arraystretch}{0.5}
\begin{tabular}{|r|c|c c c c c c c c c c} 
\toprule[1.0pt]\toprule[0.5pt]  
 rNO      & Core name         &  RA$_{\rm 2000}$ &  Dec$_{\rm 2000}$        & Sig$_{\rm 070}$ &  $S^{\rm peak}_{\rm 070}$ &  $S^{\rm peak}_{\rm 070}$/$S_{\rm bg}$ &  $S^{\rm conv,500}_{\rm 070}$ &  $S^{\rm tot}_{\rm 070}$ &  FWHM$^{\rm a}_{\rm 070}$ &  FWHM$^{\rm b}_{\rm 070}$ &  PA$_{\rm 070}$  \\ 
          & HGBS\_J*          &  (h m s)         &  (\degr~\arcmin~\arcsec) &                 & (Jy/beam)                 &                                        & (Jy/beam$_{\rm 500}$)         &  (Jy)                    &  (\arcsec)                &  (\arcsec)                &  (\degr)         \\         
 (1)      & (2)               &  (3)             &  (4)                     &  (5)            &      (6) ~ $\pm$ ~ (7)    &  (8)                                   & (9)                           &    (10) ~ $\pm$ ~ (11)   &  (12)                     &  (13)                     &  (14)            \\         
\toprule[0.8pt] 
$\cdots$  &                   &                  &                          &                 &                           &                                        &                               &                          &                           &                           &                  \\
  204     & 054126.7-014453   & 05:41:26.78	 &   --01:44:53.2	    &	  0.0	      &  --8.09e-01 ~ 8.1e-01	  &  --0.00			           &   --8.34e-02		   &    1.03e+00 ~ 1.8e+00    &  --1		          &  --1 		      &   --1	      \\
$\cdots$  &                   & 	   	 &   			    &	     	      &   	     	    	  &  	    			           &				   &		   	      &    		          &     		      &	    	      \\
  428     & 054202.9-020745   & 05:42:02.93	 &   --02:07:45.2	    &	 68.0	      &   1.13e+00 ~ 1.9e-02	  &  5.78				   &    1.26e+00		   &    1.76e+00 ~ 4.7e-02    &   9		          &   8 		      &   24	      \\
$\cdots$  &                   & 	   	 &   			    &	     	      &   	     	    	  &  	    				   &				   &		   	      &    		          &     		      &	    	      \\
  682     & 054316.0-025859   & 05:43:16.09	 &   --02:58:59.0	    &	  0.0	      &   8.37e-03 ~ 1.5e-02	  &  0.10				   &    9.79e-02		   &    2.31e-01 ~ 3.5e-02    &  42		          &  40 		      &   86	      \\
$\cdots$  &                   &                  &                          &                 &                           &                                        &                               &                          &                           &                           &                  \\

\midrule[0.5pt]
\end{tabular}
}
\scalebox{1.2}{$\sim$}
\vspace{0.2cm}
%

\scalebox{1.2}{$\sim$}
{\renewcommand{\arraystretch}{0.5}
\begin{tabular}{c c c c c c c c c  c c c c c c c} 
\toprule[1.0pt]\toprule[0.5pt]  
 Sig$_{\rm 160}$ &  $S^{\rm peak}_{\rm 160}$    &  $S^{\rm peak}_{\rm 160}$/$S_{\rm bg}$ &  $S^{\rm conv,500}_{\rm 160}$ &  $S^{\rm tot}_{\rm 160}$ &  FWHM$^{\rm a}_{\rm 160}$ &  FWHM$^{\rm b}_{\rm 160}$ &  PA$_{\rm 160}$  &  Sig$_{\rm 250}$ &  $S^{\rm peak}_{\rm 250}$ &  $S^{\rm peak}_{\rm 250}$/$S_{\rm bg}$ &  $S^{\rm conv,500}_{\rm 250}$ &  $S^{\rm tot}_{\rm 250}$ &  FWHM$^{\rm a}_{\rm 250}$ &  FWHM$^{\rm b}_{\rm 250}$ &  PA$_{\rm 250}$  \\ 
                 &  (Jy/beam)                   &                                        & (Jy/beam$_{\rm 500}$)         &  (Jy)		    &  (\arcsec)		&  (\arcsec)		    &  (\degr)         &		  & (Jy/beam)		      & 				       & (Jy/beam$_{\rm 500}$)         &  (Jy)  		  &  (\arcsec)  	      &  (\arcsec)		  &  (\degr)	     \\       
  (15)           &      (16) ~ $\pm$ ~ (17)     &   (18)                                 &  (19)   			 &    (20) ~ $\pm$ ~ (21)   &  (22)                     & (23)                      & (24)             & (25)             &    (26) ~ $\pm$ ~ (27)    &  (28)                                  &  (29)                         &   (30) ~ $\pm$ ~ (31)    &  (32)                     &  (33)                     &  (34)            \\        
\toprule[0.8pt]
$\cdots$         &                              &                                        &                               &			    &				&			    &		       &		  &			      & 				       &                               &			  &			      & 			  &		     \\
 279.3           &  9.10e+00 ~ 2.0e+00  	&  0.35 				 &  1.21e+01			 &  1.42e+01 ~ 3.2e+00	    &  16			&  14			    &  106	       &   376.9          & 1.69e+01 ~ 2.0e+00	      & 0.76				       &  1.80e+01		       &   2.33e+01 ~ 2.1e+00	  &  20 		      &  19			  &  130	  	\\
$\cdots$         &  	                	&       				 &  				 &  		     	    &     			&    			    &      	       &   	          & 		     	      &	    				       &  			       &  		    	  &     		      &    			  &  		\\
 228.6           &  8.30e+00 ~ 2.9e-01  	&  1.93 				 &  1.04e+01			 &  1.32e+01 ~ 4.2e-01	    &  16			&  14			    &  116	       &   366.2          & 1.49e+01 ~ 5.0e-01	      & 2.24				       &  1.49e+01	               &   1.94e+01 ~ 5.0e-01	  &  19 		      &  18			  &  125		\\
$\cdots$         &  	                	&        				 &  				 &  		     	    &     			&    			    &      	       &   	          & 		     	      &	    				       &  			       &  		    	  &     		      &    			  &  		\\
   6.8           &  1.65e-01 ~ 4.1e-02  	&  0.15  				 &  6.11e-01			 &  7.33e-01 ~ 9.2e-02	    &  34			&  30			    &   92	       &     9.1          & 2.88e-01 ~ 4.0e-02	      & 0.23				       &  4.38e-01		       &   6.63e-01 ~ 5.7e-02	  &  28 		      &  24			  &  --37		\\
$\cdots$         &                              &                                        &                               &			    &				&			    &		       &		  &			      & 				       &                               &			  &			      & 			  &		     \\
\midrule[0.5pt]
\end{tabular}
}
\scalebox{1.2}{$\sim$}
\vspace{0.2cm}

\scalebox{1.2}{$\sim$}
{\renewcommand{\arraystretch}{0.5}
\begin{tabular}{c c c c c c c c c c c c c c c c}  
\toprule[1.0pt]\toprule[0.5pt] 
Sig$_{\rm 350}$  &  $S^{\rm peak}_{\rm 350}$ &  $S^{\rm peak}_{\rm 350}$/$S_{\rm bg}$ &  $S^{\rm conv,500}_{\rm 350}$ &  $S^{\rm tot}_{\rm 350}$  &  FWHM$^{\rm a}_{\rm 350}$ &  FWHM$^{\rm b}_{\rm 350}$ &  PA$_{\rm 350}$   &  Sig$_{\rm 500}$ &  $S^{\rm peak}_{\rm 500}$ &  $S^{\rm peak}_{\rm 500}$/$S_{\rm bg}$ &  $S^{\rm tot}_{\rm 500}$ &  FWHM$^{\rm a}_{\rm 500}$ &  FWHM$^{\rm b}_{\rm 500}$  &  PA$_{\rm 500}$   \\
                 &  (Jy/beam)                &                                        & (Jy/beam$_{\rm 500}$)         &  (Jy)			  &  (\arcsec)  	      &  (\arcsec)		  &  (\degr)	      & 		 & (Jy/beam)		     &  				      &  (Jy)                    &  (\arcsec)                &  (\arcsec)                 &  (\degr)          \\	       
  (35)           &   (36) ~ $\pm$ ~ (37)     &  (38)                                  &  (39)                         &    (40) ~ $\pm$ ~ (41)    &  (42)                     &  (43)                     &  (44)             &  (45)            &    (46) ~ $\pm$ ~ (47)    &  (48)                                  &    (49) ~ $\pm$ ~ (50)   &  (51)                     &  (52)                      &  (53)             \\	
\toprule[0.8pt] 
$\cdots$         &                           &                                        &                               & 			  &			      & 			  &		      & 		 &			     &  				      &                          &                           &                            &                   \\
295.3		 & 1.41e+01 ~ 1.9e+00	     &	0.99				      &  1.45e+01		      &  1.77e+01 ~ 1.9e+00	  &  27 		      &  25			  &   127	      &	  196.8          &  9.92e+00 ~ 1.3e+00	     &  1.25				      &  1.26e+01 ~ 1.3e+00	 &  39  		     &  36		          &   45	  \\
$\cdots$         & 		    	     &	    				      &  			      &  		  	  &     		      &    			  &  		      &	                 &  		     	     &	    				      &	 		  	 &      		     &     		          &  		  \\
362.8		 & 1.31e+01 ~ 3.5e-01	     &	2.34				      &  1.30e+01		      &  1.48e+01 ~ 3.5e-01	  &  25 		      &  25			  &   82	      &	  249.7          &  8.80e+00 ~ 1.7e-01	     &  2.55				      &  9.19e+00 ~ 1.7e-01	 &  36  		     &  36		          &   50		  \\
$\cdots$         & 		    	     &	    				      &  			      &  		  	  &     		      &    			  &  		      &	                 &  		     	     &	    				      &	 		  	 &      		     &     		          &  		  \\
  7.8		 & 2.32e-01 ~ 5.0e-02	     &	0.20				      &  2.55e-01		      &  3.28e-01 ~ 5.3e-02	  &  31 		      &  25			  &    1	      &	    0.0          &  1.52e-01 ~ 6.8e-02	     &  0.13				      &  1.22e-01 ~ 6.8e-02	 &  36  		     &  36		          &    4		  \\
$\cdots$         &                           &                                        &                               & 			  &			      & 			  &		      & 		 &			     &  				      &                          &                           &                            &                   \\
\midrule[0.5pt]
\end{tabular}
}
\scalebox{1.2}{$\sim$}
\vspace{0.2cm}

\scalebox{1.2}{$\sim$}
{\renewcommand{\arraystretch}{0.5}
\begin{tabular}{c c c c c c c c c c c c c|}  
\toprule[1.0pt]\toprule[0.5pt] 
Sig$_{\rm N_{H_2}}$& $N^{\rm peak}_{\rm H_2}$ &  $N^{\rm peak}_{\rm H_2}$/$N_{\rm bg}$ &  $N^{\rm conv,500}_{\rm H_2}$ &  $N^{\rm bg}_{\rm H_2}$  &  FWHM$^{\rm a}_{\rm N_{H_2}}$ &  FWHM$^{\rm b}_{\rm N_{H_2}}$ & PA$_{\rm N_{H_2}}$  &  N$_{\rm SED}$  &   Core type      & {\it Spitzer}	      & SIMBAD							& Comments	\\
                   & (10$^{21}$ cm$^{-2}$)    &                                        & (10$^{21}$ cm$^{-2}$)         &  (10$^{21}$ cm$^{-2}$)   &  (\arcsec)  	          &  (\arcsec)                    &  (\degr)            &                 &  		     &  		      &								&		\\ 
 (54)              & (55)                     &  (56)                                  & (57)                          &  (58)                    &  (59)                         &  (60)                         &  (61)               &    (62)         &   (63)	     & (64)		      & (65)							& (66)		\\
\toprule[0.8pt] 																									            
$\cdots$           &                          &                                        &                               &                          &                               &                               &                     &                 &  		     &  		      &								&		\\
313.6              & 18.0		      &   1.25  			       &    5.2 		       &  14.4  		  &  22 			  &  18 			  &  127	        &	       4  &   2 	     &  0		      & [KDJ2016] J054126.7-014451				&		\\
$\cdots$           & 	 		      &         			       &        		       &        		  &     			  &     			  &  	   	        &	          &  		     &  		      &								&		\\
943.9              & 49.2		      &   3.58  			       &   13.1 		       &  13.7  		  &  20 			  &  18 			  &  --40	        &	       4  &   3 	     &  1		      & [KDJ2016] J054202.7-020745, [MSJ2009] L1630MIR-94	&		\\
$\cdots$           & 	 		      &         			       &        		       &        		  &     			  &     			  &  	   	        &	          &  		     &  		      &								&		\\
9.1                &  0.5		      &   0.24  			       &    0.2 		       &   1.9  		  &  31 			  &  18 			  &   24	        &	       3  &   1 	     &  0		      &								&		\\
$\cdots$           &                          &                                        &                               &                          &                               &                               &                     &                 &  		     &  		      &								&		\\
\midrule[0.5pt]
\end{tabular}
}
\tablefoot{Catalog entries are as follows: 
{\bf(1)} Core running number;
{\bf(2)} Core name $=$ HGBS\_J prefix directly followed by a tag created from the J2000 sexagesimal coordinates; 
{\bf(3)} and {\bf(4)}: Right ascension and declination of core center; 
{\bf(5)}, {\bf(15)}, {\bf(25)}, {\bf(35)}, and {\bf(45)}: Detection significance from monochromatic single scales, in the 70, 160, 250, 350, and 500~$\mu$m maps, respectively. 
(NB: the detection significance has the special value of $0.0$ when the core is not visible in clean single scales); 
{\bf(6)}$\pm${\bf(7)}, {\bf(16)}$\pm${\bf(17)} {\bf(26)}$\pm${\bf(27)} {\bf(36)}$\pm${\bf(37)} {\bf(46)}$\pm${\bf(47)}: Peak flux density and its error in Jy/beam as estimated by \textsl{getsources};
{\bf(8)}, {\bf(18)}, {\bf(28)}, {\bf(38)}, {\bf(48)}: Contrast over the local background, defined as the ratio of the background-subtracted peak intensity to the local background intensity ($S^{\rm peak}_{\rm \lambda}$/$S_{\rm bg}$); 
{\bf(9)}, {\bf(19)}, {\bf(29)}, {\bf(39)}: Peak flux density measured after smoothing to a 36.3$\arcsec$ beam; 
{\bf(10)}$\pm${\bf(11)}, {\bf(20)}$\pm${\bf(21)}, {\bf(30)}$\pm${\bf(31)}, {\bf(40)}$\pm${\bf(41)}, {\bf(49)}$\pm${\bf(50)}: Integrated flux density and its error in Jy as estimated by \textsl{getsources}; 
{\bf(12)}--{\bf(13)}, {\bf(22)}--{\bf(23)}, {\bf(32)}--{\bf(33)}, {\bf(42)}--{\bf(43)}, {\bf(51)}--{\bf(52)}: Major \& minor FWHM diameters of the core (in arcsec), respectively, 
as estimated by \textsl{getsources}. (NB: the special value of $-1$ means that no size measurement was possible); 
{\bf(14)}, {\bf(24)}, {\bf(34)}, {\bf(44)}, {\bf(53)}: Position angle of the core major axis, measured east of north, in degrees; 
{\bf(54)} 
Detection significance in the high-resolution column density image;  
{\bf(55)} 
Peak H$_{2}$ column density in units of $10^{21}$ cm$^{-2}$ as estimated by \textsl{getsources} in the high-resolution column density image; 
{\bf(56)} 
Column density contrast over the local background, as estimated by \textsl{getsources} in the high-resolution column density image;
{\bf(57)} 
Peak column density measured in a 36.3$\arcsec$ beam; 
{\bf(58)} Local background H$_{2}$ column density 
as estimated by \textsl{getsources} in the high-resolution column density image; 
{\bf(59)}--{\bf(60)}--{\bf(61)}: Major \& minor FWHM diameters of the core, and position angle of the major axis, respectively,
as measured in the high-resolution column density image; 
{\bf(62)} Number of {\it Herschel} bands in which the core is significant (Sig$_{\rm \lambda} >$ 5) and has a positive flux density, excluding the column density plane; 
{\bf(63)} Core type: 1-starless, 2-prestellar, 3-protostellar, or 0-tentative core. The latter is a likely extragalactic source; 
{\bf(64)} 1 if a {\it Spitzer}-identified YSO \citep{Megeath+2012} is found within 6$\arcsec$ of the {\it Herschel} peak position, 0 otherwise (see text for details);
{\bf(65)} SIMBAD counterparts, if any, up to 6$\arcsec$ from the {\it Herschel} peak position;
{\bf(66)} Comments: ``N region''.
 }
\end{sidewaystable*}

\clearpage

\begin{sidewaystable*}[htb]\tiny\setlength{\tabcolsep}{6.8pt}
\caption{Derived properties of dense cores identified in HGBS maps of Orion~B (template, full table only provided online).}
\label{tab_der_cat}
{\renewcommand{\arraystretch}{0.9}
\begin{tabular}{|r|c|c c c c c c c c c c c c|} 
\toprule[1.0pt]\toprule[0.5pt]  
 rNO     & Core name         &  RA$_{\rm 2000}$ &  Dec$_{\rm 2000}$        & $R_{\rm core}$        &  $M_{\rm core}$   &  $T_{\rm dust}$   &  $N^{\rm peak}_{\rm H_2}$ &  $N^{\rm ave}_{\rm H_2}$   &   $n^{\rm peak}_{\rm H_2}$  &  $n^{\rm ave}_{\rm H_2}$   &  $\alpha_{\rm BE}$  & Core type  & Comments  \\
         & HGBS\_J*          &  (h m s)         &  (\degr~\arcmin~\arcsec) & (pc)                  &  ($M_\odot$)      &  (K)              &  (10$^{21}$ cm$^{-2}$)    &  (10$^{21}$ cm$^{-2}$)     &   (10$^{4}$ cm$^{-3}$)      &  (10$^{4}$ cm$^{-3}$)      &                     &		  &	       \\   
 (1)     & (2)               &  (3)             &   (4)                    & (5) ~~~~~~ (6)        &  (7) $\pm$ (8)    &  (9) $\pm$ (10)   &  (11)                     &  (12) ~~~~ (13)            &   (14)                      &  (15) ~~~ (16)             &         (17)        &    (18)    &    (19)    \\   
\toprule[0.8pt] 										 
$\cdots$ &                   &                  &                          &                       &                   &                   &                           &                            &	                          &                            &                     &            &	       \\ 
 204     & 054126.7-014453   & 05:41:26.78	&   --01:44:53.2 	   &  2.1e-02 ~~~ 3.9e-02  &  8.80 ~~~ 2.07    &  11.1 ~~~ 0.6     &  59.0		       &  68.2 ~~~ 233.9	    &    24.8		          &  38.2 ~~~ 242.5	       &	   0.0       &  2         &           \\
$\cdots$ & 		     &  	  	&   	        	   &  		           &  		       &                   &  	    		       &           		    & 	     		          &  	      	     	       &	             &	          &	       \\ 
 428     & 054202.9-020745   & 05:42:02.93	&   --02:07:45.2 	   &  1.8e-02 ~~~ 3.7e-02  &  6.17 ~~~ 1.17    &  11.5 ~~~ 0.5     &  56.9		       &  54.4 ~~~ 235.6	    &    24.3		          &  32.5 ~~~ 292.7	       &	   0.1       &  3         &           \\
$\cdots$ & 		     &  	  	&   	        	   &  		           &  		       &                   &  	    		       &           		    & 	     		          &  	      	     	       &	             &	          &	       \\ 
 682     & 054316.0-025859   & 05:43:16.09	&   --02:58:59.0 	   &  2.9e-02 ~~~ 4.6e-02  &  0.05 ~~~ 0.02    &  15.7 ~~~ 1.7     &   0.2		       &   0.3 ~~~   0.7	    &     0.1		          &   0.1 ~~~   0.5	       &	  11.4       &  1         &           \\
$\cdots$ &                   &                  &                          &                       &                   &                   &                           &                            &	                          &                            &                     &            &	       \\ 
\midrule[0.5pt]
\end{tabular}
}
\tablefoot{Table entries are as follows: {\bf(1)} Core running number; {\bf(2)} Core name $=$ HGBS\_J prefix directly followed by a tag created from the J2000 sexagesimal coordinates; 
{\bf(3)} and {\bf(4)}: Right ascension and declination of core center; 
{\bf(5)} and {\bf(6)}: Geometrical average between the major and minor FWHM sizes of the core (in pc), as measured in the high-resolution column density map 
after deconvolution from the 18.2$\arcsec$ HPBW resolution of the map and before deconvolution, respectively.
(NB: Both values provide estimates of the object's outer radius when the core can be approximately described by a Gaussian distribution, as is the case 
for a critical Bonnor-Ebert spheroid); 
{\bf(7)} Estimated core mass ($M_\odot$) assuming the dust opacity law advocated by \citet{Roy+2014}; 
{\bf(9)} SED dust temperature (K); {\bf(8)} \& {\bf(10)} Statistical errors on the mass and temperature, respectively, including calibration uncertainties, but excluding dust opacity uncertainties; 
{\bf(11)} Peak H$_2$ column density, at the resolution of the 500$~\mu$m data, derived from a graybody SED fit to the core peak flux densities measured in a common 36.3$\arcsec$ beam at all wavelengths; 
{\bf(12)} Average column density, calculated as $N^{\rm ave}_{\rm H_2} = \frac{M_{\rm core}}{\pi R_{\rm core}^2} \frac{1}{\mu m_{\rm H}}$, 
          where $M_{\rm core}$ is the estimated core mass (col. {\bf 7}), $R_{\rm core}$ the estimated core radius prior to deconvolution (col. {\bf 6}), and $\mu = 2.8$;
{\bf(13)} Average column density calculated in the same way as for col. {\bf 12} but using the deconvolved core radius (col. {\bf 5});  
{\bf(14)} Beam-averaged peak volume density at the resolution of the 500~$\mu$m data, derived from the peak column density (col. {\bf 11}) assuming a Gaussian spherical distribution: 
          $n^{\rm peak}_{\rm H_2} = \sqrt{\frac{4 \ln2}{\pi}} \frac{N^{\rm peak}_{\rm H_2}}{\overline{FWHM}_{\rm 500}}$; 
{\bf(15)} Average volume density, calculated as
          $n^{\rm ave}_{\rm H_2} = \frac{M_{\rm core}}{4/3 \pi R_{\rm core}^3} \frac{1}{\mu m_{\rm H}}$, using the estimated core radius prior to deconvolution; 
{\bf(16)} Average volume density, calculated in the same way as for col. {\bf 15} but using the deconvolved core radius (col. {\bf 5}); 
{\bf(17)} Bonnor-Ebert mass ratio: $\alpha_{\rm BE} = M_{\rm BE,crit} / M_{\rm obs} $ (see text for details); 
{\bf(18)} Core type: 1-starless, 2-prestellar, 3-protostellar, or 0-tentative core. The latter is a likely extragalactic source (see comments);
{\bf(19)} Comments: ``no SED fit'', ``tentative bound'', ``N region''.
}

\end{sidewaystable*}

\clearpage
\newpage

\begin{figure*}
 \begin{center}
  \begin{minipage}{0.9\linewidth}
   \resizebox{1\hsize}{!}{\includegraphics[angle=0]{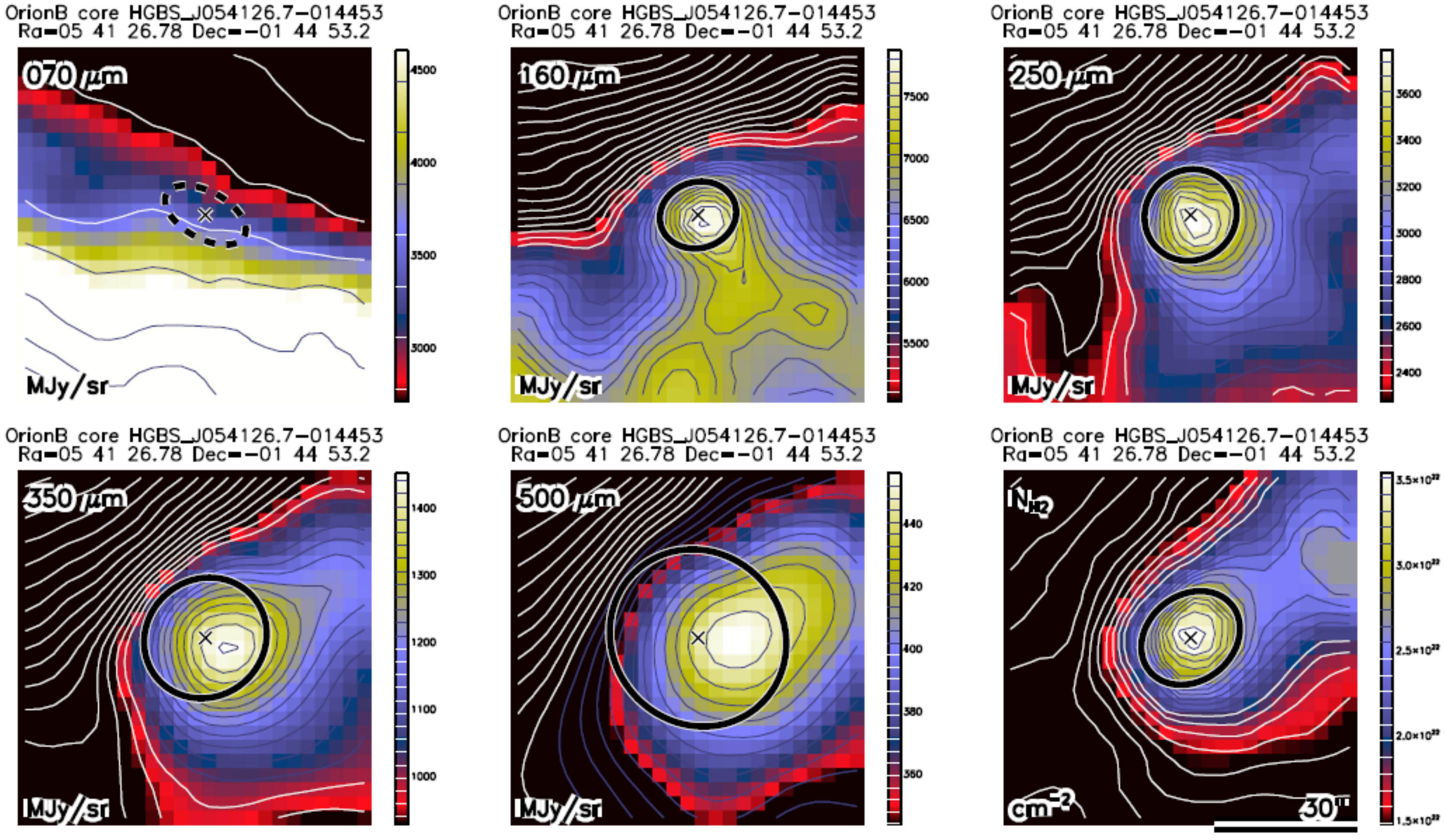}} 
  \end{minipage}
 \end{center}
   \caption{Example blow-up {\it Herschel} images at 70--160--250--350--500~$\mu$m and high-resolution column density map for a (bound) prestellar core. 
           Ellipses represent the estimated major and minor FWHM sizes of the core at each wavelength; they are shown as solid or dashed curves
           depending on whether the core is significantly detected or not, respectively, at a particular wavelength.
	   See Table \ref{tab_der_cat} for the physical radius of the core and other derived properties. 
	   An angular scale of $30\arcsec $ ($\sim 0.058$~pc at $d=400$~pc) is shown at the bottom right. 
	   North is up, east is left. 
	   Similar image cutouts are provided online for all selected starless cores. 
           }
   \label{fig_zooms1}%
\end{figure*}
%
\begin{figure*}
 \begin{center}
  \begin{minipage}{0.9\linewidth}
   \resizebox{1\hsize}{!}{\includegraphics[angle=0]{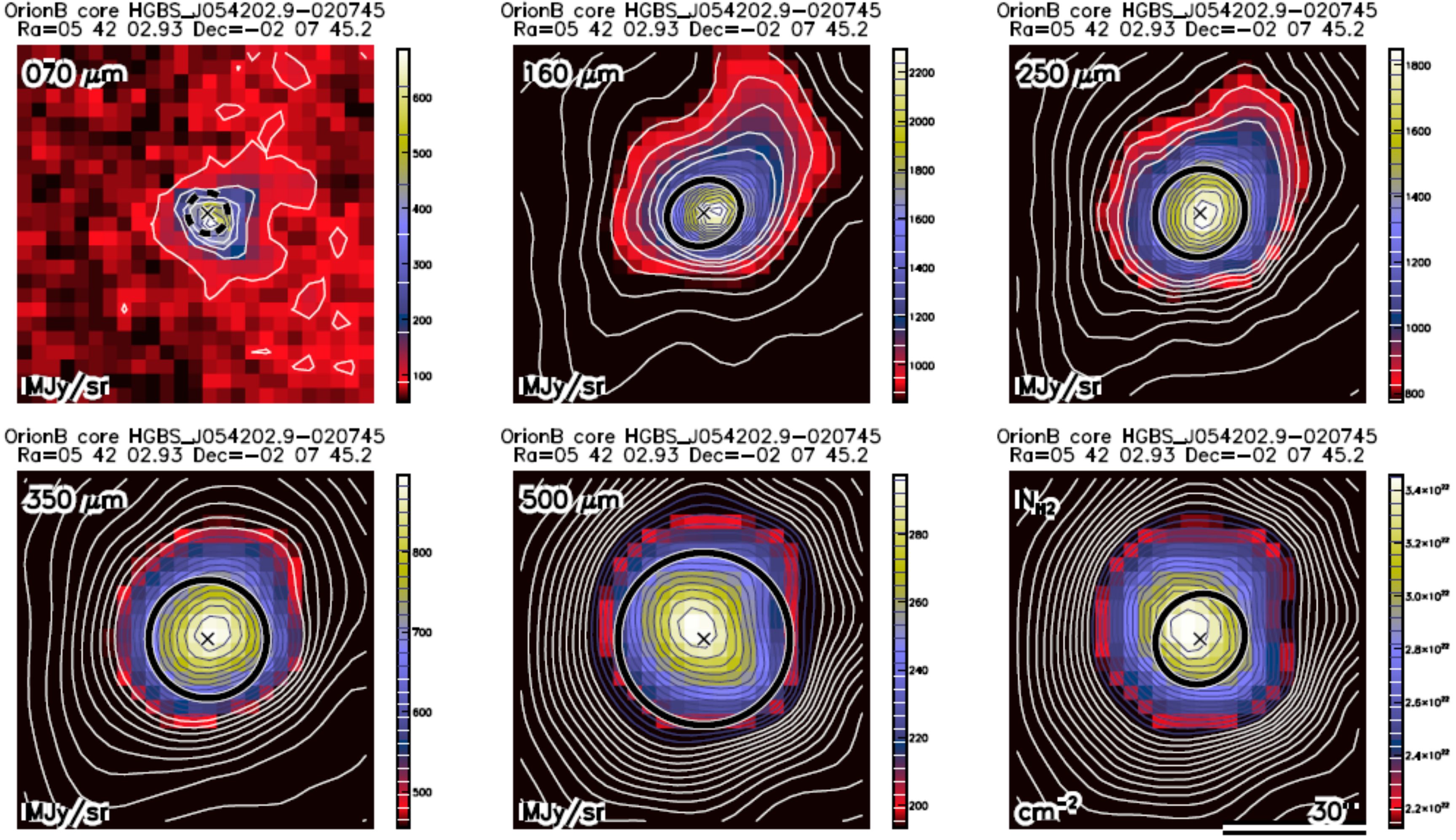}}
  \end{minipage}
 \end{center}
   \caption{Same as Fig.~\ref{fig_zooms1} for a protostellar core. See Table \ref{tab_der_cat} for its derived properties.  
            Similar image cutouts are provided online for all selected protostellar cores.}
   \label{fig_zooms2}%
\end{figure*}

\clearpage
\newpage

\section{Automated post-selection of reliable core candidates}\label{sec:appendix_postSel}

The \textsl{getsources} source extraction algorithm \citep{Menshchikov2012}, which analyzes fine spatial decompositions of the original images across 
a wide range of scales and at all observed wavebands, is a powerful technique to extract most candidate dense cores in spatially-varying backgrounds. 
At the same time, however, this technique tends to extract irregularities in the emission maps used for detection which must be discarded by the user at 
the post-selection stage, that is after the initial selection criteria described in Sect.~\ref{sec:init_selection} 
have been automatically applied.

The criteria of Sect.~\ref{sec:init_selection} are applied to the raw output catalog from \textsl{getsources}. 
At this stage, the properties of candidate cores are derived directly from \textsl{getsources} measurements without going back to the observed maps themselves. 
After the initial selection steps, there may still be a significant number of spurious sources in the selected core list which do not correspond to real 
emission peaks or do not appear like genuine cores in the maps. 
These spurious sources can only be eliminated by revisiting their locations in the {\it Herschel} maps, which is done here with a separate automated script. 
This ``post-selection'' script is run on the same input observed maps with 3$\arcsec$ pixels as used for the \textsl{getsources} extractions.

\begin{figure}[!hb]
 \begin{center}
  \begin{minipage}{1.0\linewidth}
   \resizebox{1.0\hsize}{!}{\includegraphics[angle=0]{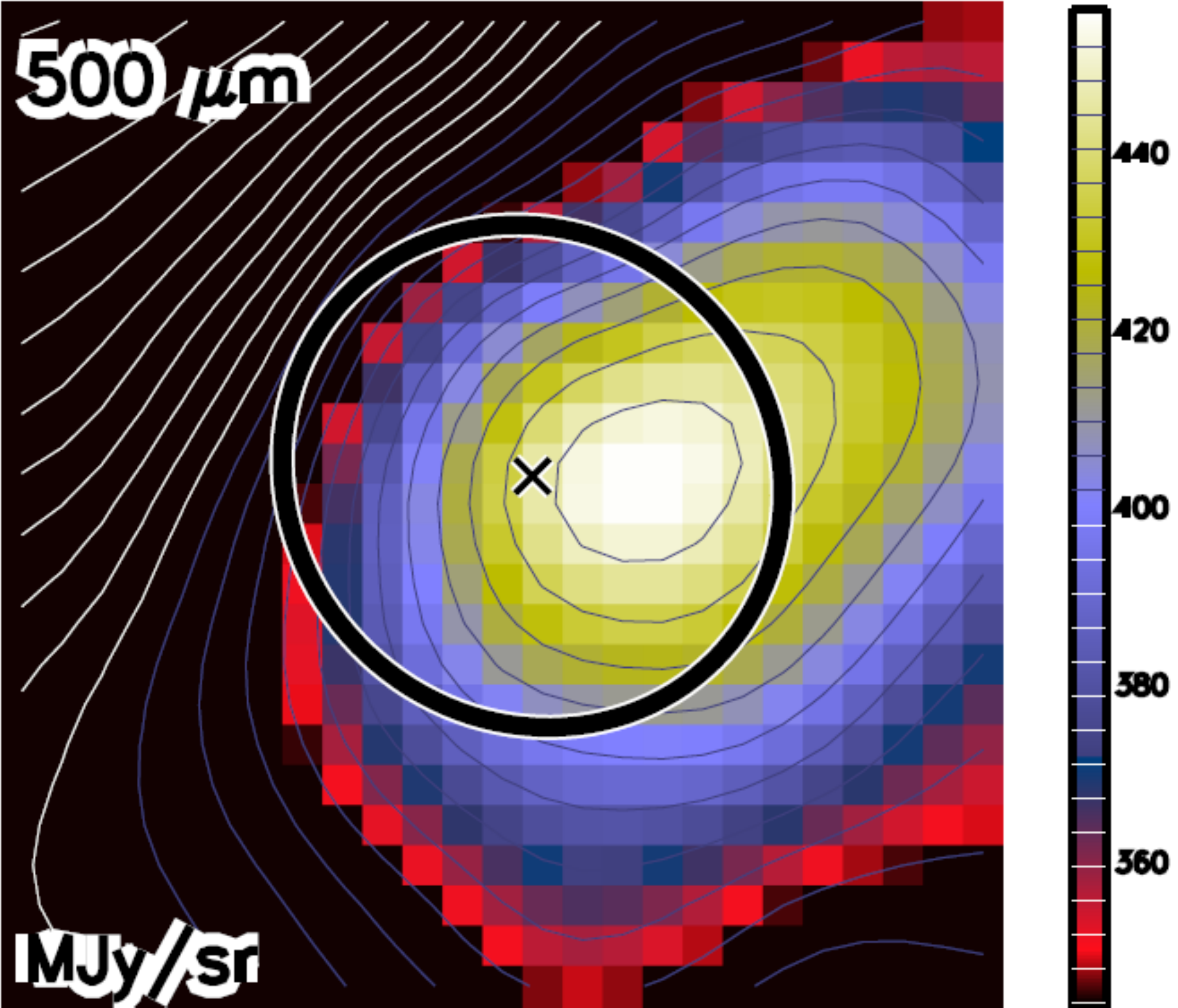}}
  \end{minipage}
 \end{center}
   \caption{
            Close-up view of a post-selected prestellar core at 500~$\mu$m (shown in Fig.~\ref{fig_zooms1}) illustrating the post-selection procedure.
            At each wavelength, the map colorscale is optimized for the immediate neighborhood of the core; the lowest and highest map values 
	    in the considered box are blanked by black and white colors, respectively. 
	    The source center position is marked by a cross, and the FWHM ellipse automatically derived by  \textsl{getsources} 
	    is plotted.
	    }
   \label{fig:postSel1}%
\end{figure}

\vspace{2mm}
\noindent
The automatic post-selection script, written in IDL, works without any free parameter. 
It performs further source selection and elimination according to the following steps:

\vspace{2mm}
\noindent
\textbf{1.}
\vspace{0mm}
  The script prepares a close-up view of each core from the input source list at the standard HGBS wavelengths (70, 160, 250, 350, 500~$\mu$m),  
  and in the high-resolution column density plane (see Figs.~\ref{fig_zooms1}, and \ref{fig_zooms2} for examples). 
  Another example of a core at 500~$\mu$m is provided in Fig.~\ref{fig:postSel1}. 
  The input source list of the post-selection stage is the output of the core selection procedure detailed in Sect.~\ref{sec:init_selection}.

\begin{figure*}[!!ht]
 \begin{center}
  \begin{minipage}{1.0\linewidth}
   \resizebox{0.413\hsize}{!}{\includegraphics[angle=0]{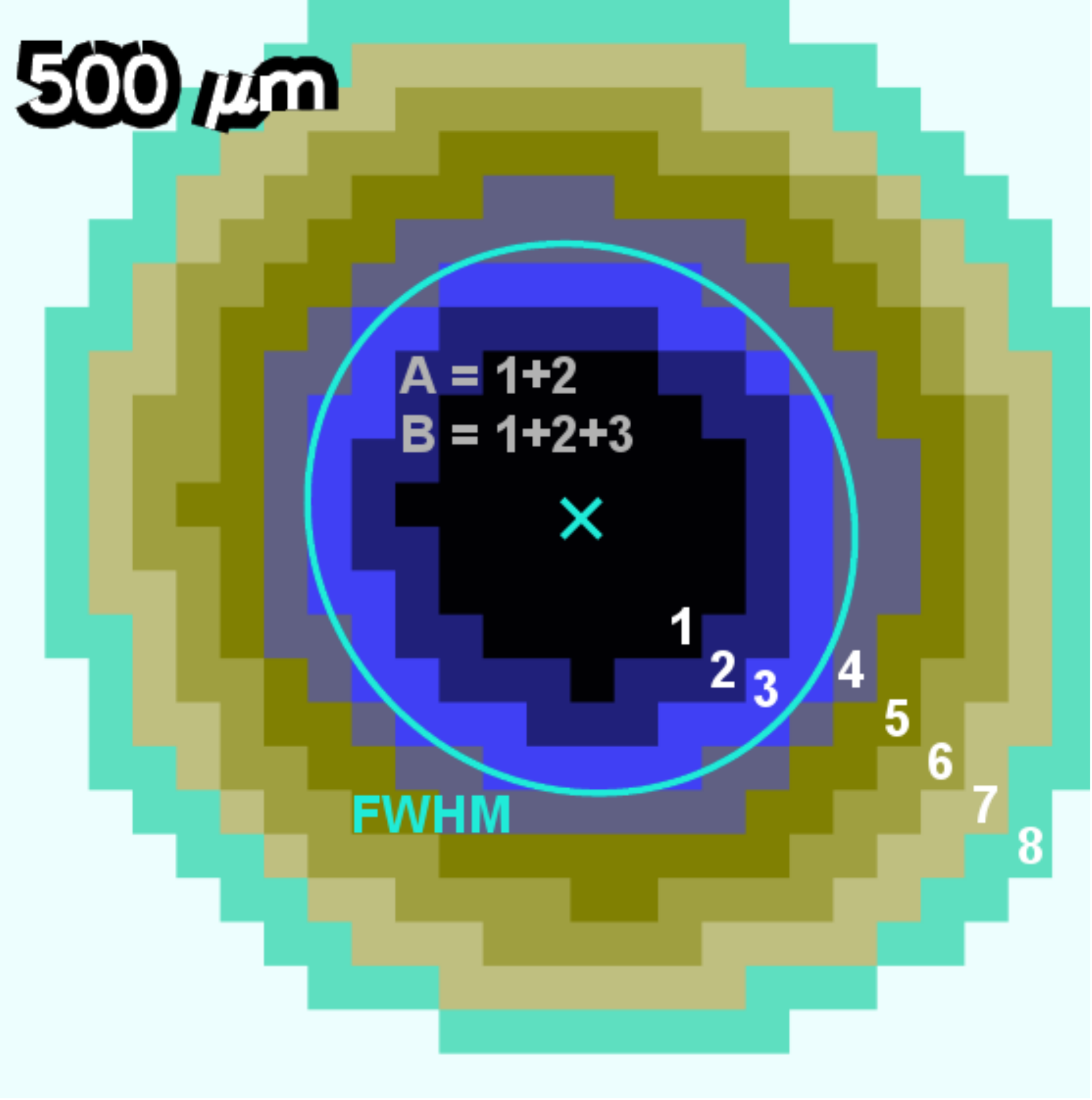}}
   \hspace{1.5cm}
   \resizebox{0.487\hsize}{!}{\includegraphics[angle=0]{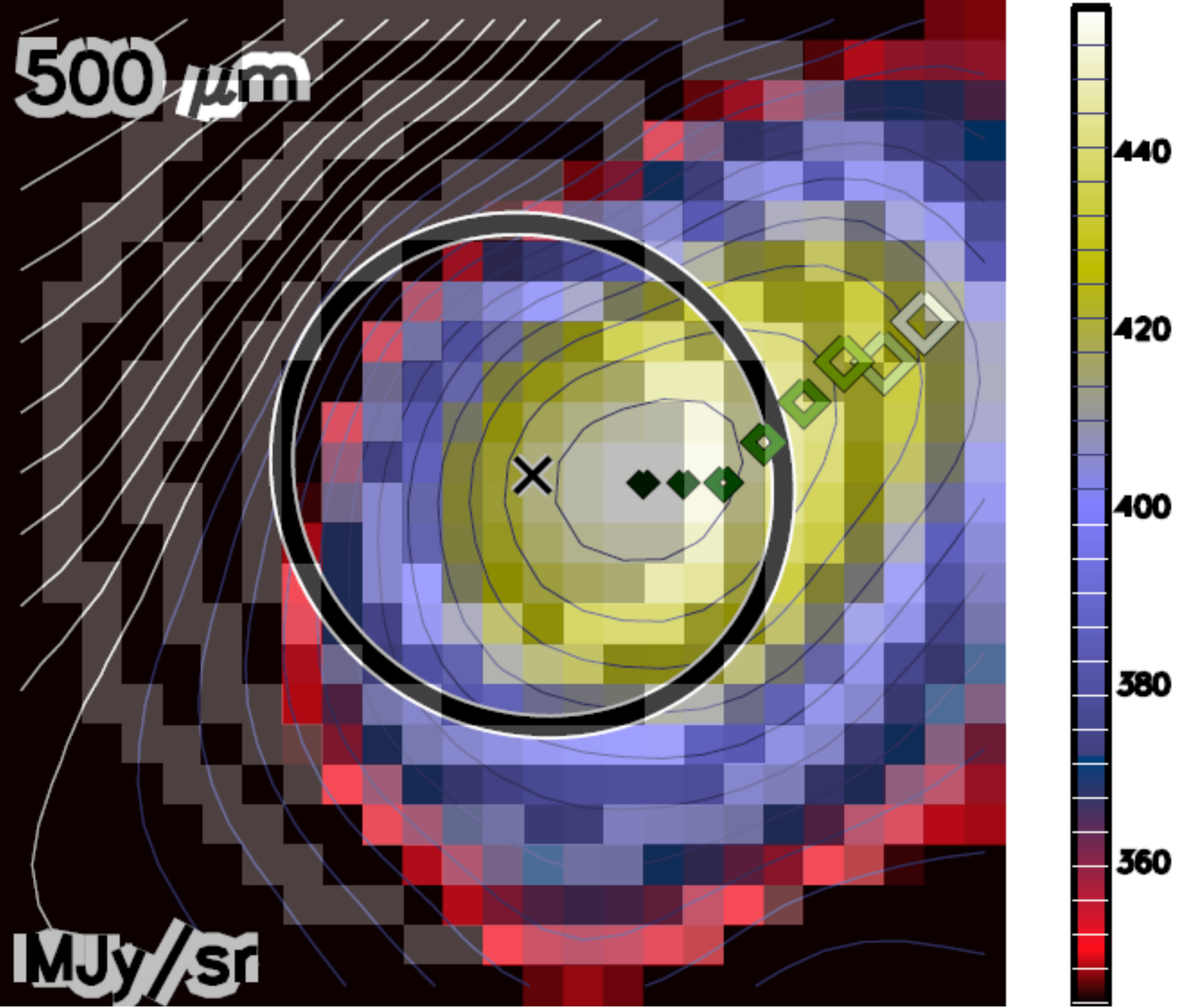}}
  \end{minipage}
 \end{center}
   \caption{{\bf Left:} Examples of annular masks created inside and outside of the FWHM ellipse of each core at each wavelength. 
            The example mask image shown here is at 500~$\mu$m and belongs to the prestellar core of Fig.~\ref{fig:postSel1}. 
	    Inside out, the division of the masks along the core radius is the following, considering the FWHM ellipse as 100\%: Region 1: 
	    $<$60\% interior of the FWHM radius, region 2: 60--80\%, 3: 80--100\%, 4: 100--120\%, 5: 120--140\%, 6: 140--160\%, 
	    7: 160--180\%, 8: 180--200\%. Region A: $<$80\% (regions 1$+$2), B: $<$100\% (1$+$2$+$3). 
            {\bf Right:}
	    The first maximum (i.e., the highest intensity pixel) in each mask is shown by a green-shaded diamond in the example image. 
            The panel size, marked source center, and FWHM sizes are the same for each pair of mask image and observed image. 
           }
   \label{fig:postSel2}%
\end{figure*}

\vspace{2mm}
\noindent
\textbf{2.}
\vspace{0mm}
  Annular masks are created inside and outside of the FWHM ellipse of a given source at each wavelength, which are expanded (or shrunk) versions of 
  the FWHM elliptical footprint, scaling with radius, not with area. 
  The division and numbering of these masks are illustrated in the left panel of Fig.~\ref{fig:postSel2}.
  We split the immediate neighborhood of each candidate core into the following masks, labeled by their linear sizes measured in units of the FWHM 
  size (taken to be 100\%). 
  Region {\bf 1}: elliptical area corresponding to a shrunk version of the FWHM footprint, within its 60\% linear size; 
  {\bf 2}: region between the 60\% and the 80\% ellipse; {\bf 3}: 80--100\% mask; {\bf 4}: 100--120\% mask, 
  {\bf 5}: 120--140\% mask; {\bf 6}: 140--160\% mask; {\bf 7}: 160--180\% mask; {\bf 8}: 180--200\% mask. 
  Region {\bf A}: elliptical area enclosed within the $80$\% ellipse ({\bf 1}$+${\bf 2}); 
  region {\bf B}: area enclosed within the FWHM elliptical contour (i.e., $<$100\% mask $=$ {\bf 1}$+${\bf 2}$+${\bf 3}).  

\vspace{2mm}
\noindent
\textbf{3.}
\vspace{0mm}
  In the next step, maxima are found in the observed images (at 160--250--350--500\,$\mu$m, and at $N_{\rm H_2}$) under the corresponding rings of masks 
  (see right panel of Fig.~\ref{fig:postSel2}). 
  The panel size, marked source center and FWHM sizes are the same for a pair of observed image and mask image at each wavelength. 
  The location of the maximum value (i.e., the highest intensity pixel) in each consecutive annular mask 
  map sections under the masks are 
  is marked by a green-shaded symbol whose size increases with radius from nominal source center. 
  At the same time, the script also locates the second and third intensity maximum points within each annular mask (not shown in Fig.~\ref{fig:postSel2}). 

\vspace{2mm}
\noindent
\textbf{4.}
\vspace{0mm}
  At this stage, when a core is post-selected at a certain observed wavelength, its {\it counter flag ($\rm cntCore \lambda$)} 
  at that wavelength is assigned a value of $1$, that is, $\rm cntCore \lambda = 1$. The default $0$ value of cntCore$\lambda$ flips to $1$ if:
\begin{itemize}
  \item [I.] the first maximum intensity values under the consecutive masks are decreasing outward from  core center,
\end{itemize}
  OR
\begin{itemize} 
  \item [II.] the maximum intensity inside the FWHM area (region {\bf B}) is higher than in the next mask outward (region {\bf 4}),
\end{itemize}
  OR
\begin{itemize} 
  \item [III.] at least two of the secondary maxima in regions {\bf 1}, {\bf 2}, and {\bf 3} are larger than the first maximum in region {\bf 4},
\end{itemize}
  OR
\begin{itemize} 
  \item [IV.] the value of the third maximum in region {\bf 1} is larger than the first maxima in regions {\bf 2} and {\bf 3}.
\end{itemize}
  AND
\begin{itemize} 
  \item [V.] the aspect ratio of the core ellipse is $\leq$ 3.
\end{itemize}
More precisely, at a certain wavelength, a candidate core is post-selected (kept) if at least one of the criteria from [I.] to [IV.] is satisfied, 
{\it and} if criterion [V.] is fulfilled at the same time.   

\vspace{2mm}
\noindent
\textbf{5.}
\vspace{0mm}
 The script then loops over the list of candidate cores to be post-selected, and analyzes each wavelength in turn. 
 The final post-selection of a core is granted if $\rm cntCore \lambda = 1$ in at least three bands among the 160--250--350--500~$\mu$m images 
 plus the high-resolution column density map.

In the case of Orion~B, the fraction of discarded sources after applying the above procedure was $\sim$30\% of the input source list.

Our primary intention was to automate the post-selection checks of cores directly from the information contained in the observed {\it Herschel} maps, 
mimicking what a human eye would do.
This script greatly facilitates the post-selection evaluation, but a final visual inspection is still recommended.
As it was primarily designed for resolved dense cores, this procedure is not optimal for point-like sources or protostars. 

\section{Completeness of HGBS prestellar core extractions in Orion~B}\label{sec:appendix_sim}

To assess the completeness of the present HGBS census of prestellar cores in Orion~B, we used both simulated data 
and the simple model described in Appendix B.2 of \citet{Konyves+2015}.

To simulate core extractions, we constructed clean maps of the background cloud emission at {\it Herschel} wavelengths (plus a 
column density map), by subtracting the compact sources identified with \textsl{getsources} from the observed images 
(see Sects.~\ref{sec:getsources} \& \ref{sec:core_selection}). 
We then injected model Bonnor-Ebert-like cores within the clean-background 
maps to produce synthetic {\it Herschel} and column density images of the region. 
We used a population of 711 model starless cores 
with the following piece-wise power-law mass distribution: 
$\Delta N/\Delta$log$M$ $\propto$ $M^{+1.35}$ from $0.1\, M_\odot$ to $0.4\, M_\odot$, 
$\Delta N/\Delta$log$M$ $\propto$ $M^{0}$ from $0.4\, M_\odot$ to $0.9\, M_\odot$,
and $\Delta N/\Delta$log$M$ $\propto$ $M^{-1.35}$ from $0.9\, M_\odot$ to $20\, M_\odot$. 
The positions of the model cores in a mass versus size diagram similar to that shown in Fig.~\ref{fig:massSize} were consistent with critical Bonnor-Ebert 
isothermal spheres at effective gas temperatures $\sim \, $7--20~K.
The far-infrared and submillimeter continuum emission from the synthetic Bonnor-Ebert cores at all {\it Herschel} wavelengths 
was simulated using an extensive grid of spherical dust radiative 
transfer models generated with the MODUST code \citep[e.g.,][]{Bouwman+2000, Bouwman2001PhDT}.
Accordingly, each synthetic prestellar core had a realistic dust temperature profile with a significantly lower dust temperature in its central inner part, 
as observed in resolved starless cores \citep[cf.][]{Roy+2014}. 
The synthetic cores were randomly distributed within the portions of the observed field where $N_{\rm H_2}^{\rm bg} \ge 5 \times 10^{21}$ cm$^{-2}$  
(or $A_{\rm V} \ga 5$), without any mass segregation. 
Compact source extraction and core selection and classification were subsequently performed with \textsl{getsources} 
in the same manner as for the real observations (see Sects.~\ref{sec:getsources} \& \ref{sec:core_selection}). 

\begin{figure*}[!ht]
 \centering
  \begin{minipage}{1.0\linewidth}
   \hspace{0.5cm}
   \resizebox{0.45\hsize}{!}{\includegraphics[angle=0]{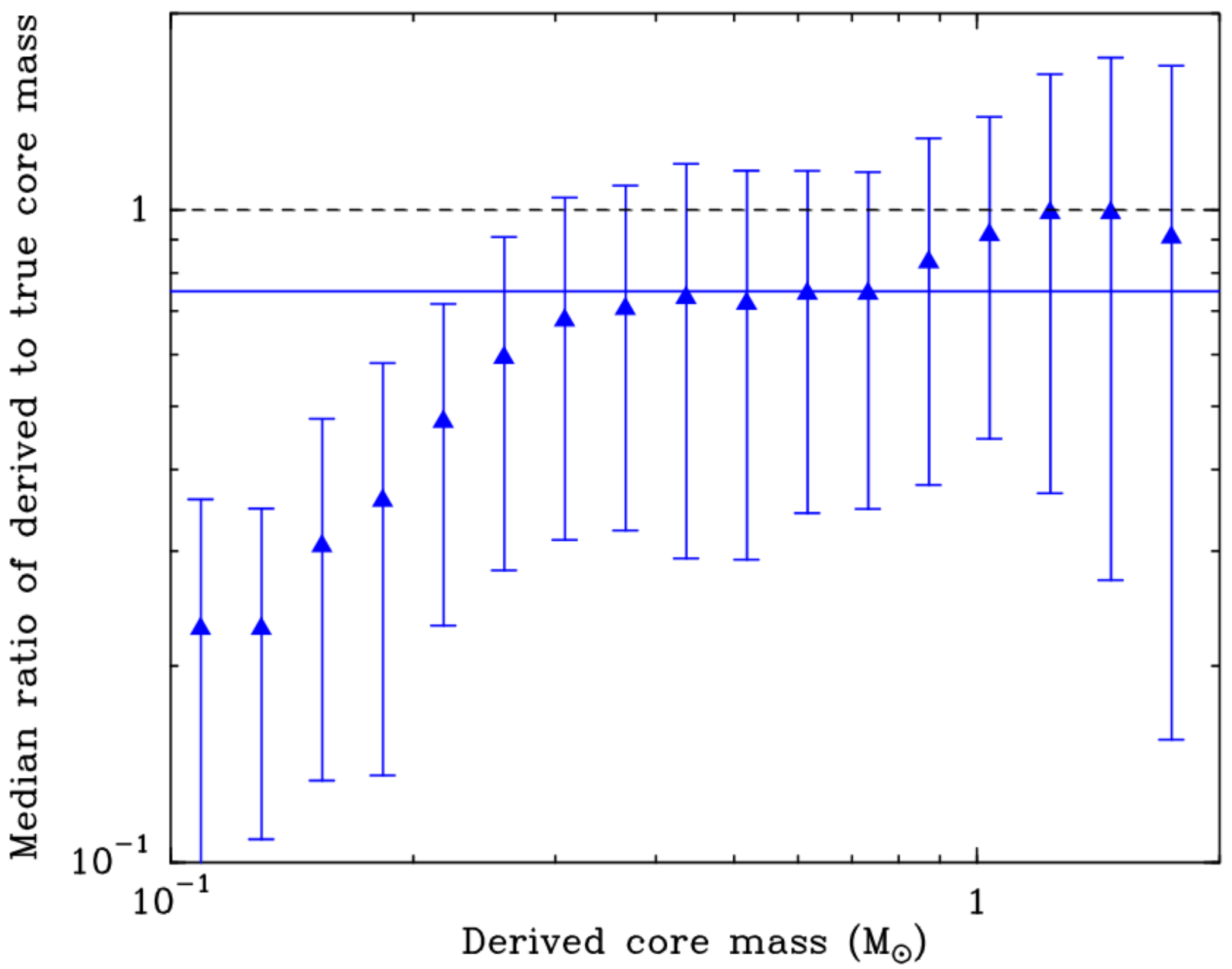}}
   \hspace{0.75cm}
   \resizebox{0.43\hsize}{!}{\includegraphics[angle=0]{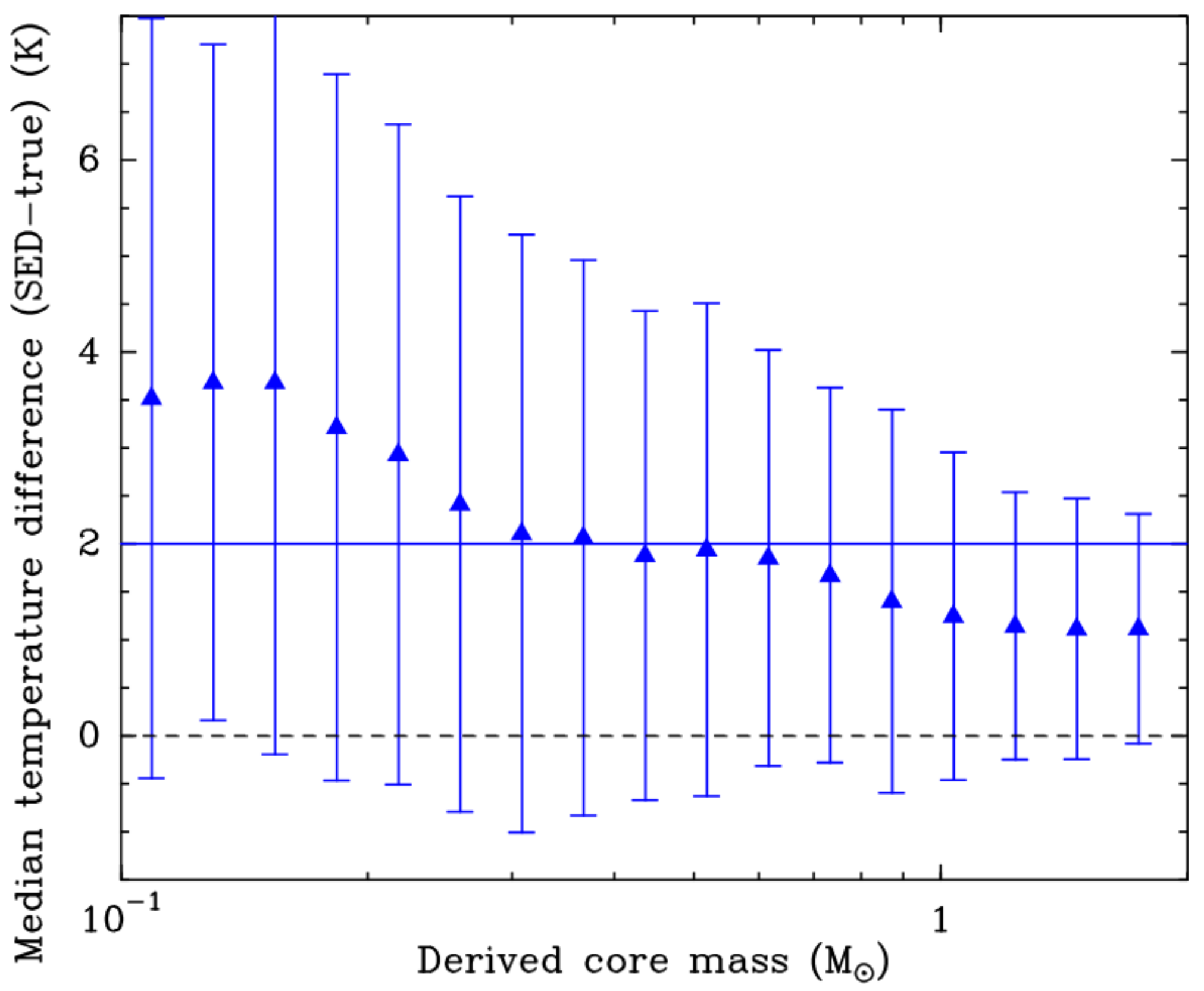}}
  \end{minipage}
\caption{
         {\bf Left:} Ratio of measured to true core mass as function of measured core mass for the simulated core extractions. 
         The error bars are $\pm 1\sigma$ where $\sigma$ is the standard deviation of the mass ratio in each mass bin. 
         The median mass ratio is $\sim 0.75$ for observed core masses between $\sim 0.3\, M_\odot $ and $\sim 0.8\, M_\odot $ (as indicated by the horizontal blue line).  
         The horizontal dashed line indicates a mass ratio of 1, as expected for perfect core mass estimates.
         {\bf Right:} Difference between measured SED temperature and true mass-averaged dust temperature as function of measured core mass 
         for the simulated core extractions. 
         The error bars are $\pm 1\sigma $ where $\sigma$ is the standard deviation of the temperature difference in each mass bin. 
         The median temperature difference is approximately $+2$~K between $\sim 0.3\, M_\odot $ and $\sim 0.8\, M_\odot $ (as marked by the horizontal blue line), 
         close to the fiducial completeness limit.
         The horizontal dashed line indicates no temperature difference, as expected for perfect core temperature estimates.
        }
        \label{fig:completeMT}       
\end{figure*}

As stated in Sect.~\ref{sec:completeness} and illustrated in Fig.~\ref{fig:complete}, the results of these Monte-Carlo tests 
indicate that the present {\it Herschel} census of prestellar cores in Orion~B 
is $\sim 80\% $ complete down to $\sim 0.5\, M_\odot $ in {\it true} core mass. 

\begin{figure}[!hb]
 \centering
  \resizebox{1.0\hsize}{!}{\includegraphics[angle=0]{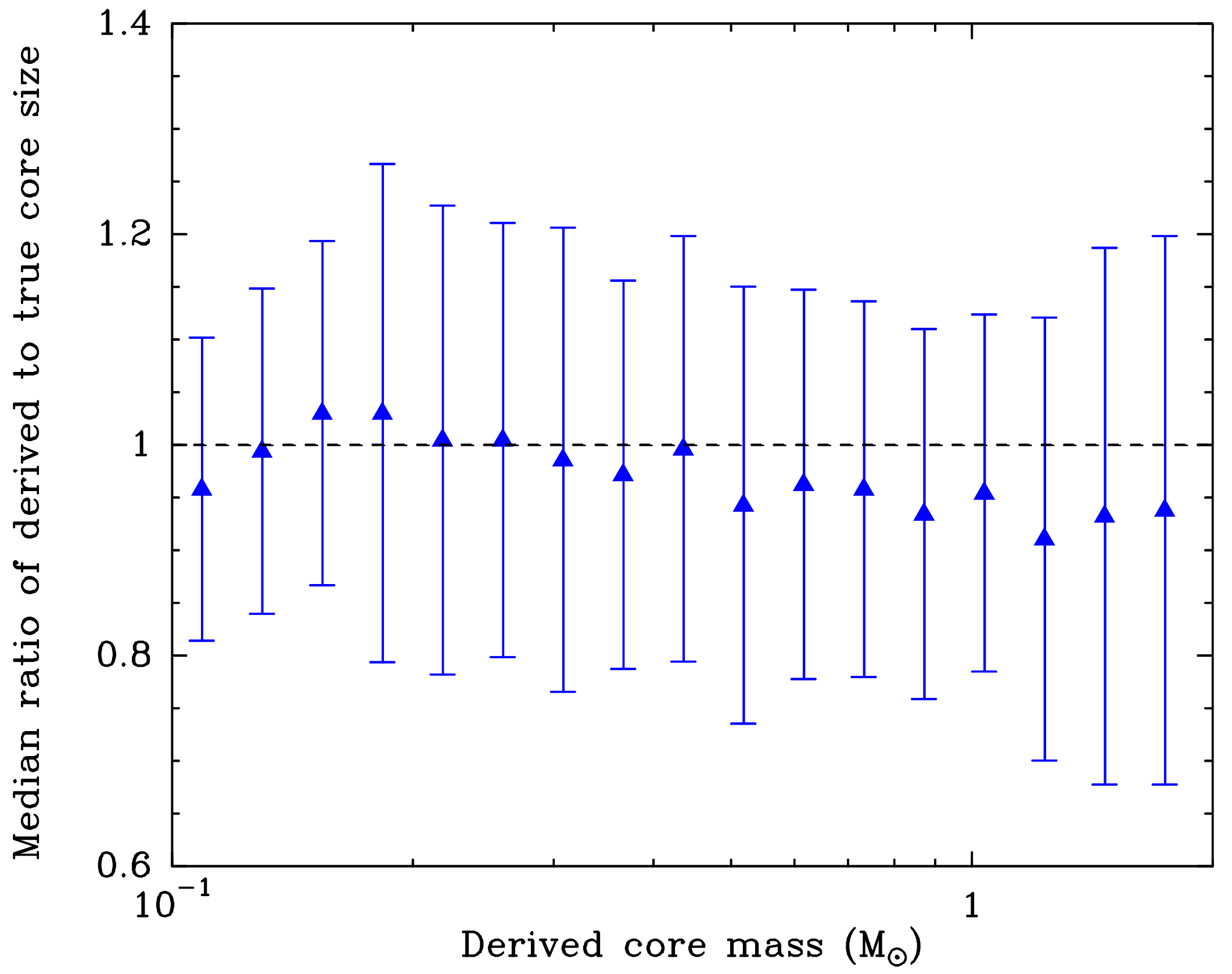}}
\caption{
         Ratio of measured to (convolved) intrinsic core size as function of measured core mass for simulated core extractions.
         The error bars are $\pm 1\sigma$ where $\sigma$ is the standard deviation of the size ratio in each mass bin. 
         The horizontal dashed line indicates a size ratio of 1, as expected for perfect core size estimates.
        }
\label{fig:completeR}       
\end{figure}

The above Monte-Carlo simulations were used to estimate the accuracy of the derived parameters such as core mass, radius, and dust temperature 
by comparing the measured values after core extraction to the intrinsic values of the model cores. 
Figure~\ref{fig:completeMT} (left) shows that the measured core masses typically underestimate the true core masses by $\sim \, $25\%  on average, and 
Fig.~\ref{fig:completeMT} (right) suggests that the derived SED temperatures typically overestimate the true mass-averaged dust temperatures of the 
cores by typically $\sim 2$~K, around the fiducial completeness limit of $\sim 0.4\, M_\odot $ in observed mass. 
A similar comparison for the core sizes (Fig.~\ref{fig:completeR}) indicates that the measured sizes (prior to deconvolution) are within $\sim 5\% $ of 
the true core sizes on average.  
Since the core masses are underestimated when the dust temperatures are overestimated, the slight bias of the derived masses (Fig.~\ref{fig:completeMT} left) 
can be interpreted as a direct consequence of the temperature overestimation (Fig.~\ref{fig:completeMT} right). 
The latter is because the dust temperature derived from SED fitting exceeds the mass-averaged dust temperature of 
starless cores owing to the presence of an internal dust temperature gradient rising outward \citep[see, e.g.,][]{Roy+2014}. 
Accounting for the $\sim \, $25\% effect on the derived masses, we estimate that the $\sim 80\% $ completeness limit at $\sim 0.5\, M_\odot $ in {\it true} core 
mass is located at $\sim 0.4\, M_\odot $ in {\it observed} core mass. 

%

\begin{figure}[!hb]
 \begin{center}
  \begin{minipage}{1.0\linewidth}
   \resizebox{1.0\hsize}{!}{\includegraphics[angle=0]{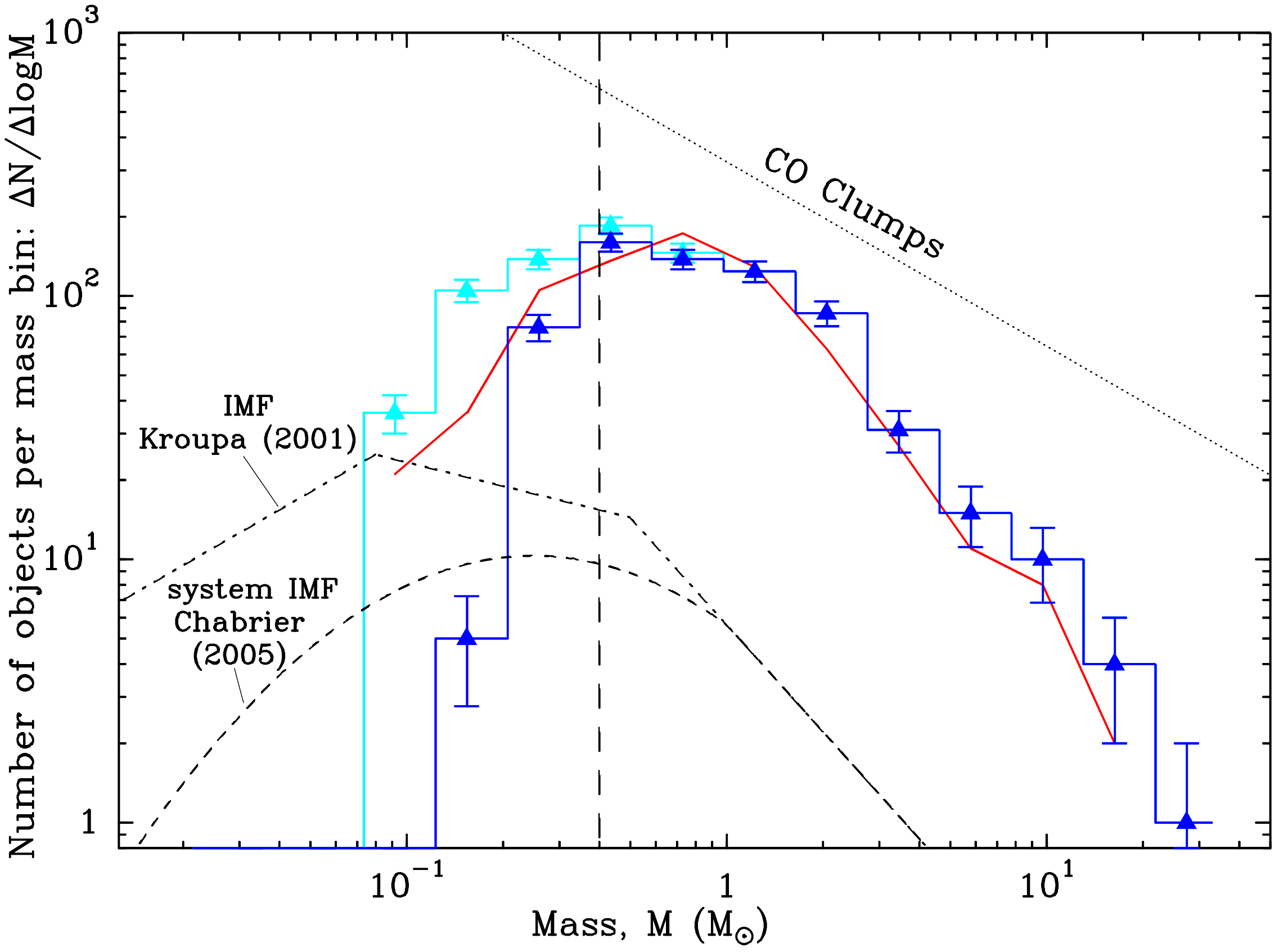}}
  \end{minipage}
 \end{center}
   \caption{Synthetic prestellar core mass function (CMF) derived from simulated source extractions (blue histograms) 
            compared to the input mass function of 711 model cores (red curve) constructed as described in the text. 
            The light blue and dark blue histograms show the mass functions of the extracted {\it candidate} and {\it robust} prestellar 
	    cores, respectively. 
            The estimated 80\% completeness level of our {\it Herschel} census of prestellar cores in Orion~B is indicated 
	    by the vertical dashed line at $0.4\, M_\odot $ (in observed core mass).
            }
   \label{fig_simu_CMF}%
\end{figure}

Figure~\ref{fig_simu_CMF} confirms that the prestellar CMF can be reliably determined down to a fiducial completeness limit of $\sim 0.4\, M_\odot $ 
in observed mass. 
A Kolmogorov-Smirnov (K-S) test shows that the derived CMF (blue histograms in Fig.~\ref{fig_simu_CMF}) is statistically indistinguishable (at the 
$\sim 90\% $ confidence level) from the input mass function (red curve in Fig.~\ref{fig_simu_CMF}) above $0.4\, M_\odot $. 
We also point out  that the mass function of extracted {\it candidate} prestellar cores (light blue histogram in Fig.~\ref{fig_simu_CMF}) starts 
to deviate from the mass function of {\it robust} prestellar cores (dark blue histogram) below $\sim 0.4\, M_\odot $, implying that the selection of bona 
fide self-gravitating starless cores becomes quite uncertain below the fiducial completeness limit. Indeed, the sample of {\it candidate} prestellar cores 
includes a significant number of non-self-gravitating or spurious objects below this limit. Conversely, the sample of {\it robust} prestellar cores 
becomes severely incomplete.

\begin{figure}[!h]
 \begin{center}
  \begin{minipage}{1.0\linewidth}
   \resizebox{1.0\hsize}{!}{\includegraphics[angle=0]{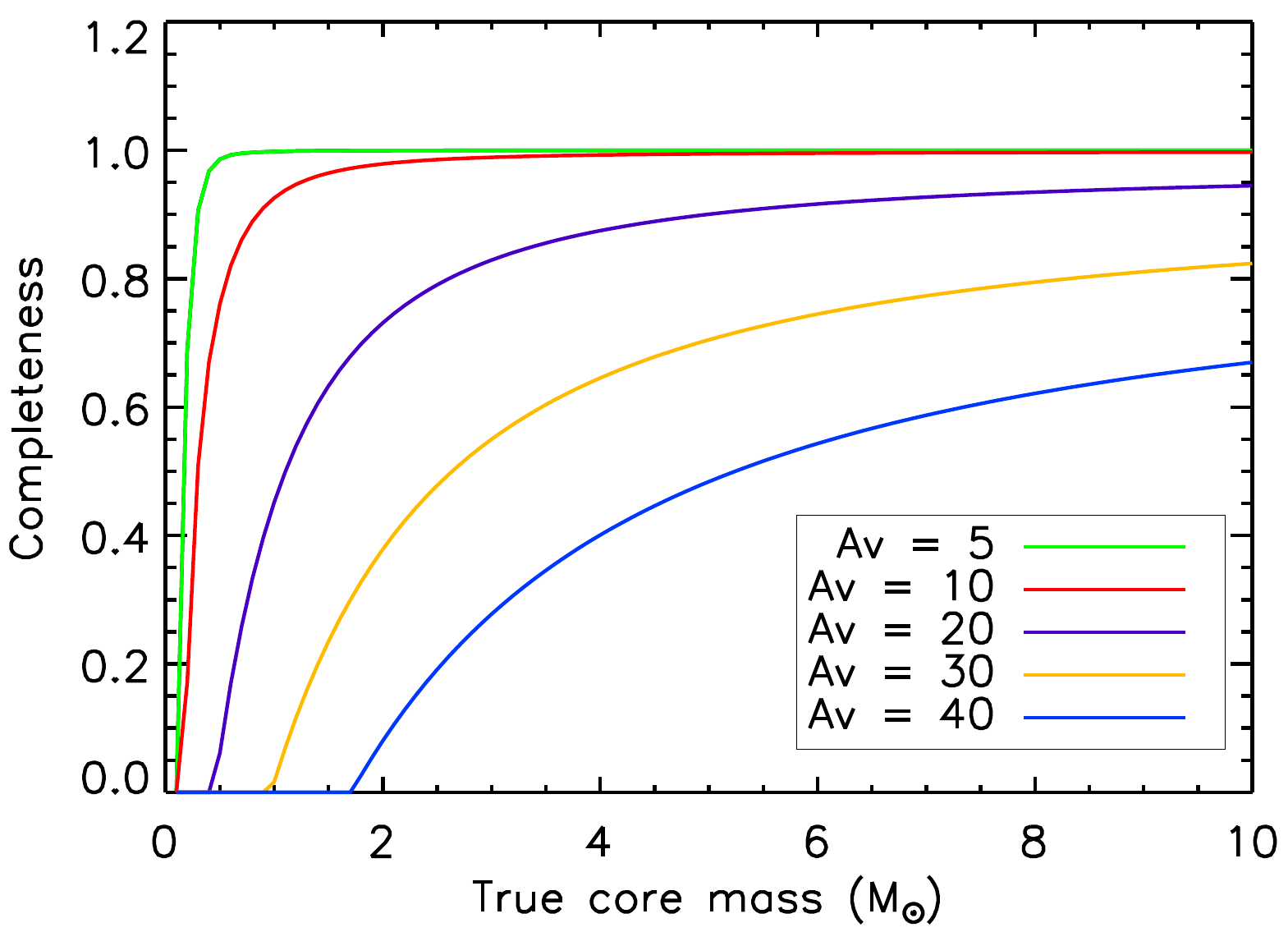}} 
   \end{minipage}
  \end{center}
  \caption{Model completeness curves of {\it Herschel} prestellar core extractions in Orion~B for five values of background cloud column 
  density expressed in units of visual extinction from $A_{\rm V}^{\rm bg} = 5$ to $A_{\rm V}^{\rm bg} = 40$.
           }
  \label{fig_completeness}%
\end{figure}

As the rms level of background emission fluctuations (often referred to as ``cirrus confusion noise'') generally increases with column density 
(cf. \citealp{Kiss+2001} and Fig.~B.4 in \citealp{Konyves+2015}), the completeness level of our census of prestellar cores is background dependent 
and becomes worse in areas with higher background cloud emission.
To estimate the variations of the completeness level in the observed field, we used the simple model of the core extraction process and completeness 
problem presented in Appendix B.2 of \citet{Konyves+2015} for the Aquila cloud.
This model, scaled to the distance of Orion~B, allowed us to estimate individual completeness curves as a function of background column density, 
as shown in Fig.~\ref{fig_completeness}. 
A global completeness curve (shown by the dashed blue curve in Fig.~\ref{fig:complete}) was then computed as a weighted average of the individual 
completeness curves using the observed distribution of mass in the cloud as a function of background column density (cf. Fig.~\ref{fig:cdPDF} here 
and Appendix B.2 of \citealp{Konyves+2015} for further details).

\section{Spatial clustering of prestellar cores in different mass regimes}\label{sec:appendix_massSegreg}

\subsection{Clustering of dense cores}\label{sec:cluster}

\begin{figure*}[!!!ht]
 \centering
  \resizebox{1.0\hsize}{!}{\includegraphics[angle=0]{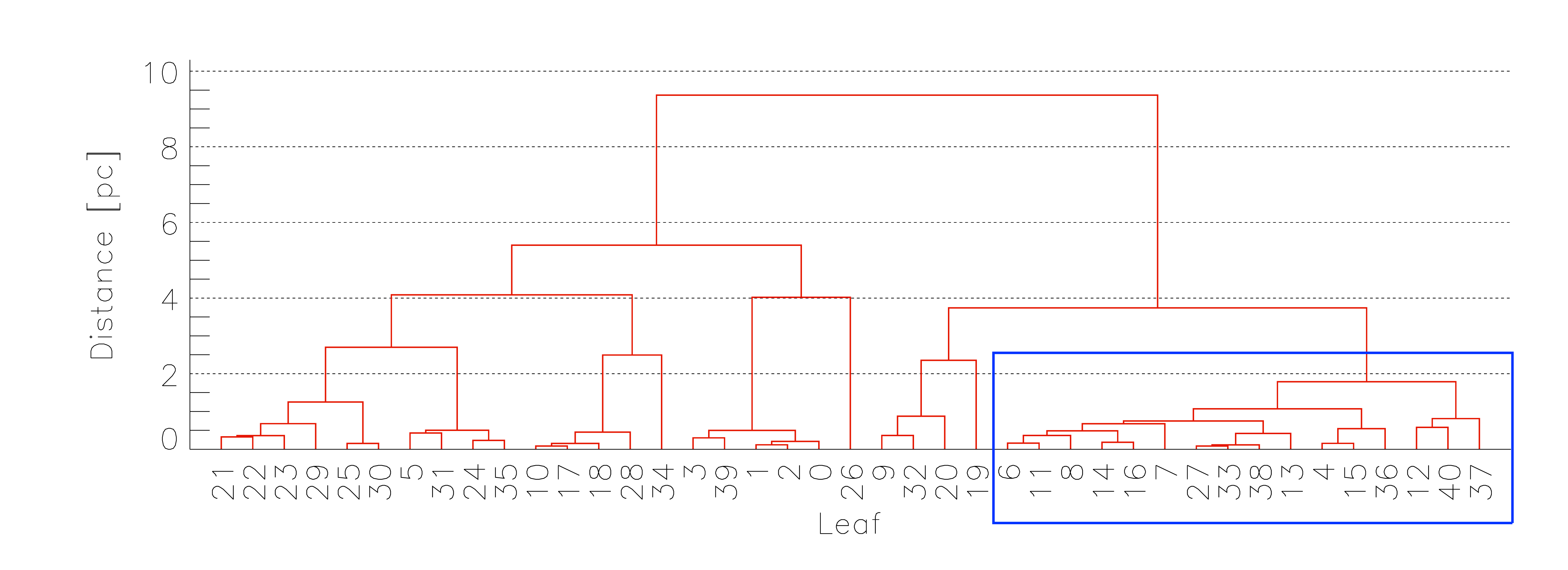}}
   \caption{Dendrogram tree used for selecting cores in L1622. The leaves represent the list of cores in the northern part of 
   Orion~B (Dec$_{\rm 2000} > 1$\degr 31$'$55$''$). 
   The 16 cluster members in this case were selected from leaves 6 to 37, marked by the blue rectangle. 
   The red vertical connections projected on the Y axis show the distance between leaves and branches (see text for details).}
   \label{fig:dendro}%
\end{figure*}

The spatial clustering of dense cores can be defined and investigated with many methods \citep[e.g.,][]{HKirk+2016b, Gutermuth+2009}. 
Here, we used dendrograms \citep{Rosolowsky+2008} to define groups or clusters of cores. 
We utilized the IDL functions CLUSTER\_TREE and DENDRO\_PLOT separately for the {\it robust} prespellar core sample in the northern, middle, and southern 
tiles, separated by the declination lines at Dec$_{\rm 2000}=1$\degr 31$'$55$''$, and Dec$_{\rm 2000}=-0$\degr 28$'$38$''$. Based on the column density 
distribution in the entire field, this seems to be a valid choice.
A clear and simple example of a hierarchical clustering is shown in Fig.~\ref{fig:dendro} for the northern region L1622.

In this figure, the labeled individual cores are arranged in the bottom of the dendrogram. In the jargon they are referred to as leaves, marked by the 
lowest vertical lines, and they have no substructures. The horizontal lines connect leaves, then higher hierarchy levels, making branches. The point 
where a vertical line originates from a horizontal one is called a ``node''. Its height represents the distance (in parsec) between the left and right 
branches. The elements (i.e., cores) of these branches can also be called clusters.  

\begin{figure*}[!ht]
 \begin{center}
  \begin{minipage}{0.35\linewidth}
   \resizebox{1.0\hsize}{!}{\includegraphics[angle=0]{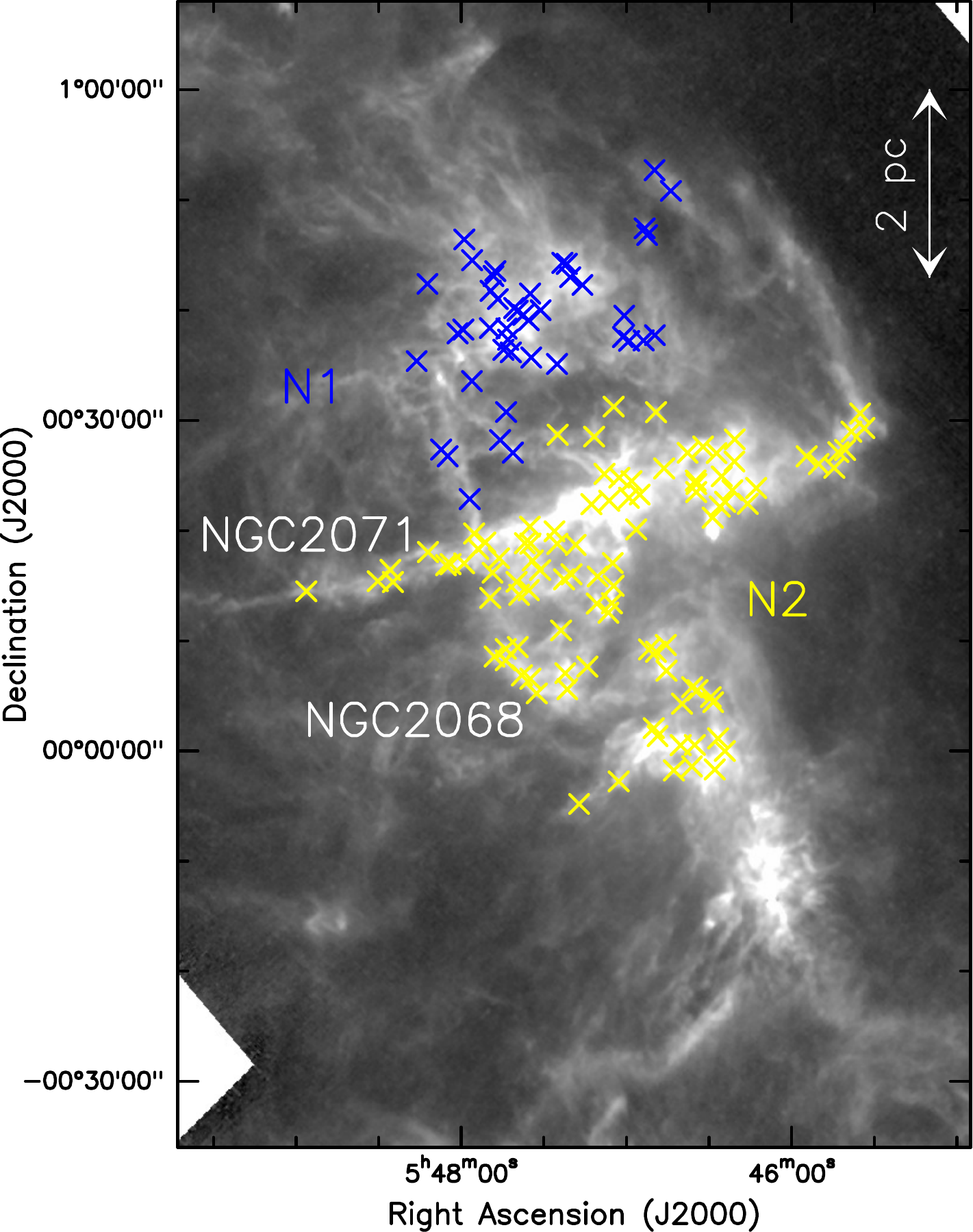}}
  \end{minipage}
  \begin{minipage}{0.287\linewidth}
   \resizebox{1.0\hsize}{!}{\includegraphics[angle=0]{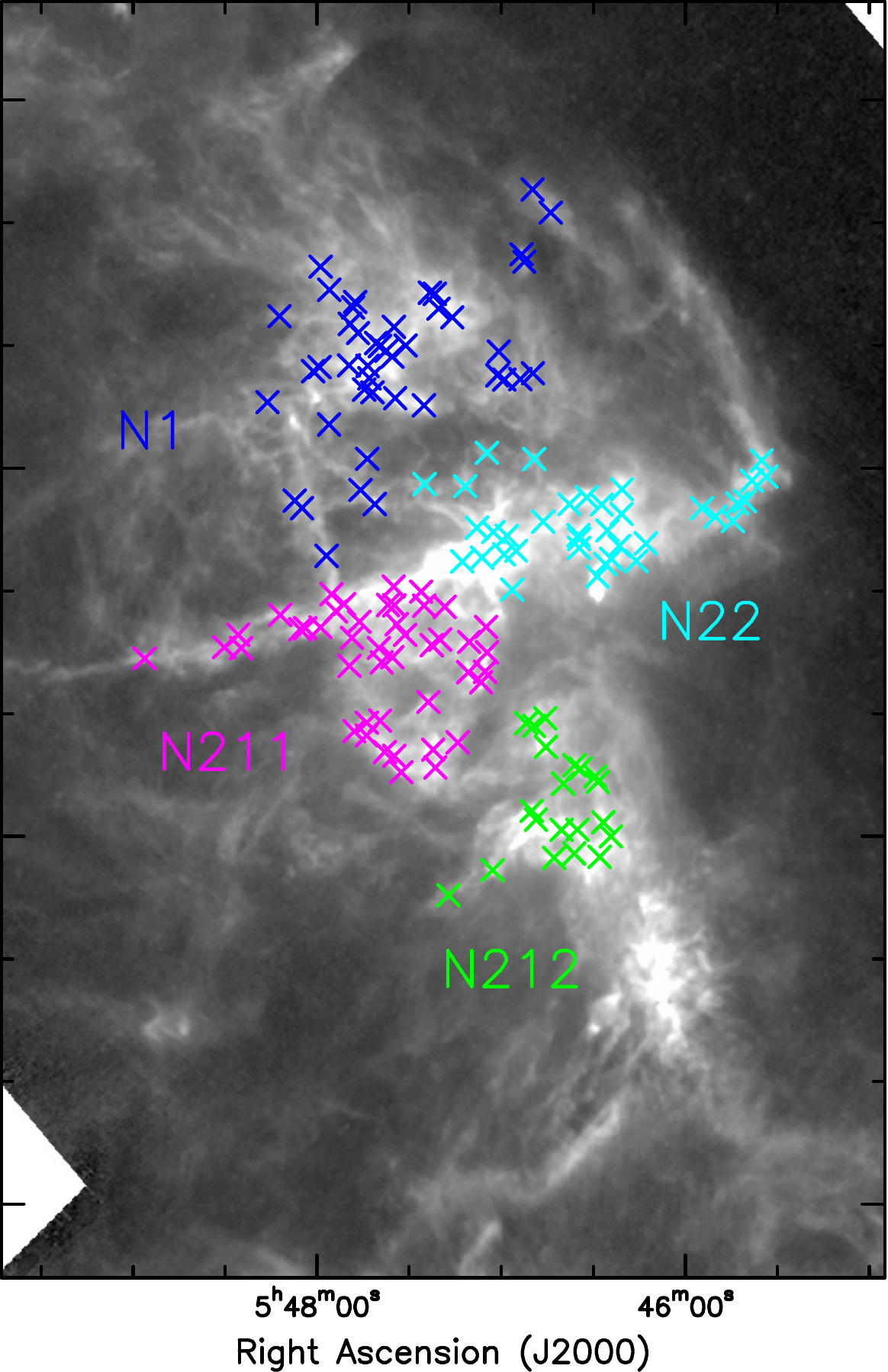}}
  \end{minipage} 
  \begin{minipage}{0.33\linewidth}
   \resizebox{1.0\hsize}{!}{\includegraphics[angle=0]{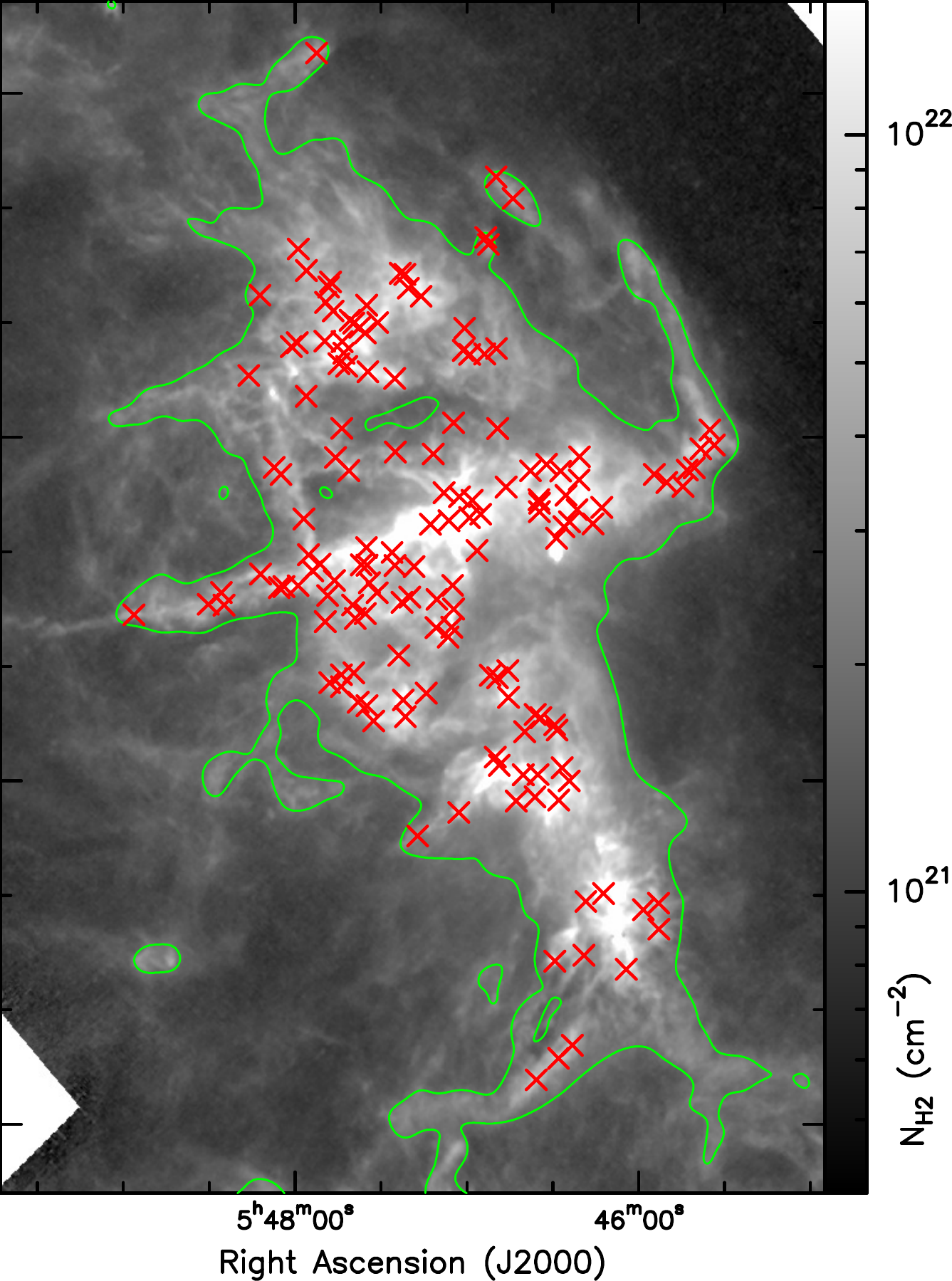}}
  \end{minipage} 
 \end{center}
   \caption{ 
   Clusterings of {\it robust} prestellar cores in NGC~2068 and 2071 of Orion~B, defined mainly using dendrograms. 
   The selected clusters are displayed in different colors, overplotted on the column density map of NGC~2068 and 2071.
   The {\bf left} panel shows a bigger cluster (N2 in yellow), where cores were selected within 3\,pc distances from one another in a branch
   of the dendrogram tree around NGC~2068 and 2071 (see text). A similar dendrogram tree is shown in Fig.~\ref{fig:dendro} for L1622.  
   The {\bf middle} panel displays subclusters of the N2 group. The cores within these smaller clusters are closer to each other 
   than $\sim 1.5-2$\,pc. In the {\bf right} image, an arbitrary column density level of about 2$\times 10^{21}$ cm$^{-2}$ defines 
   a group of cores in the region of NGC~2068 and 2071.   
            }
  \label{fig:cluster}%
\end{figure*}

\begin{figure*}[!hhh]
 \begin{center}
  \begin{minipage}{0.35\linewidth}
   \resizebox{1.0\hsize}{!}{\includegraphics[angle=0]{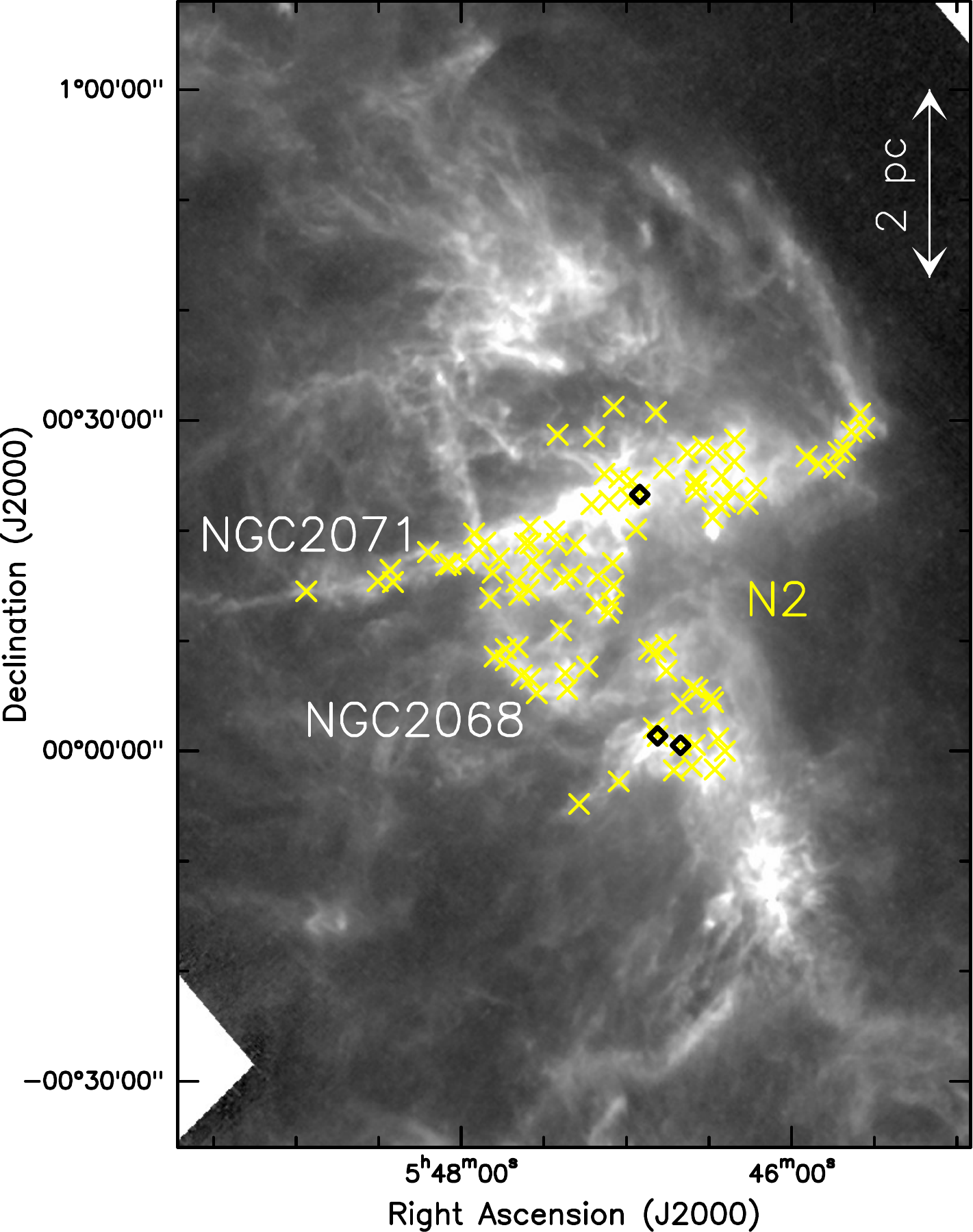}}
  \end{minipage}
  \begin{minipage}{0.287\linewidth}
   \resizebox{1.0\hsize}{!}{\includegraphics[angle=0]{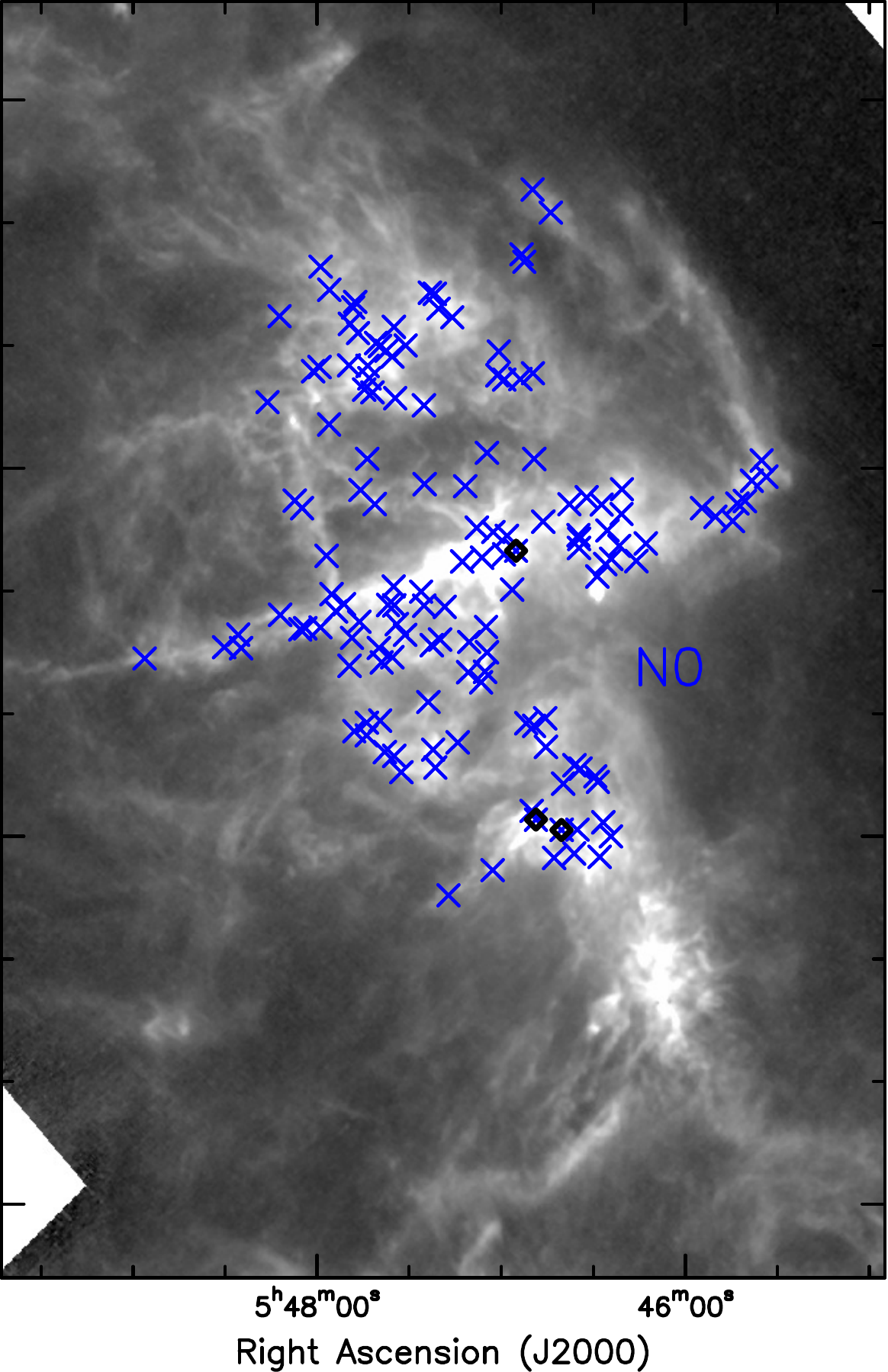}}
  \end{minipage} 
  \begin{minipage}{0.33\linewidth}
   \resizebox{1.0\hsize}{!}{\includegraphics[angle=0]{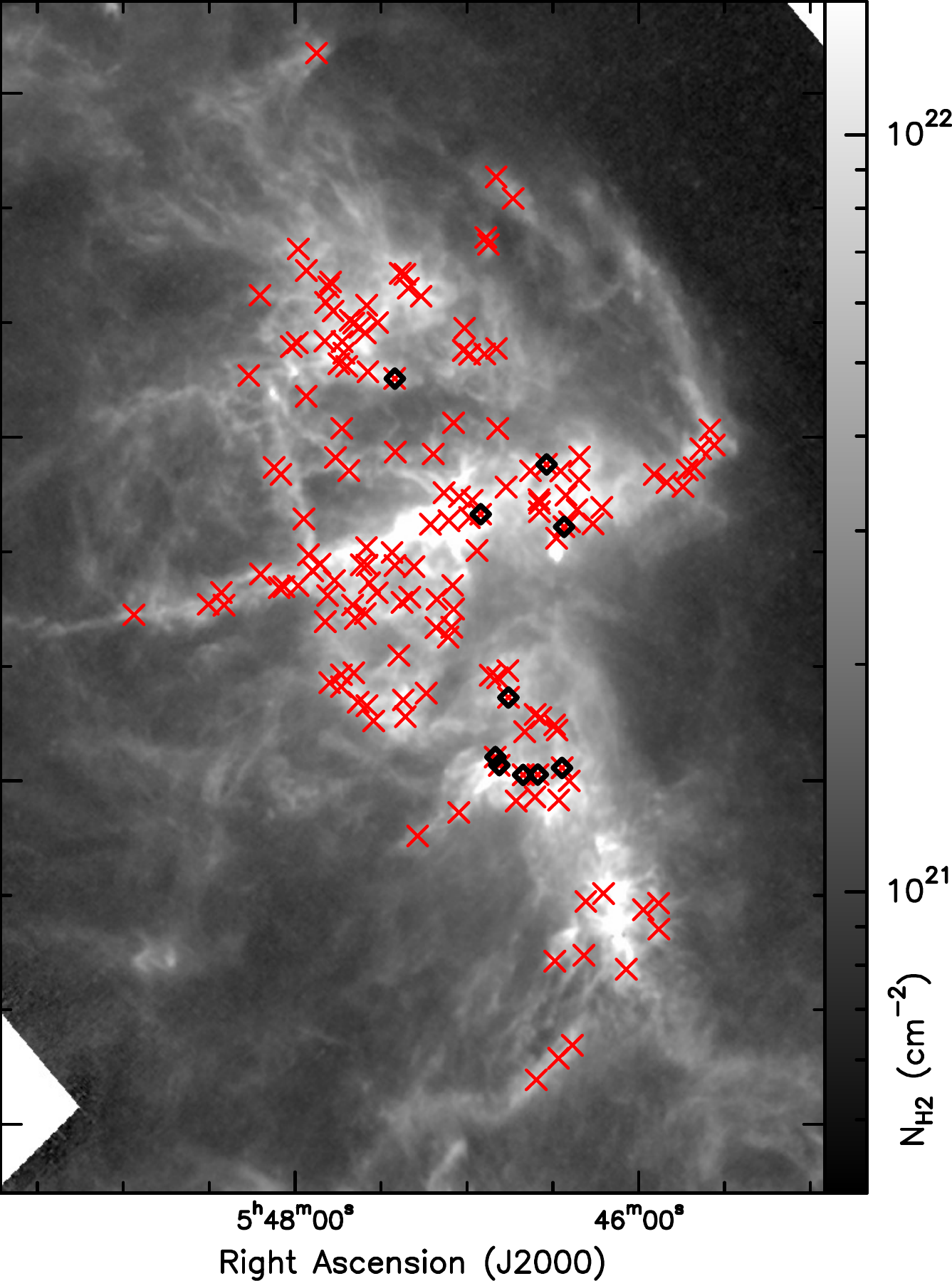}}
  \end{minipage} 
 \end{center}
   \caption{Alternative groupings of {\it robust} prestellar cores in NGC~2068 and 2071.
   The {\bf left} panel shows the N2 cluster, the {\bf middle} panel the N0 $=$ N1 $+$ N2 clusters (see left panel of Fig.~\ref{fig:cluster}),
   and the {\bf right} panel all cores lying above $\sim$2$\times 10^{21}$ cm$^{-2}$ in this area (see right panel of Fig.~\ref{fig:cluster}).  
   Black open diamonds mark the 3 (left and middle panels) or 10 (right panel) most massive prestellar cores in these clusters. 
            }
  \label{fig:coreMS}%
\end{figure*}

In order to select clusters for our analysis, the dendrogram trees were inspected, and cores within 3\,pc separation of each other in a branch 
were selected. We required a minimum number of 12 core elements. If a bigger cluster had significant sub-branches, we split it again into smaller clusters, 
typically at $\sim 1.5-2$\,pc distances. Their properties can be studied separately, or together in larger clusters. This way, we found one cluster containing 
16 {\it robust} prestellar cores around L1622 in the north of Orion~B (see Fig.~\ref{fig:dendro}). 
The left and middle panels of Fig.~\ref{fig:cluster} (and Fig.~\ref{fig:coreMS}) display the various clusters of {\it robust} prestellar cores in the NGC~2068 and 2071 regions based on the 
above selection from their dendrogram trees. The number of core elements in these clusters are as follows: cluster N1: 43 cores, cluster N2: 100, 
N22: 36, N211: 44, N212: 20. 

The CLUSTER\_TREE function was run with the LINKAGE keyword ($=2$) which uses a weighted pairwise average. The distance between two leaves and clusters 
is defined as the average distance for all pairs of objects between each cluster, weighted by the number of objects in each cluster. We chose this method 
because it works well both for homogeneous clusters and for chain-like clusters. We indeed trace rather elongated, and more compact (sub)clusters as well. 

Although these clustering properties stem only from the extracted core positions, the resulted clusters correlate very well with underlying 
column density features, including filaments, along which most of the cores are found. If we compare Fig.~\ref{fig:fils_cores} (left panel) 
and Fig.~\ref{fig:cluster}, we find eye-catching chains of cores along filaments.    
Using this idea, we also set an arbitrary column density level of about 2$\times 10^{21}$ cm$^{-2}$ to select another group of cores in the region 
of NGC~2068 and 2071 (right panel of Fig.~\ref{fig:cluster}). 
With various clusters over the same region we can test if certain trends of core properties hold on different spatial scales.

\subsection{Mass segregation of dense cores with the MST method}\label{sec:appendix_MST}

Figure~\ref{fig:medMcore_bgCD} in Sect.~\ref{sec:massSegreg} indicates that the most massive prestellar cores in our sample are spatially segregated in 
the highest column density portions of the Orion~B complex. 
This mass segregation can also be quantified using the minimum spanning tree (MST) technique \citep[e.g.,][]{Allison+2009, ParkerGoodwin2015}. 
We applied this technique to the clusters of dense cores defined with the help of dendrograms in Appendix~\ref{sec:cluster}. 
The MST method compares the minimum spanning tree of a given number $N$ of the most massive cores with that of a randomly distributed population of cores. 
The observed and random MSTs are constructed from graphs where the shortest possible separations are calculated between cores, without allowing closed loops. 
If mass segregation is present, the MST length of the most massive cores in the sample should be significantly shorter than that of a random population of cores. 
The degree of mass segregation can be quantified via the mass segregation ratio, $\Lambda_{\rm MSR}$, defined as the ratio of the average MST length for 
$N$ randomly chosen cores to the MST length for the $N$ most massive cores in the sample \citep{Allison+2009, Parker2018}. 
The $\Lambda_{\rm MSR}$ should significantly exceed 1 when mass segregation is present. 

\begin{figure}[!!!!!hh]
 \centering
  \resizebox{0.95\hsize}{!}{\includegraphics[angle=0]{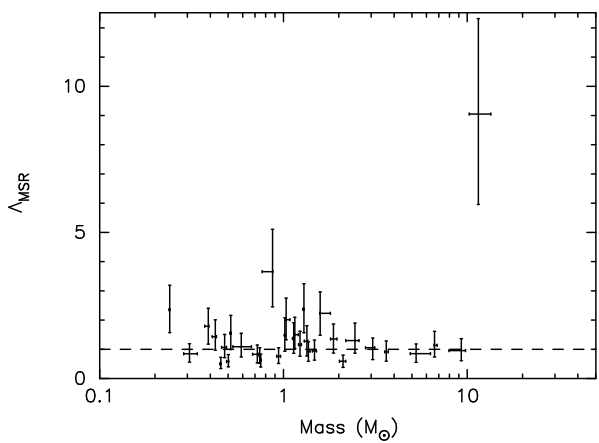}}
   \caption{ 
   Mass segregation ratio, $\Lambda_{\rm MSR}$, as function of core mass 
   in the N2 cluster of NGC~2068 and 2071, including 100 {\it robust} prestellar cores (see left panel of Fig.~\ref{fig:cluster}).
   The tool used to derive $\Lambda_{\rm MSR}$ is described in \citet{Allison+2009}. 
   A sliding window of three cores ($N=3$) was set in this case.
   }
  \label{fig:MSR}%
\end{figure}

Figure~\ref{fig:MSR} shows a mass segregation ratio plot, in which MST calculations were applied to the N2 cluster of NGC~2068 and 2071 (see left panel of 
Fig.~\ref{fig:cluster}).
With a sliding window of three cores, the N2 cluster shows an explicit positive mass segregation ($\Lambda_{\rm MSR} > 6$), where the most massive cores 
in this region with $M > 10\, M_\odot $ are more concentrated spatially than cores with lower masses (left panel of Fig.~\ref{fig:coreMS}). 
However, comparing and contrasting the panels of Fig.~\ref{fig:coreMS} with the summary table of mass segregation properties in various core clusters 
(Table~\ref{tab_MS}), it does not appear obvious that the three most massive cores in the N2 cluster 
are much more concentrated than the --same-- three most massive cores in the N0 (=N1$+$N2) cluster (see middle panel of Fig.~\ref{fig:coreMS}). 
After carefully inspecting the values in Table~\ref{tab_MS}, the overall result in NGC~2068 and 2071 provides only marginal evidence for other than a random 
distribution of core masses, even though in one case the segregation of the most massive cores is strong (i.e., Fig.~\ref{fig:MSR}). 
Further caveats and assumptions of the $\Lambda_{\rm MSR}$ method are discussed in detail in, for example, \citet{Parker2018}. 

Finding some degree of mass segregation at the prestellar core stage in Orion~B would lend support to the idea of primordial mass segregation in stellar 
protoclusters. Mass segregation at birth may be expected since protostellar sources embedded in denser filaments can accrete longer before feedback 
effects stop further gas inflow. 
This effect differs from the dynamical mass segregation quickly affecting evolved stars in revealed star clusters \citep[e.g.,][]{Kroupa2008}. 

Using published dense core samples from both the HGBS and ALMA (Atacama Large Millimeter Array) observations, \citet{DibHenning2018} performed structure 
and mass segregation studies that they linked to the star formation activity of their set of clouds. They claim that the level of core mass segregation is 
positively correlated with the surface density of the star formation rate in the cloud. 
They attribute this correlation to a dependence on the physical conditions that have been imprinted by the large scale environment at the time these 
clouds started to form.

Based on dense cores in Orion~B from JCMT's SCUBA-2 observations, \citet{HKirk+2016b} and \citet{Parker2018} analyzed the clustering and mass segregation 
properties by various methods. While \citet{HKirk+2016b} found segregation of massive cores in NGC~2068 and 2071, and in NGC~2023 and 2024, according to 
\citet{Parker2018}, only a mild mass segregation is present in the NGC~2068 and 2071 regions. Our above findings seem to approach these latter results.
However, we issue here a word of caution: the conclusions on mass segregation may depend somewhat on the chosen cluster and the exact method of analysis.

\begin{table}\small\setlength{\tabcolsep}{4.5pt}
\caption{Mass segregation results with $\Lambda_{\rm MSR}$ method \citep[based on][]{Allison+2009} in various core clusters of NGC~2068 and 2071.}
\label{tab_MS}      
{\renewcommand{\arraystretch}{1.2}        
\begin{tabular}{l|c|c|c|c}  
 \hline
 \hline
                                            & $N=2$      & $N=3$        & $N=5$     &  $N=10$  \\
 \hline
 N11                                        & {\it 0}    & {\it 0}      &           &          \\
 \hline
 N1                                         &            & {\it 0}      & {\it 0}   &          \\
 \hline
 N2                                         &            & {\it 4}      & {\it 2}   & {\it 2}  \\
 \hline
 N22                                        &            & {\it 0}      & {\it 0}   &          \\
 \hline
 N211                                       &            & {\it 0}      & {\it 0}   &          \\
 \hline
 N212                                       & {\it 0}    & {\it 1}      & {\it 0}   &          \\
 \hline
 N0 (N1$+$N2)                               &            & {\it 0}      & {\it 1}   & {\it 2}  \\
 \hline
 N22$+$N211                                 &            & {\it 1}      & {\it 1}   & {\it 0}  \\
 \hline
 {\it robust} pre $> 2 \times 10^{21}$ cm$^{-2}$  &            & {\it 0}      & {\it 2}   & {\it 3}  \\
 \hline
 {\it cand.} pre $> 2 \times 10^{21}$ cm$^{-2}$  &            & {\it 0}      & {\it 2}   & {\it 3}  \\
 \hline
\end{tabular}
}
\tablefoot{This table summarizes results on mass segregation using various core samples that are listed in the left column 
(see Fig.~\ref{fig:cluster} and text for details).   
The assigned numbers in the table from 0 to 4 correspond to a range of {\it No} to {\it Strong} in properties of 
mass segregation. In a given cluster we analyzed the spatial distribution of cores against the $N$=2, 3, 5, 10 highest mass cores
in that cluster, provided that this was reasonable based on the size of the cluster (table cells are empty if not). 
} 
\end{table}

\subsection{Core masses versus core surface density}\label{sec:appendix_Simon}

As we have shown in Sect.~\ref{sec:massSegreg}, the presence of prestellar cores is strongly correlated with the local column density in the ambient cloud. 
This relationship is perhaps not surprising as dense cores require sufficient gas (at high column- and volume density) to form in the first place. 
In this appendix we present some initial results on the distributions of cores relative to each other (which will be investigated in more detail in a later paper).

We consider the sample of 423 {\it robust} prestellar cores, with masses $> 0.4 M_\odot$ above the completeness limit (see Sect.~\ref{sec:completeness}).
The core surface number density, $\Sigma_{\rm cores}$, is defined from the radius containing the 6 nearest neighbors, $r_6$, as $\Sigma_{\rm cores} = 6/r_6^2$ 
\citep[cf.][]{CasertanoHut1985}.
Very low values of $\Sigma_{\rm cores}$ indicate that cores are relatively isolated, very high values of $\Sigma_{\rm cores}$ that there are many cores close-by. 

\begin{figure}[]
 \centering
  \begin{minipage}{0.9\linewidth}
   \resizebox{0.95\hsize}{!}{\includegraphics[angle=0]{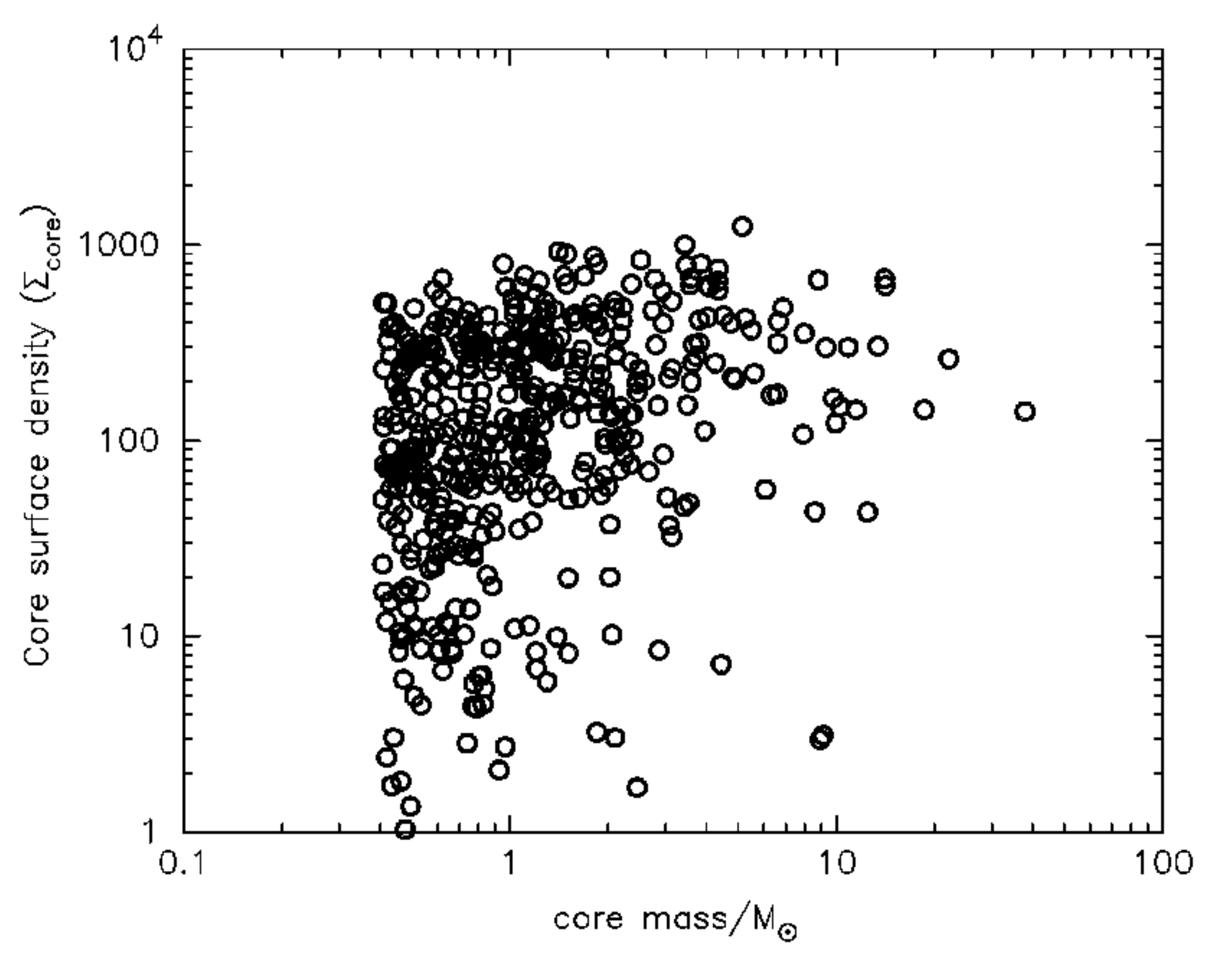}}
  \end{minipage}
\caption{ 
   Distribution of core surface number densities (in angular area units) against core mass (in $M_\odot$). The sample of {\it robust} prestellar cores 
   was used above $0.4 M_\odot$. 
   }
\label{fig:SimonFig1}%
\end{figure}

From the results of Sect.~\ref{sec:massSegreg}, we know that the most massive cores are in the areas with the highest background column density.
An interesting related question to ask is whether the highest-mass cores are also found in regions of highest {\rm core} surface density. 
In Fig.~\ref{fig:SimonFig1}, we show a plot of core surface number density (using arbitrary angular area units) against core mass. 
It is immediately obvious that there is no strong correlation between core mass and core surface density. 

We can analyze Fig.~\ref{fig:SimonFig1} more quantitatively using a Gini-like statistic \citep[][see the latter for an astrophysical application]{Gini1912, Ward-Duong+2015}.  
In an ordered list of cores by surface density we can quantify if a subset of cores (e.g., those with the highest masses) are consistent with being 
randomly distributed in the list (i.e., looking for a bias in what surface density those cores are found at).

\begin{figure}[!h]
 \centering
  \begin{minipage}{0.9\linewidth}
   \centering
    \resizebox{0.87\hsize}{!}{\includegraphics[angle=0]{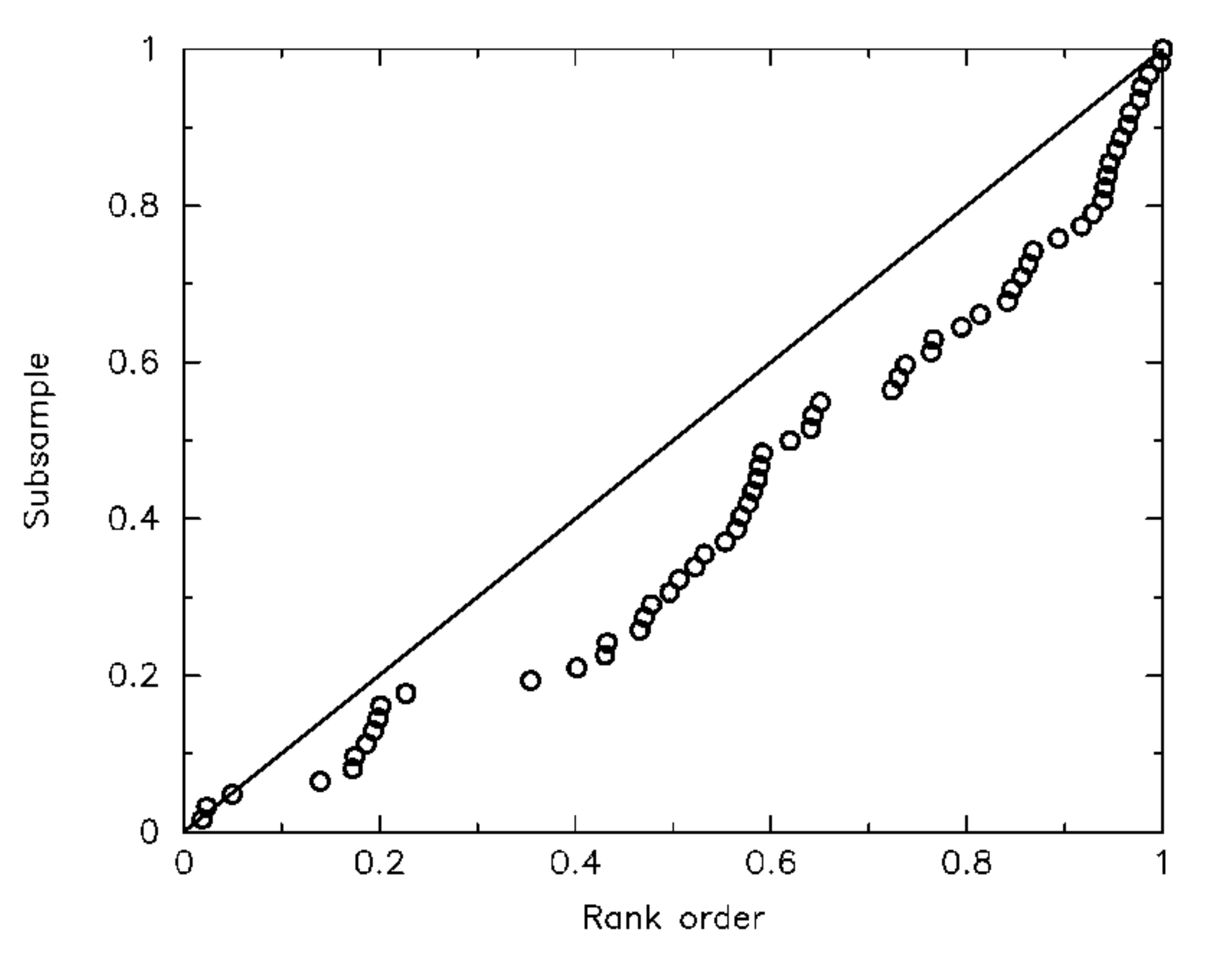}}
    \vspace{2mm}
    \resizebox{0.87\hsize}{!}{\includegraphics[angle=0]{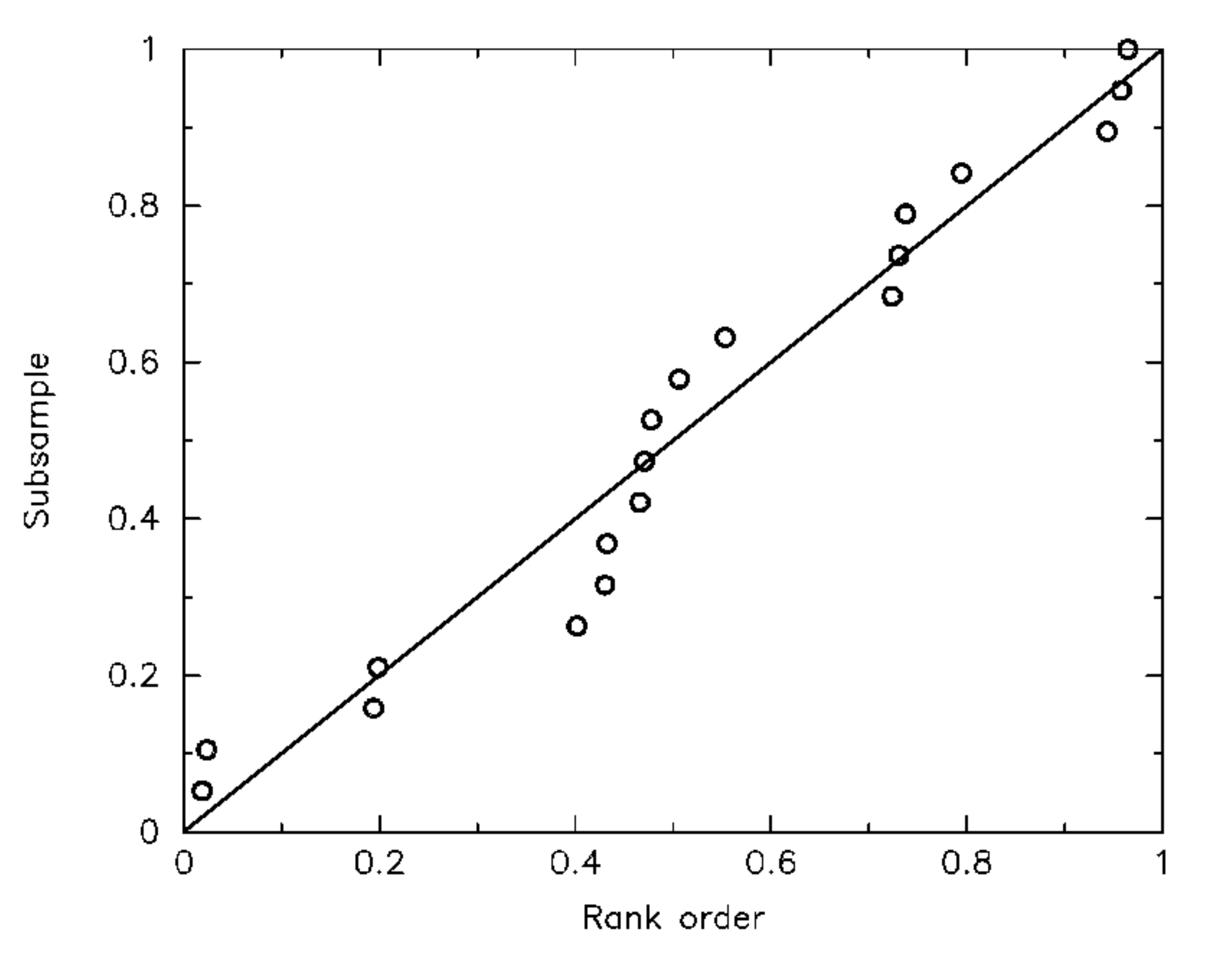}}
  \end{minipage}
\caption{
Cumulative distribution function of core positions (open circles) in ordered list of 423 core surface densities.
{\bf Top:} The positions of 62 cores with masses $> 3 M_\odot$ are shown.
{\bf Bottom:} The 19 most massive cores ($> 7 M_\odot$) are plotted. See text for explanation.  
        }
\label{fig:SimonFig23}       
\end{figure}

As an example, the upper panel of Fig.~\ref{fig:SimonFig23} shows the positions in an ordered list of all 423 core surface densities (the $x$-axis), 
and a cumulative distribution function (CDF) of the positions of the 62 cores with masses $> 3 M_\odot$ (open circles). If the cores with masses 
$> 3 M_\odot$ were randomly distributed in the list then we would expect them to roughly follow the diagonal line (i.e., be equally likely to be 
anywhere in the list).  
The clear deviation from randomness can be quantified by calculating the area under the actual distribution and comparing it to a large ensemble of 
samples of 62 cores taken randomly from the list.  
If the distribution were random we would expect the area under the CDF to be $0.51 \pm 0.03$ (1-$\sigma$ error), while the actual area is 0.38. 
More usefully dividing the true by the expected values we have a quantity $\Sigma^+_{\rm cores} = 0.76 \pm 0.03$ (formally 5-$\sigma$ from random, although the 
distribution of random areas is not normally distributed in the tails). More qualitatively, this discrepancy is because cores $> 3 M_\odot$ are very 
unlikely to be at low core surface densities (only 25\% of them are in the lowest 50\% of core surface densities), and very likely to be at high core surface 
density (a third of them are in the top 20\% of core surface densities).
A strong correlation between core mass and local core surface density was also found based on the SCUBA-2 data in Orion~B \citep{HKirk+2016b}.

However, if we take only the 19 most massive $Herschel$ cores $> 7 M_\odot$, as shown in the lower panel of Fig.~\ref{fig:SimonFig23}, we find that 
the most massive cores are completely consistent with being randomly distributed in core surface density with $\Sigma^+_{\rm cores} = 0.95 \pm 0.05$ (where unity 
is consistent with random).  
This is telling us that all cores with $> 3 M_\odot$ have a strong bias to be in regions of high core surface density, the trend is not following a simple 
higher-mass higher-core surface density relationship, and the very highest-mass cores could be ``different'' in some way and certainly show no signatures that 
could be easily interpreted as ``mass segregation''.

\section{Additional figures}\label{sec:appendix_addfigs}

\begin{figure*}[!!]
 \centering
  \begin{minipage}{0.9\linewidth}
   \resizebox{0.505\hsize}{!}{\includegraphics[angle=0]{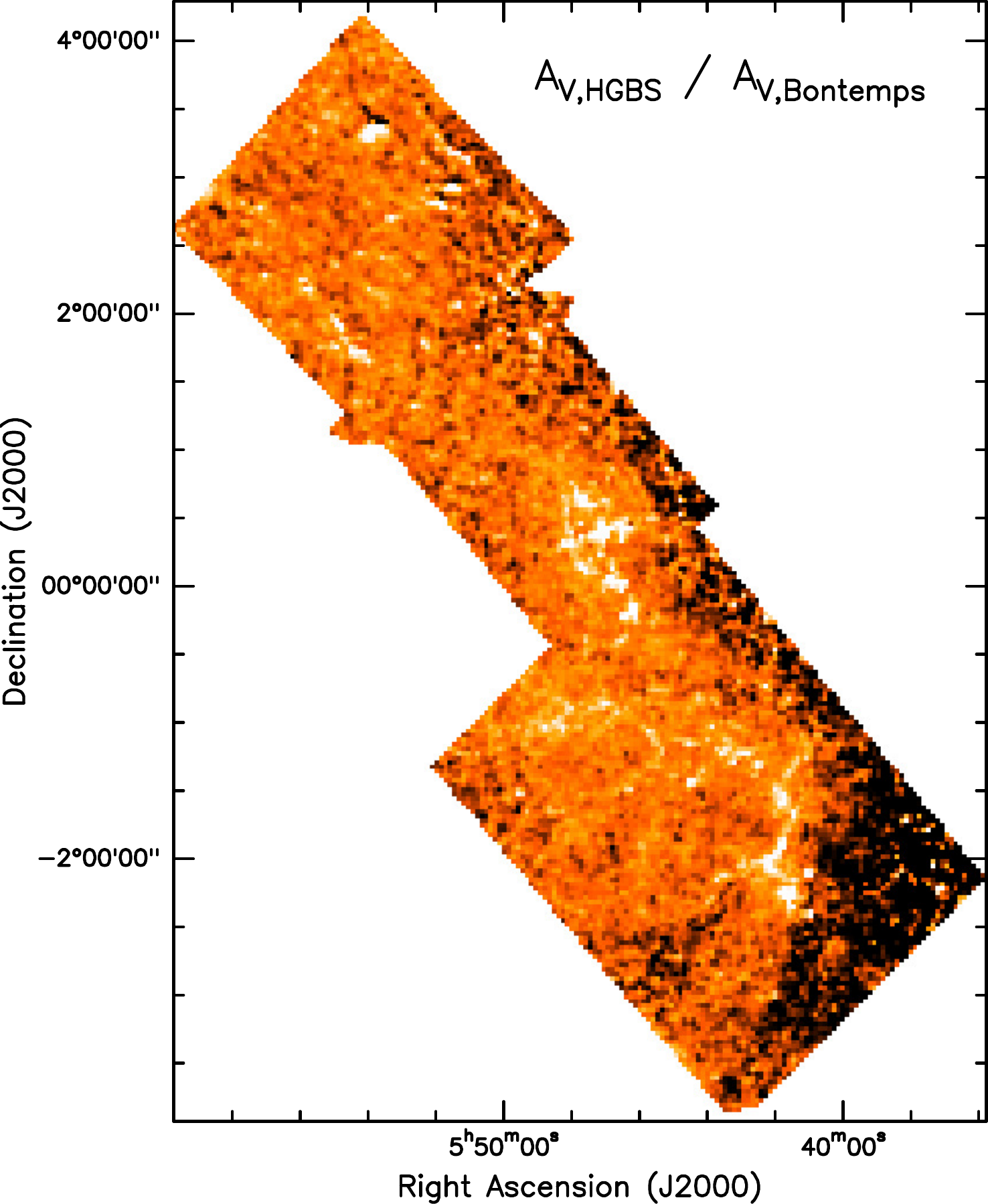}}
   \resizebox{0.47\hsize}{!}{\includegraphics[angle=0]{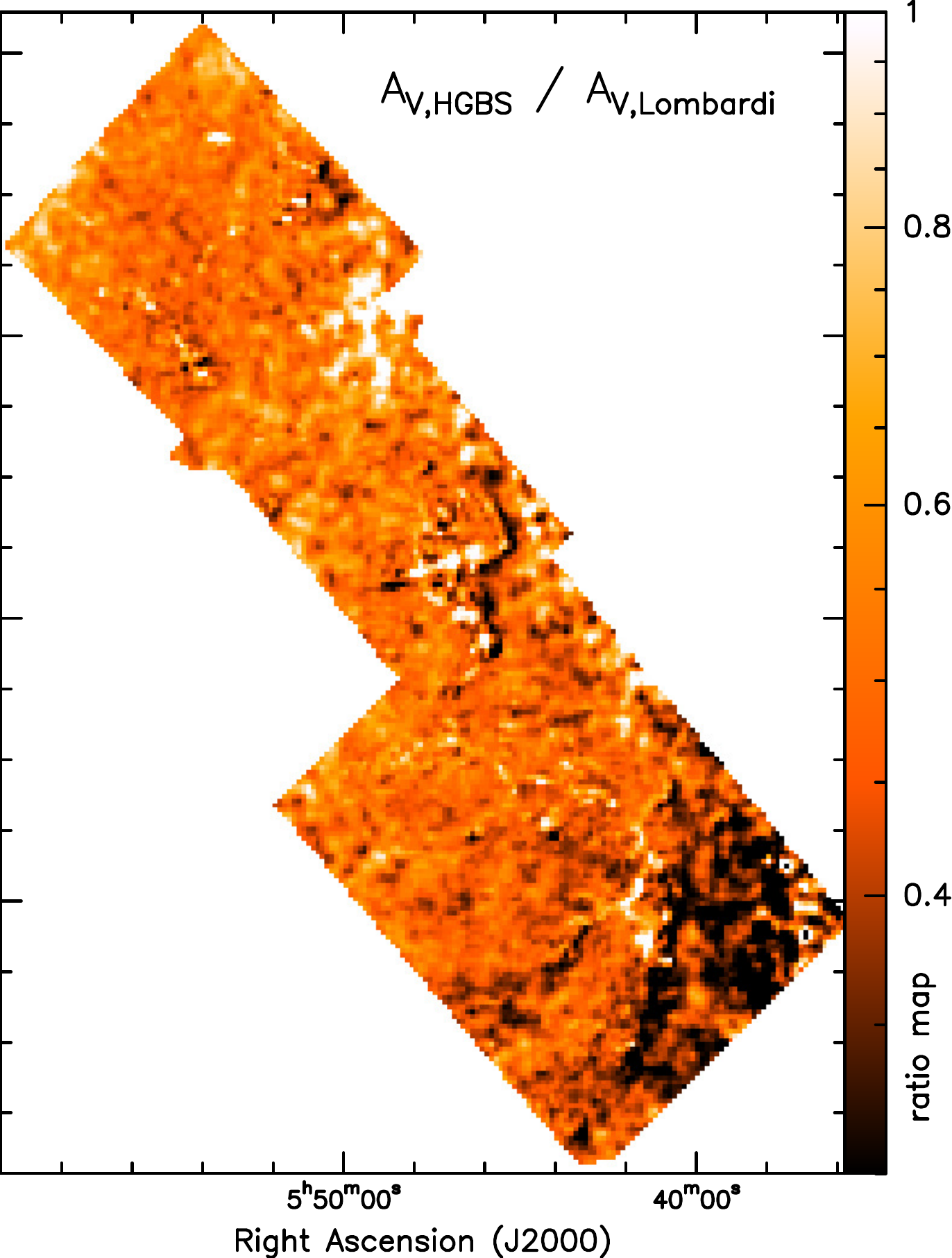}}
  \end{minipage}
   \caption{
   {\bf Left:} Map of the ratio of $A_{\rm V}$-converted  column density from HGBS data to visual extinction from Bontemps et al.
   {\bf Right:} Map of the ratio of $A_{\rm V}$-converted column density from HGBS data to visual extinction from Lombardi et al. 
   The median values of both ratio maps are $\sim$0.5 (see Sect.~\ref{sec:cd_t_maps} for details). 
   }
  \label{fig:ratioHGBSvsSylLom}%
\end{figure*}

\begin{figure*}[!ht]
 \centering
  \begin{minipage}{1.0\linewidth}
   \resizebox{0.48\hsize}{!}{\includegraphics[angle=0]{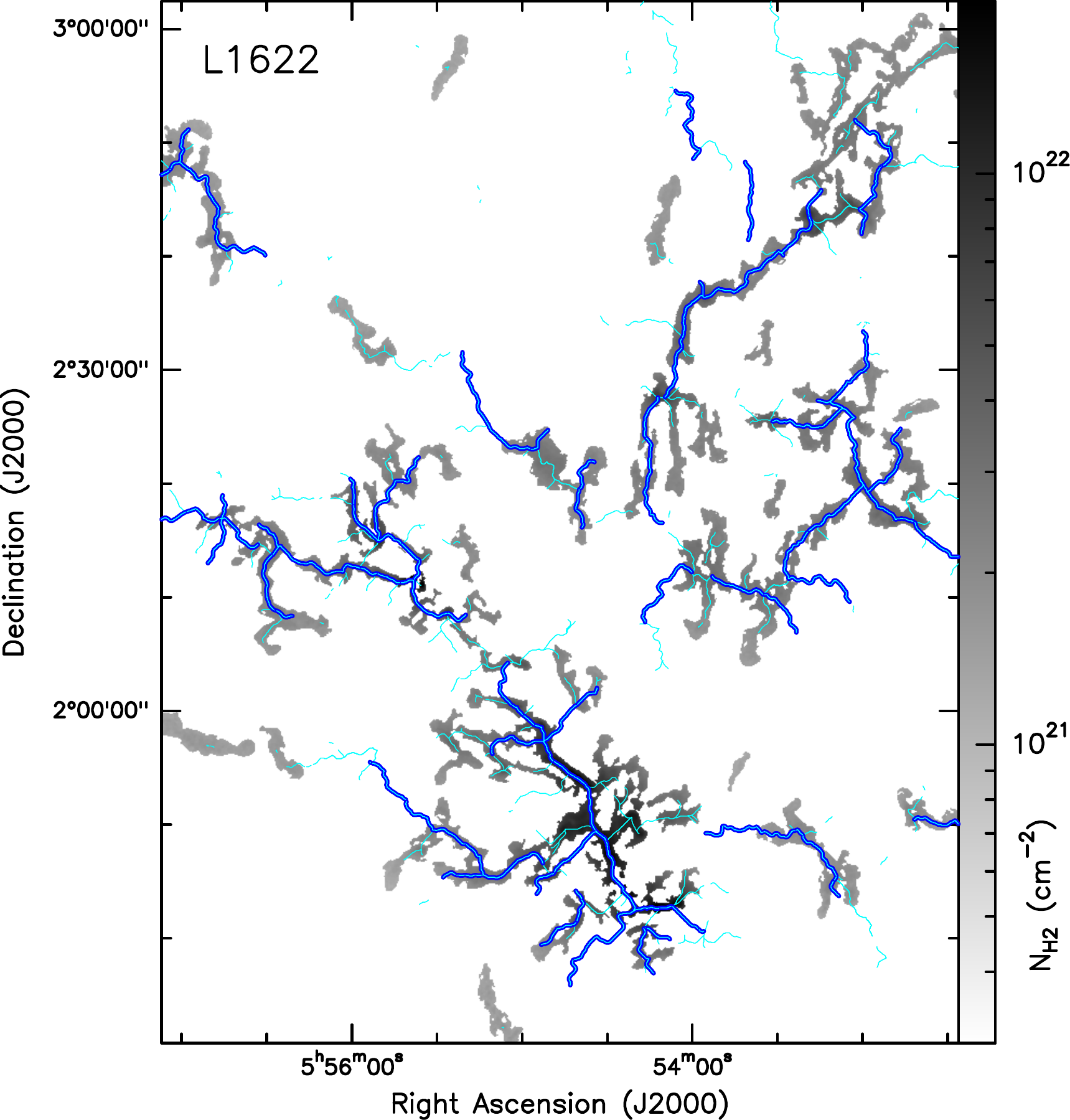}}
   \hspace{3mm}
   \resizebox{0.48\hsize}{!}{\includegraphics[angle=0]{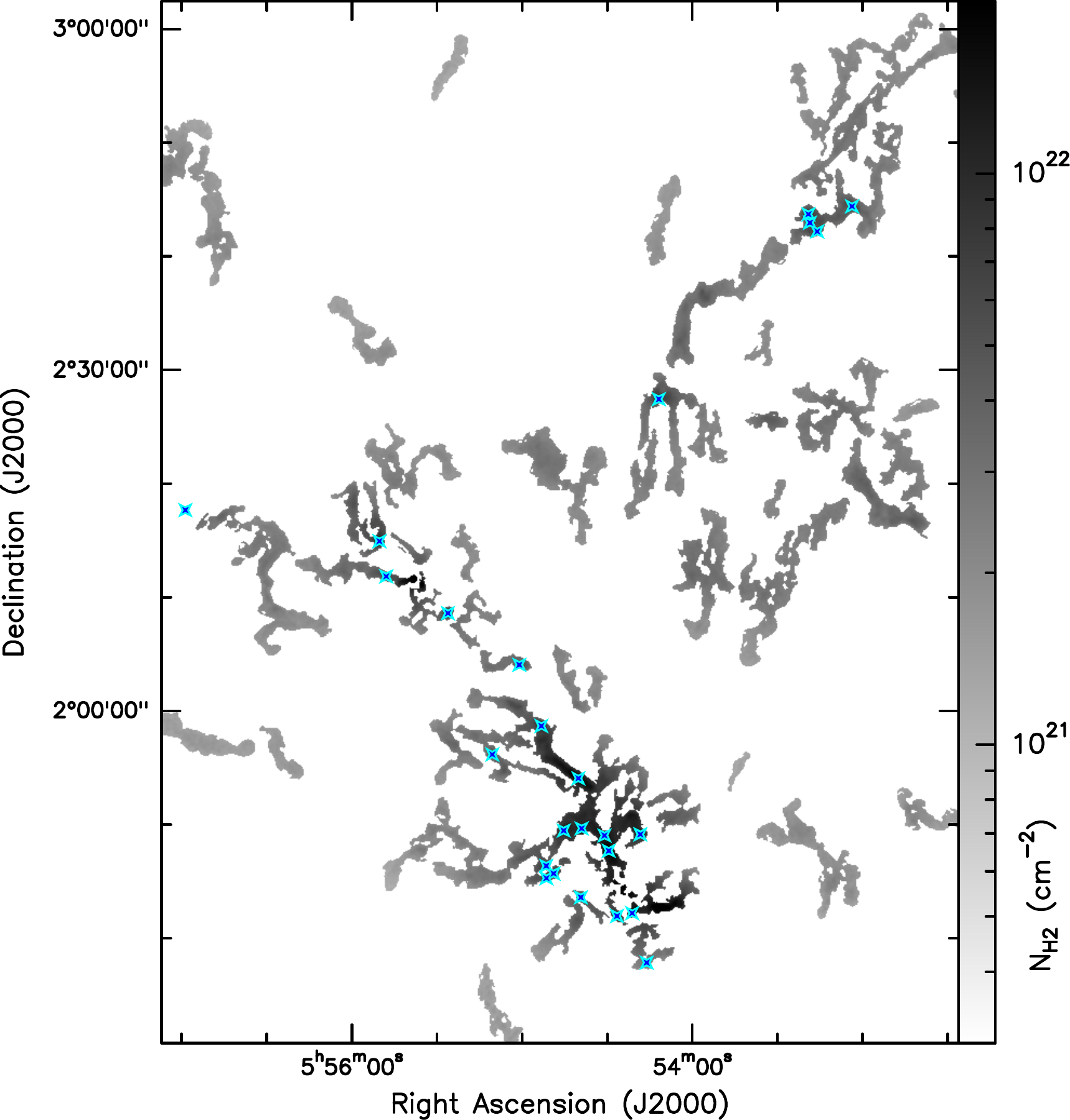}}
  \end{minipage}
   \caption{   
   L1622 cloud and its surroundings. The gray background image shows the mask of filamentary network traced by \textsl{getfilaments} \citep{Menshchikov2013}. 
   Within the \textsl{getfilaments} mask the color scale represents column density values in the column density map 
   (Fig.~\ref{fig_T_NH2_maps} left). Angular scales up to $100\arcsec$ are shown, which corresponds to $\sim$0.2~pc at $d = 400$~pc.
   {\bf Left:} Overplotted thicker blue skeletons mark the robust filament crests from the DisPerSE \citep{Sousbie2011} sample (see Sect.~\ref{sec:filam} for details). 
   The latter sample together with the thin cyan crests make the DisPerSE raw sample of filaments. 
   {\bf Right:} Overplotted blue crosses mark {\it robust} prestellar cores in the northern region.
   }
  \label{fig:fils_coresL1622}%
\end{figure*}

\begin{figure}[!!h]
 \centering
  \resizebox{1.0\hsize}{!}{\includegraphics[angle=0]{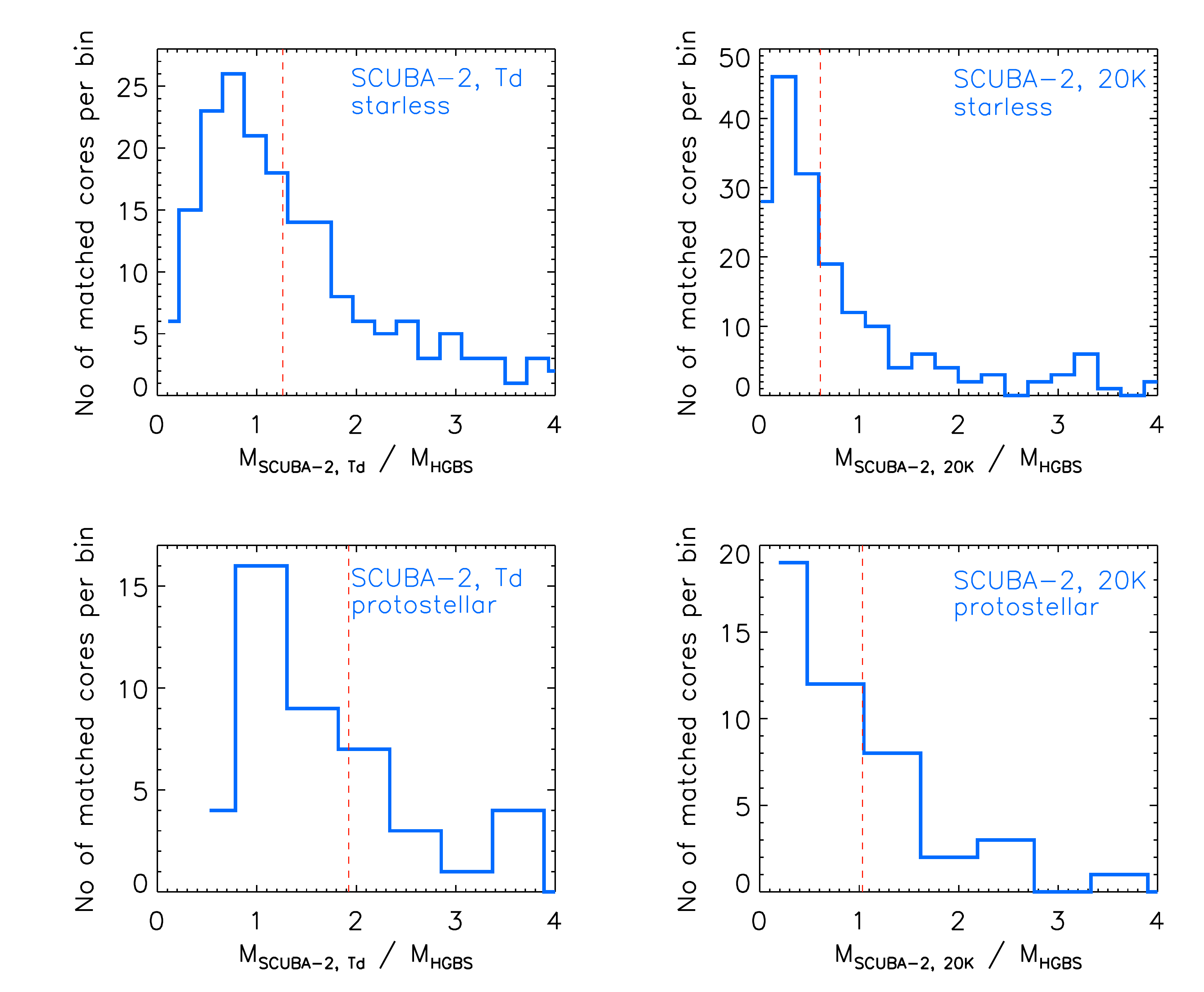}}
   \caption{Ratios of SCUBA-2 to HGBS core mass estimates for matched starless and protostellar pairs within 6\arcsec. 
   The SCUBA-2 core masses were re-derived assuming either 1) the dust temperatures estimated from {\it Herschel} SED fitting ($T_{\rm d}^{\rm SED}$), 
   or 2) a uniform dust temperature of 20~K for all cores.
   In each panel, the median ratio is marked by a red dashed line (see details in Sect.~\ref{sec:deriv_core_prop}). 
   }
  \label{fig:massComp}%
\end{figure} 

\end{appendix}

\end{document}